\documentclass[11pt]{article}
\usepackage[dvipsnames]{xcolor}
\usepackage[utf8]{inputenc}
\usepackage{tikz}

\usetikzlibrary{shapes,arrows,positioning}
\usetikzlibrary{automata,arrows,positioning,calc}
\usepackage{amsmath}
\usepackage{amssymb}
\usepackage{amsthm}

\usepackage{dutchcal}
\usepackage{hyperref}

\usepackage{bbm}
\usepackage{blkarray}
\usepackage{caption}
\usepackage{subcaption}
\usepackage{float}
\usepackage{stmaryrd}
\usepackage[normalem]{ulem}
\usepackage{enumitem}

\usepackage{fancyhdr}
\renewcommand{\headrulewidth}{0pt}
\usepackage{amsopn}

\usepackage{booktabs}

\hypersetup{colorlinks,
	citecolor=blue,
	citebordercolor=blue,
	linkcolor=blue,
	urlcolor=blue
}

\usepackage{natbib}

\usepackage{algorithmic}
\usepackage[linesnumbered,ruled,vlined]{algorithm2e}

\numberwithin{equation}{section}
\allowdisplaybreaks

\usepackage{todonotes}

\newtheorem{theorem}{Theorem}[section]
\newtheorem{remark}[theorem]{Remark}
\newtheorem{assumption}{Assumption}
\newtheorem{definition}[theorem]{Definition}

\newtheorem{proposition}[theorem]{Proposition}

\DeclareMathOperator*{\argmax}{arg\,max}

\newcommand{\EE}{\mathbb{E}}

\newcommand{\RR}{\mathbb{R}}

\usepackage{bm}
\usepackage{enumitem}
\usepackage{mathtools}

\newcommand\pj{\mathrm{Proj}}
\newcommand\opi{\bar{\bm{\pi}}_*}
\newcommand\ckpi{{\check{\bpi}_*}}
\newcommand\sa{\bar{s},\bar{a}^1\dots\bar{a}^m}

\newcommand\oa{\bar{a}_*^1,\dots,\bar{a}_*^{m}}
\newcommand\csa{\check{s},\check{a}^1,\dots,\check{a}^m}
\newcommand\ckoa{\check{a}_*^1,\dots,\check{a}_*^m}
\newcommand\psa{\pj_{\check{S}}(\bar{s}),\pj_{\check{A}^1}(\bar{a}^1),\dots,\pj_{\check{A}^m}(\bar{a}^m)}

\newcommand\poa{\pj_{\check{A}^1}(\bar{a}_*^1),\dots,\pj_{\check{A}^m}(\bar{a}_*^m)}
\newcommand\states{S}
\newcommand\actions{A}
\newcommand{\cE}{\mathcal{E}}

\newcommand{\cD}{\mathcal{D}}

\newcommand{\bpi}{{\bm{\pi}}}

\newcommand\Nash{\mathrm{Nash}}

\newcommand\defi[1]{{\textbf{#1}}}

\title{\vspace{-2.0cm}Reinforcement Learning for Finite Space Mean-Field Type Games}

\author{
    Kai Shao{$^{*,}$}\footnote{These authors contributed equally.}\footnote{Shanghai Frontiers Science Center of Artificial Intelligence and Deep Learning; NYU Shanghai, Shanghai, 200126, People’s Republic of China
		(\href{mailto:kaishao@nyu.edu}{kaishao@nyu.edu}).}
    \and Jiacheng Shen{$^{*,}$}\footnotemark[1]\footnote{NYU Center for Data Science, NY 10011, United States; NYU Shanghai,
    Shanghai, 200126, People’s Republic of China.
		(\href{mailto:shen.patrick.jiacheng@nyu.edu}{shen.patrick.jiacheng@nyu.edu}).}
	\and Mathieu Lauri\`ere\footnote{Shanghai Frontiers Science Center of Artificial Intelligence and Deep Learning; NYU-ECNU Institute of Mathematical Sciences, NYU Shanghai, Shanghai, 200126, People’s Republic of China 
		(\href{mailto:mathieu.lauriere@nyu.edu}{mathieu.lauriere@nyu.edu}).}
}
\date{}
\begin{document}
\maketitle
\begin{abstract}
Mean field type games (MFTGs) describe Nash equilibria between large coalitions: each coalition consists of a continuum of cooperative agents who maximize the average reward of their coalition while interacting non-cooperatively with a finite number of other coalitions. Although the theory has been extensively developed, we are still lacking efficient and scalable computational methods. Here, we develop reinforcement learning methods for such games in a finite space setting with general dynamics and reward functions. We start by proving that the MFTG solution yields approximate Nash equilibria in finite-size coalition games. We then propose two algorithms. The first is based on the quantization of mean-field spaces and Nash Q-learning. We provide convergence and stability analysis. We then propose a deep reinforcement learning algorithm, which can scale to larger spaces. Numerical experiments in 4 environments with mean-field distributions of dimension up to $200$ show the scalability and efficiency of the proposed method.
\end{abstract}

\section{Introduction}

Game theory has found a large number of applications, from economics and finance to biology and epidemiology. 
The most common notion of solution is the concept of Nash equilibrium, in which no agent has any incentive to deviate unilaterally~\citep{nash1951non}. At the other end of the spectrum is the concept of social optimum, in which the agents cooperate to maximize a total reward over the population. 
These notions have been extensively studied for finite-player games, see e.g.~\citep{fudenberg1991game}. Computing exactly Nash equilibria in games with a large number of players is known to be a very challenging problem~\citep{daskalakis2009complexity}.

To address this challenge, the concept of mean field games (MFGs) has been introduced in~\citep{MR2295621,MR2346927}, relying on intuitions from statistical physics. The main idea is to consider an infinite population of agents, replacing the finite population with a probability distribution, and to study the interactions between one representative player with this distribution. Under suitable conditions, the solution to an MFG provides an approximate Nash equilibrium for the corresponding finite-player game. While MFGs typically focus on the solution concept of Nash equilibrium, mean field control (MFC) problems focus on the solution concept of social optimum~\citep{bensoussan2013mean}. The theory of these two types of problems has been extensively developed, in particular using tools from stochastic analysis and partial differential equations, see e.g.~\citep{bensoussan2013mean,gomes2014mean,carmona2018probabilisticI-II} for more details. 

However, many real-world situations involve agents that are not purely cooperative or purely non-cooperative. In many scenarios, the agents form coalitions: they cooperate with agents of the same group and compete with other agents of other groups. In the limit where the number of agents is infinite while the number of coalitions remains finite, this leads to the concept of mean-field type games (MFTGs)~\citep{tembine2017mean}. Various applications have been developed, such as blockchain token economics~\citep{barreiro2019blockchain}, risk-sensitive control~\citep{tembine2015risk} or more broadly in engineering~\citep{barreiro2021meanbook,djehiche2017mean}. Similar problems have been studied under the terminology of mean field games among teams~\citep{subramanian2023mean} and team-against-team mean field problems~\citep{sanjari2023nash,yuksel2024information}. 
The case of zero-sum MFTG has received special interest~\citep{basar2021zero,cosso2019zero,guan2024zero}, but the framework of MFTGs also covers general sum games with more than two (mean-field) coalitions.
MFTGs are different from MFGs because the agents are cooperative within coalitions, while MFGs are about purely non-cooperative agents. They are also different from MFC problems, in which the agents are purely cooperative. As a consequence, computational methods and learning algorithms for MFGs and MFC problems cannot be applied to compute Nash equilibria between mean-field coalitions in MFTGs. { MFTGs are also different from multi-population MFGs and MFC problems (see~\citep[Section 3]{bensoussan2018mean}). Last, graphon games~\citep{caines2019graphon} and mixed mean field control games~\citep{angiuli2023reinforcementmixed} correspond to limit scenarios with infinitely many mean-field groups. In such games, each player has a negligible impact on the rest of the population, which is not the case in MFTGs, see~\citep{tembine2017mean}, so new methods are required for MFTGs.}

Inspired by the recent successes of RL in two-player games such as Go~\citep{silver2016mastering} and poker~\citep{brown2020combining},
RL methods have been adapted to solve MFGs and MFC problems, see e.g.~\citep{SubramanianMahajan-2018-RLstatioMFG,guo2019learning,elie2020convergence,cui2021approximately} and~\citep{gu2021meanQ,carmona2023model,angiuli2023deep} respectively, among many other references. We refer to~\citep{lauriere2022learning} and the references therein for more details. Such methods compute the solutions to mean field problems. A related topic is mean field multi-agent reinforcement learning (MFMARL)~\citep{yang2018mean}, which studies finite-agent systems and replaces the interactions between agents with the mean of neighboring agents' states and actions. Extensions include situations with multiple types and partial observation~\citep{subramanian_multi-type-MFG, subramanian_POMFGRL}.  However, the MFMARL setting differs substantially from MFTGs: (1) it does not take into account a general dependence on the mean field (i.e., the whole population distribution), (2) it aims directly for the finite-agent problem while using a mean-field approximation in an empirical way, and (3) it is not designed to tackle Nash equilibria between coalitions. The works most related to ours applied RL to continuous space linear-quadratic MFTGs by exploiting the specific structure of the equilibrium policy in these games~\citep{carmona2020policyCDC,zaman2024independent,zaman2024robust}. In these settings, policies can be represented exactly with a small number of parameters. In contrast, we focus on finite space MFTGs with general dynamics and reward functions, for which there has been no RL algorithm thus far to the best of our knowledge.

{\bf Main contributions.} 
Our main contributions are as follows: %
\begin{enumerate}[leftmargin=0.5cm,topsep=0pt,itemsep=0pt,partopsep=0pt, parsep=0pt]
    \item We prove that solving an MFTG provides an $\epsilon$-Nash equilibrium for a game between finite-size coalitions (Theorem~\ref{thm:approx-Nash}), { which justifies studying MFTGs for finite-player applications.}
    \item We propose a tabular RL method based on quantization of the mean-field spaces and Nash Q-learning~\citep{hu2003nash}. We prove the convergence of this algorithm and analyze the error due to discretization (Theorem~\ref{thm:discrete-Nash-Q}). 
    \item We propose a deep RL algorithm based on DDPG~\citep{lillicrap2016continuousDDPG}, which does not require quantization and hence is more scalable to problems with a large number of states. 
    \item We illustrate both methods in $4$ environments with distributions in dimensions up to $200$. Since this paper is the first to propose RL algorithms for (finite space) MFTGs with general dynamics and rewards, there is no standard baseline to compare with. We thus carry out a comparison with two baselines inspired by independent learning. 
\end{enumerate}

The rest of the paper is organized as follows. In Section~\ref{sec:def-model}, we define the finite-agent problem with coalitions, its mean-field limit, and establish their connection. We then reformulate the MFTG problem in the language of mean field MDPs. In Section~\ref{sec:NashQ-tabular}, we present an algorithm based on the idea of Nash Q-learning, and we analyze it. Section~\ref{sec:deepRL-algo} gives our deep RL algorithm for MFTG, without mean-field discretization. Numerical results are given in Section~\ref{sec:num-expe}. Section~\ref{sec:conclusion} is dedicated to a summary and a discussion. The appendices contain proofs and additional numerical results.

\section{Definition of the model}
\label{sec:def-model}
In this section, we define the finite-population $m$-coalition game and the limiting MFTG with $m$ (central) players. We will use the terminology \defi{agent} for an individual in a coalition and \defi{central player} for the player who chooses the policy to be used by her coalition. We will sometimes write player instead of central player.

\subsection{Finite-population $m$-coalition game}
We consider a game between $m$ groups of many agents. Each group is called a \defi{coalition} and behaves cooperatively within itself. Alternatively, we can say that there are $m$ central players, and each of them chooses the behaviors to be used in their respective coalition. For each $i \in [m]$, let $\states^i$ and $\actions^i$ be respectively the finite state space and the finite action space for the individual agents in coalition $i$. Let $N_i$ denote the number of individual agents in coalition $i$. Let $\Delta(S^i)$ and $\Delta(A^i)$ be the sets of probability distributions on $S^i$ and $A^i$, respectively. Agent $j$ in coalition $i$ has a state $x^{ij}_t$ at time $t$. The state of coalition $i$ is characterized by the empirical distribution {$\mu^{i,\bar{N}}_t = \frac{1}{N_i} \sum_{j=1}^{N_i}\delta_{x^{ij}_t} \in \Delta(\states^i)$}, and the state of the whole population is characterized by the joint empirical distribution:  $\mu^{\bar{N}}_t = (\mu^{1,\bar{N}}_t, \dots , \mu^{m,\bar{N}}_t)$.
The state of every agent $j \in [N_i]$ in coalition $i$ evolves according to a transition kernel $p^i: \states^i \times \actions^i \times \prod_{i'=1}^m \Delta(\states^{i'}) \to \Delta(\states^i)$. If the agent takes action $a^{ij}_t$ and the distribution is $\mu^{\bar{N}}_t$, then:
$x^{ij}_{t+1} \sim p^i(\cdot|x^{ij}_t, a^{ij}_t, \mu^{\bar{N}}_t)$.
We assume that the states of all agents in all coalitions are sampled independently. 
During this transition, the agent obtains a reward $r^i(x^{ij}_t, a^{ij}_t, \mu^{\bar{N}}_t)$ given by a function $r^i: \states^i \times \actions^i \times \prod_{i'=1}^m \Delta(\states^{i'}) \to \RR$. All the agents in coalition $i$ independently pick their actions according to a common policy $\pi^i: \states^i \times \Delta(S^1) \times \dots \times \Delta(S^m) \to \Delta(\actions^i)$, i.e., $a^{ij}_t$ for all $j \in [N_i]$ are i.i.d. with distribution $\pi^i(\cdot|x^{ij}_t, \mu^{\bar{N}}_t)$. Notice that the arguments include the individual state and the distribution of each coalition. We denote by $\Pi^i$ the set of such policies. The average social reward for the central player of population $i$ is defined as: $J^{i,\bar{N}}(\pi^1,\dots,\pi^m) = \frac{1}{N_i} \sum_{j=1}^{N_i} \EE[ \sum_{t \ge 0} \gamma^t r^{ij}_t]$,
where $\gamma \in [0,1)$ is a discount factor and the one-step reward at time $t$ is $r^{ij}_t = r^i_t(x^{ij}_t, a^{ij}_t, \mu^{\bar{N}}_t)$. We focus on the solution corresponding to a Nash equilibrium between the central players.

\begin{definition}[Nash equilibrium for finite-population $m$-coalition type game]
A policy profile $(\pi_*^1,\dots,\pi_*^m) \in \Pi^1\times\dots\times\Pi^m$ is a \defi{Nash equilibrium} for the above finite-population game if: for all $i \in [m]$, for all $\pi^i \in \Pi^i$,
$
    J^{i,\bar{N}}(\pi^i; \pi_*^{-i}) \le J^{i,\bar{N}}(\pi_*^i; \pi_*^{-i}),
$
where $\pi_*^{-i}$ denotes the vector of policies for central players in other coalitions except $i$.
\end{definition}
In a Nash equilibrium, there is no incentive for unilateral deviations at the coalition level.
When each $N_i$ goes to infinity, we obtain a game between $m$ central players in which each player controls a population distribution. Such games are referred to as \defi{mean-field type games} (MFTG for short).

\subsection{Mean-field type game}
Informally, as $N_i \to +\infty$, the state $\mu^{i,\bar{N}}_t$ of coalition $i$ has a limiting distribution $\mu^{i}_t \in \Delta(\states^i)$ for each $i \in [m]$, and the state $\mu^{\bar{N}}_t$ of the whole population converges to $\mu_t = (\mu^{1}_t, \dots, \mu^{m}_t) \in \Delta(\states^1) \times \dots \times \Delta(\states^m)$. We will refer to the limiting distributions as the \defi{mean-field} distributions. Based on propagation-of-chaos type results~\citep{mckean1966,sznitman1991topics}, we expect all the agents' states to evolve independently, interacting only through the mean-field distributions. It is thus sufficient to understand the behavior of one representative agent per coalition. A representative agent in mean-field coalition $i$ has a state $x^i_t \in \states^i$ which evolves according to: $x^{i}_{t+1} \sim p^i(\cdot|x^{i}_t, a^{i}_t, \mu_t)$, $a^i_t \sim \pi^i(\cdot|x^{i}_t,\mu_t)$, where $\pi^i \in \Pi^i$ is the policy for coalition $i$. We consider that this policy is chosen by a \defi{central player} and then applied by all the infinitesimal \defi{agents} in coalition $i$. %
The total reward for coalition $i$ is: $J^{i}(\pi^1,\dots,\pi^m) = \EE\Big[ \sum_{t \ge 0} \gamma^t r^i(x^{i}_t, a^{i}_t, \mu_t)\Big]$,
where, intuitively, the expectation takes into account the average over all the agents of coalition $i$. Then, the goal is to find a Nash equilibrium between the $m$ central players.
\begin{definition}[Nash equilibrium for $m$-player MFTG]
\label{def:Nash-mftg}
A policy profile $(\pi_*^1,\dots,\pi_*^m) \in \Pi^1\times\dots\times\Pi^m$ is a \defi{Nash equilibrium} for the above MFTG if: for all $i \in [m]$, for all $\pi^i \in \Pi^i$,
$
    J^{i}(\pi^i; \pi_*^{-i}) \le J^{i}(\pi_*^i; \pi_*^{-i}),
$ 
where $\pi_*^{-i}$ denotes the vector of policies for players in other coalitions except $i$.
\end{definition}
In other words, in a Nash equilibrium, the central players have no incentive to deviate unilaterally. This can also be expressed through the notion of exploitability, which quantifies to what extent a policy profile is far from being a Nash equilibrium, see~\citep{heinrich2015fictitious,perrin2020fictitious}.

\begin{definition}[Exploitability]
    The \defi{exploitability} of a policy profile $(\pi^1,\dots,\pi^m) \in \Pi^1\times\dots\times\Pi^m$ is
    $
        \cE(\pi^1,\dots,\pi^m) = \sum_{i=1}^m \cE^i(\pi^1,\dots,\pi^m),
    $
    where the $i$-th central player's exploitability is:
    $
        \cE^i(\pi^1,\dots,\pi^m) = \max_{\tilde\pi^i \in \Pi^i} J^{i}(\tilde\pi^i; \pi^{-i}) - J^{i}(\pi^i; \pi^{-i})
    $
and $\pi^{-i}$ denotes the vector of policies for central players in other coalitions except $i$.
\end{definition}
Notice that $\cE^i(\pi^1,\dots,\pi^m)$ quantifies how much player $i$ can be better off by playing an optimal policy against $\pi^{-i}$ instead of $\pi^i$. In particular $\cE(\pi^1,\dots,\pi^m) = 0$ if and only if $(\pi^1,\dots,\pi^m)$ is a Nash equilibrium for the MFTG. More generally, we will use the exploitability to quantify how far $(\pi^1,\dots,\pi^m)$ is from being a Nash equilibrium. %

The main motivation behind the MFTG is that its Nash equilibrium 
provides an approximate Nash equilibrium in the finite-population $m$-coalition game, and the quality of the approximation increases with the number of agents. In particular, we can show that solving an MFTG provides an $\epsilon$-Nash equilibrium for a game
between finite-size coalitions. 
{The following assumptions are classical in the literature on MFC and MFTGs, see e.g.~\citep{cui2024learning,guan2024zero}.}
\begin{assumption}\label{assm-lip}
{\bf (a)} For each $i\in[m]$, the reward function $r^i(x, a, \mu)$ is bounded by a constant $C_r>0$ and Lipschitz w.r.t. $\mu$ with constant $L_r$.\\
{\bf (b)} The transition probability $p(x'|x,a,\mu)$ satisfies the following Lipschitz bound:
$\|p(\cdot|x,a,\mu)-p(\cdot|x,a,\tilde{\mu})\|_{1}\le L_p d(\mu,\tilde{\mu})$
for every $x\in\states^i$, $a\in A^i$, and $\mu$, $\tilde{\mu}\in\Delta(\states^i)$.\\
{\bf (c)} The policies $\pi(a|x,\mu)$ {satisfy} the following Lipschitz bound:
$\|\pi(\cdot|x,\mu)-\pi(\cdot|x,\tilde{\mu})\|_{1}\le L_{\pi}d(\mu,\tilde{\mu})$
for every $x\in\states^i$, and $\mu$, $\tilde{\mu}\in\Delta(\states^i)$.
\end{assumption}
\begin{theorem}[Approximate Nash equilibrium]\label{thm:approx-Nash}
    Suppose that Assm.~\ref{assm-lip} holds. Let $(\pi_*^1,\dots,\pi_*^m) \in \Pi^1\times\dots\times\Pi^m$ be a Nash equilibrium for the MFTG. When the discount factor $\gamma$ satisfies 
$
    \gamma(1+L_{\pi}+L_p)<1, 
$ 
then 
     $\displaystyle \max_{\tilde\pi^i} J^{i,\bar{N}}(\tilde\pi^i; \pi_*^{-i}) \le J^{i,\bar{N}}(\pi_*^i; \pi_*^{-i}) + \varepsilon(N), 
    $
    for all $i \in [m]$, with $\varepsilon(N) = C \max_{i\in[m]}\left\{|S^i|\sqrt{|A^i|} / \sqrt{N_i}\right\}$,  where $C$ is a constant that depends on the Lipschitz constants and the bound on the reward in Assm.~\ref{assm-lip}. %
\end{theorem}
In other words, if all the agents use the policy coming from the MFTG corresponding to their coalition, then each coalition can increase its total reward only marginally (at least when the number of agents is large enough).  In contrast with e.g.~\cite[Theorem 4.1]{saldi2018markov}, our result provides not only asymptotic convergence but also a rate of convergence.

\begin{proof}(sketch)
The proof consists of three main steps. First, we show that the distance between $\mu_{t}$ and $\mu^{\bar{N}}_{t}$ 
 for any $t\ge0$ can be controlled by the distance between $\mu_{0}$ and $\mu^{\bar{N}}_{0}$, and $\max_{i\in[m]}\left\{|S^i|\sqrt{|A^i|} / \sqrt{N_i}\right\}$, using the idea of propagation of error and analyzing the state-action distribution. Second, based on the Lipschitz conditions, we use the derived bound to control $|J^{i,\bar{N}}(\pi^1,\dots,\pi^m) -J^{i}(\pi^1,\dots,\pi^m)|$. Lastly, we prove the approximated Nash equilibrium by rewriting $\max_{\tilde\pi^i} J^{i,\bar{N}}(\tilde\pi^i; \pi_*^{-i})-J^{i,\bar{N}}(\pi_*^i; \pi_*^{-i})$ and the triangle inequality. More details of the proof are provided in Suppl.~\ref{app:approx-Nash-finite}.
\end{proof}

\subsection{Reformulation with Mean-Field MDPs}

Our next step towards RL methods is to rephrase the MFTG in the framework of Markov decision processes (MDPs). Since the game involves the population's states represented by probability distributions, the MDPs will be of mean-field type. We will thus rely on the framework of mean-field Markov decision processes (MFMDP)~\citep{motte2022mean,carmona2023model}. But in contrast with these prior works, we consider a game between MFMDPs, which is more challenging than a single MFMDP.
The key remark is that, since $x^i_t$ has distribution $\mu^i_t$ and $a^i_t$ has distribution $\pi^i(\cdot|x^i_t,\mu_t)$, the expected one-step reward can be expressed as a function $\bar{r}^i$ of the $i$-th policy and the distributions: 
\[
    \bar{r}^i(\mu_t, \bar{a}^i_t) = \sum_{x \in \states^i} \mu^i_t(x) \sum_{a \in \actions^i} \bar{a}^i_t(a|x) r^i(x, a, \mu_t),
\]
where $\bar{a}^i_t = \pi^i_t(\cdot|\cdot,\mu_t)$. 
This will help us to rewrite the problem posed to the central player $i$, as an MDP. Before doing so, we introduce the following notations:
    $\bar{S} = \bigtimes_{i=1}^m\bar S^i$ is the (mean-field) state space, where $\bar{S}^i = \Delta(\states^i)$ is the (mean-field) state space of population $i$. The (mean-field) state is $\bar{s}_t = \mu_t \in \bar{S}$; 
    $\bar{A}^i=\Delta(\actions^i)^{|\states^i|}$ is the (mean-field) action space; 
    $\bar{r}^i: \bar{S} \times \bar{A}^i \to \RR$ is as defined above; 
    $\bar{p}: \bar{S} \times \bar{A}^1 \times \dots \times \bar{A}^m \to \bar{S}$ is defined such that: $\bar{p}(\bar{s}_t, \bar{a}^1_t,\dots,\bar{a}^m_t) = \bar{s}_{t+1}$ where, if $\bar{s}_t = (\mu^1_t,\dots,\mu^m_t)$ and $\bar{a}^i_t = \pi_t^i(\cdot|\cdot,\mu_t)$, then $\bar{s}_{t+1} = (\mu^{1}_{t+1},\dots,\mu^{m}_{t+1})$, where we recall that $\mu^{i}_{t+1}$ is the distribution of $x^i_{t+1}$. In other words, $\bar{p}$ encodes the transitions of the mean-field state, which depends on all the central players' (mean-field) actions. To stress the fact that the transitions are deterministic, we will sometimes use the notation $\bar{F} = \bar{p}$ to stress that this is a transition function (at the mean-field level). %
A \defi{(mean-field) policy} is now a function $\bar\pi^i: \bar{S} \to \bar{A}^i$. In other words, the central player first chooses a function $\bar\pi^i$ of the mean field. When applied on $\mu_t$, $\bar\pi^i(\mu_t)$ returns a policy for the individual agent, i.e., $\bar\pi^i(\mu_t)=\bar{a}^i(\mu_t): \states^i \ni x^i_t \mapsto \bar{a}^i(\mu_t)(\cdot|x_t) = \pi^i(\cdot|x^i_t,\mu_t) \in \Delta(\actions^i)$.
Although this approach may seem quite abstract, it allows us to view the problem posed to the $i$-th central player as a ``classical'' MDP (modulo the fact that the state consists of the distributions of all coalitions). We can then borrow tools from reinforcement learning to solve this MDP.  

\begin{remark}
    Notice that an action for central player $i$, i.e., an element $\bar{a}^i$ of $\bar{A}^i$. From the point of view of an agent in coalition $i$, it is a decentralized policy. Then $\bar\pi^i$ is a mean-field policy for the central player, whose input is a mean field. This generalizes the approach proposed in~\citep{carmona2023model} to the case of multiple controllers. It is different from, e.g.~\citep{yang2018mean}, in which there is no central player and no mean-field policies. This allows us to represent the behaviors of coalitions that react to the mean fields of other coalitions.
\end{remark}

\subsection{Stage game equilibria}

We now rephrase the notion of MFTG equilibrium using the value function, which will lead to a connection with the concept of the stage game. To make the model more general, we also assume that the reward of coalition $i$ could be a function of the actions of all central players.

The central player of coalition $i$ aims to choose a policy $\bar\pi^i$ to maximize the discounted sum of reward: 
    $\bar{v}^i_{\bar{\bpi}}(\bar{s})= \bar{v}^i(\bar{s}, \bar{\bpi})
    \coloneqq
    \EE_{\bm{\bar{\pi}}}\Big[\sum^{\infty}_{t=0}\mathbb \gamma^t \bar{r}^i(\bar s_t,\bar a^i_t)\Big]$,
where $\bm{\bar{\pi}}=(\bar\pi^1,\dots,\bar\pi^m)$ is the policy profile and $\bar s_0 = \bar s$, $\bar s_{t+1}\sim\bar p(\cdot|\bar s_t, \bar a^1_t,\dots,\bar a^m_t)$, $\bar a^j_t\sim\bar\pi^j(\cdot|\bar s_t)$, $j=1,\dots,m,\ t\ge0$.

We can now rephrase the notion of Nash equilibrium for the MFTG (Def.~\ref{def:Nash-mftg}) in this framework. %
\begin{definition}[Nash equilibrium for MFTG rephrased]
An MFTG \defi{Nash equilibrium} $\bar{\bpi}_*=(\bar\pi_*^1,\dots, \bar\pi_*^m)$ is a policy profile such that for all $i=1,\dots,m$, we have
    $
    \bar{v}^i(\bar{s}, \bar{\bpi}_*)\ge \bar{v}^i(\bar{s}, (\bar{\pi}^i, \bar{\bpi}_*^{-i})),$ $\forall \bar s\in\bar S, \forall \bar{\pi}^i\in\bar\Pi^i.
    $

\end{definition}
To simplify the notation, we let $\bar{\bm a} = (\bar{a}^1,\dots,\bar{a}^{m})$, $\bar{\bpi}^{-i}(\mathrm{d}\bar{\bm a}^{-i}| \bar{s}) = \prod_{j \neq i}\bar{\pi}^j(\mathrm{d}\bar{a}^j|\bar{s})$, $\bar{\bm a}^{-i} \in \bar{A}^{-i}=\prod_{j \neq i}\bar{A}^j$.
The Q-function for central player $i$ is defined as:\linebreak
$\bar Q^i_{\bm{\bar{\pi}}}(\bar s, \bar{\bm a})=\mathbb E_{\bm{\bar{\pi}}}\Big[{\sum_{t=0}^{\infty}}\gamma^t\bar{r}^i(\bar{s}_t, \bar{a}_t^i)|\bar{s}_0=\bar{s},\bar{\bm a}_0=\bar{\bm a}\Big].
$ We now introduce an MFMDP for the central player $i$ when the policies of the other players are fixed. We define the following MDP, denoted by MDP($\bar{\bpi}^{-i}$).
\begin{definition}[MDP($\bar{\bpi}^{-i}$)]
An MDP for a central player $i$ against fixed policies of other players is a tuple 
$(\bar{S},\bar{A}^i, \bar{p}_{\bar\bpi^{-i}},\bar{r}_{\bar\bpi^{-i}}, \gamma)$,
where
$\displaystyle \bar{p}_{\bar\bpi^{-i}}(\bar{s}'|\bar{s},\bar{a}^i)
    =\int_{\bar{A}^{-i}}\bar{p}(\bar{s}'|\bar{s},\bar{\bm a}) \bar{\bpi}^{-i}(\mathrm{d}\bar{\bm a}^{-i}| \bar{s})$,$\quad\bar{r}_{\bar\bpi^{-i}}(\bar{s},\bar{a}^i)
    =\bar{r}^i(\bar{s},\bar{a}^i).
$

\end{definition}

Next, we define the notion of a stage game, which is a Nash equilibrium for a one-step game. This serves as an intermediate goal in Nash Q-learning, to learn a global-in-time Nash equilibrium.
\begin{definition}[Stage game and stage Nash equilibrium]
\label{defi:stage-game-Nash}
Given a (mean-field) state $\bar{s} \in \bar{\states}$ and a policy profile $\bar{\bpi}=(\bar{\pi}^1,\dots,\bar{\pi}^m)$, the (mean-field) \defi{stage game} induced by $\bar{s}$ and $\bar\bpi$ is a static game in which the player $i$ takes an action $\bar{a}^i \in \bar{\actions}^i$, $i=1,\dots,m$ and gets the reward $\bar{Q}^i_{\bar{\bpi}}(\bar{s}, \bar{a}^1,\dots,\bar{a}^m)$. Player $i$ is allowed to use a mixed strategy $\sigma^i \in \Delta(\bar{A}^i)$. 
A \defi{Nash equilibrium} for this stage game is a strategy profile $\bm\sigma_* = (\sigma_*^1, \dots, \sigma_*^m)$ such that, for all $\sigma^i\in\Delta(\bar\actions^i)$,
\begin{align*}
    &\sigma_*^1\cdots\sigma_*^m\bar{Q}^i_{\bar{\bpi}}(\bar{s})
    \ge \sigma_*^1\cdots\sigma_*^{i-1}\sigma^i\sigma_*^{i+1}\cdots\sigma_*^{m}\bar{Q}^i_{\bar{\bpi}}(\bar{s})
\end{align*}
where we define
$
    \displaystyle \sigma^1\cdots\sigma^m\bar{Q}^i_{\bar{\bpi}}(\bar{s})
    \coloneqq\bar{r}^i(\bar{s},\sigma^i)
    +\gamma\int_{\bar{S}}\int_{\bar{\bm\actions}}\bar{v}^i(\bar{s}',\bar{\bpi})\bar{p}(\mathrm{d}\bar{s}'|\bar{s},\bm{\bar{a}})\bm{\sigma}(\mathrm{d}\bm{\bar{a}}|\bar{s}), 
$
with $\bar{\bm \actions} \coloneqq \bar\actions^1\times\dots\times\bar\actions^m$, $\bm{\sigma}(\mathrm{d}\bm{\bar{a}}|\bar{s}) \coloneqq \prod_{i=1}^m \sigma^i(\mathrm{d}\bar{a}^i|\bar{s})$, 
and 
$\bar{r}^i(\bar{s},\sigma^i)\coloneqq\mathbb{E}_{\bar{a}^i\sim\sigma^i}\bar{r}^i(\bar{s},\bar{a}^i)$.
\end{definition}

We now define a mean-field version of the NashQ function introduced by~\cite{hu2003nash}. Intuitively, it quantifies the reward that player $i$ receives when the system starts in a given state, all players use the equilibrium strategies of the stage game for the first action, and then play according to a fixed policy profile for all remaining time steps. 
\begin{definition}[NashQ function]
Given a Nash equilibrium $(\sigma_*^1,\dots,\sigma_*^m)$, the \defi{NashQ function} of player $i$ is defined as:
$
    \Nash\bar{Q}^i_{\bar{\bpi}}(\bar{s})
    \coloneqq\sigma_*^1\cdots\sigma_*^m\bar{Q}^i_{\bar{\bpi}}(\bar{s}). 
$
\end{definition}

We conclude by showing the link between Defs.~\ref{def:Nash-mftg} and~\ref{defi:stage-game-Nash} (the proof is in Suppl.~\ref{app:equiv-Nash-stage}).
\begin{proposition}\label{prop:Nash-equiv-stage}
    The following statements are equivalent:
{\bf (i)} $\bar{\bpi}_* =(\bar\pi_*^1,\dots, \bar\pi_*^m)$ is a Nash equilibrium of the MFTG with the equilibrium payoff $(\bar{{v}}^1_{\bar{\bpi}_*},\dots,\bar{{v}}^m_{\bar{\bpi}_*})$;
    {\bf (ii)} For every $\bar{s}\in\bar{S}$, $(\bar{\pi}_*^1(\bar{s}),\dots,\bar{\pi}_*^m(\bar{s}))$ is a Nash equilibrium in the stage game induced by the state $\bar{s}$ and the policy profile $\bar{\bpi}_*$.
    
\end{proposition}

\section{Nash Q-learning and Tabular Implementation} 
\label{sec:NashQ-tabular}

In this section, we present an adaptation of the Nash Q-learning of~\cite{hu2003nash} to solve MFTGs. It should be noted that the original algorithm in~\citep{hu2003nash} is for finite state and action spaces, and to the best of our knowledge, extensions to continuous spaces have been proposed only in special cases, such as~\cite{vamvoudakis2015non,casgrain2022deep}. Still,  there is no extension to continuous spaces for general games that could be applied to MFTGs. The main difficulty lies in computing the solution to the stage game at each iteration, which relies on the fact that the action space is finite.  So, this algorithm cannot be applied directly to solve MFTGs. 

In order to implement this method using tabular RL, we will start by discretizing the simplexes following the idea in~\citep{carmona2023model}. This allows us to fully analyze the algorithm. However, this approach is not scalable in terms of the number of states, which is why in Section~\ref{sec:deepRL-algo}, we will present a deep RL method that does not require simplex discretization.

\subsection{Discretized MFTG}

Since $S^i$ and $A^i$ are finite, $\bar S^i = \Delta(S^i)$ and $\Delta(A^i)$ are (finite-dimensional) simplexes. We endow $\bar S$ and $\Delta(A^i)$ with the distances 
$
    d_{\bar{S}}(\bar{s},\bar{s}')=\sum_{i\in[m]}d(\bar{s}^i,{\bar{s'}}^i)=\sum_{i\in[m]}\sum_{x\in S^i}|\mu^i(x)-{\mu'}^i(x)|,
$
    and 
$ 
    d_{A^i}(\bar{a}^i(\bar{s}), \bar{a'}^i(\bar{s}))=\sum_{x,a}|\pi^i(a|x,\bar{s})-{\pi'}^i(a|x,\bar{s})|,
$
where $\bar{s}^i=\mu^i$, $\bar{a}^i(\bar{s})=\pi^i(\cdot|\cdot, \bar{s})$. In the action space $\bar{A}^i$, we define the distance $d_{\bar{A}^i}(\bar{a}^i, \bar{a'}^i)=\sup_{\bar{s}\in\bar{S}}d_{A^i}(\bar{a}^i(\bar{s}), \bar{a'}^i(\bar{s}))$.  However, $\bar S$ and $\bar A^i$ are not finite. To apply the tabular Q-learning algorithm, we replace $\bar S$ and $\bar A^i$ with finite sets. For $i=1,\dots,m$, let $\check S^i\subset\bar S^i$ and $\check\Delta(A^i)\subset\Delta(A^i)$ be finite approximations of $\bar{S}^i$ and $\Delta(A^i)$. 
We then define the (mean-field) \defi{finite state space and action space}  $\check S=\Pi_{i=1}^m\check S^i\subset\bar{S}$ and $\check{A}^i=\check\Delta(A^i)^{|S_i|}$. Let $\epsilon_{S} = \max_{\bar{s} \in \bar{S}} \min_{\check{s} \in \check{S}} d_{\bar S}(\bar s, \check s)$ and $\epsilon_{A} = \max_{i} \max_{\bar{a}^i \in \bar{A}^i} \min_{\check{a}^i \in \check{A}^i} d_{\bar{A}^i}(\bar{a}^i,\check {a}^i)$,
which characterize the fineness of the discretization.The policy space of each player $i$ is $\check{\Pi}^i=\{\check{\pi}^i:\check{S}\rightarrow\Delta({\check A}^i)\}$. 
We will also use the projection operator $\pj_{\check{S}}:\bar{S}\rightarrow\check{S}$, which maps $\bar{s}$ to the closest point in $\check{S}$ (ties broken arbitrarily). 
This will ensure that the state takes value in $\check{S}$. Specifically, given a state $\check{s}_t$ and a joint action $(\check a^1_t,\dots,\check a^m_t)$, we generate $\bar s_{t+1}=\bar{F}(\check s_t, \check a^1_t,\dots\check a^m_t)$. Then, we project $\bar{s}_{t+1}$ back to $\check{S}$ and denote the projected state by $\check{s}_{t+1}=\pj_{\check{S}}(\bar{s}_{t+1})$.  This finite space setting can be regarded as a special case of a stochastic game of $m$ players, and Theorem 2 in \citep{fink1964equilibrium}  guarantees the existence of a Nash equilibrium.
\subsection{Nash Q-learning algorithm} 
We briefly describe the tabular Nash Q-learning algorithm, which is similar to the algorithm presented in~\citep{hu2003nash}. The main idea is that, instead of using classical $Q$-learning updates, which involve only the player's own $Q$-function, the players will use the $\Nash Q$ function for a stage game. 

At each step $t$, the players use their current estimate of the $Q$-functions to define a stage game. They compute the Nash equilibrium, say $(\check{\sigma}^1,\dots, \check{\sigma}^m) \in \prod_{i=1}^m\check\Pi^i$, and deduce the associated $\Nash Q$ function, which is then used to update their estimates of the $Q$-functions. Namely, at each step $t$, player $i$ observes $\check{s}$ and takes an action according to a behavior policy chosen to ensure exploration. Then, she observes the reward, actions of each player, and the next state $\check{s}'$. She then solves the stage game with rewards $(\check{Q}^1_t(\check{s}'),\dots, \check{Q}^m_t(\check{s}'))$, %
where $\check{Q}^i_t(\check{s}'): (\bar{a}^1,\dots,\bar{a}^m) \mapsto \check{Q}^i_t(\check{s}',\bar{a}^1,\dots,\bar{a}^m)$. Let $(\check{\pi}^{i,1}_{*}(\check{s}'),\dots, \check{\pi}^{i,m}_{*}(\check{s}'))$ be the Nash equilibrium obtained on player $i$'s belief. The NashQ function of player $i$ is defined as:
$\Nash\check{Q}^i_t(\check{s}')=\check{\pi}_{*}^{i,1}\cdots\check{\pi}^{i,m}_{*}\check{Q}^i_t(\check{s}')$.
From here, she updates the Q-values according {to the following rule, where $\alpha_t$ is a learning rate}: 
\begin{equation}\label{eq:NashQ-updates}
    \check{Q}^i_{t+1}(\check{s},\check a^1,\dots, \check a^m)
    = (1-\alpha_t)\check{Q}^i_t(\check{s},\check a^1,\dots, \check a^m)+\alpha_t(\bar{r}^i_t+\beta \Nash\check{Q}^i_t(\check{s}')).
\end{equation}
It is noted that in each iteration, the Q-values of each player are updated asynchronously based on the observation. The detailed algorithm is described in Suppl.~\ref{app:nashq-algo}, Algo.~\ref{alg:nashq-main-algo}.

\subsection{Nash Q-learning analysis} 
We will see that $\check{Q}^i_{t}$ from Algo.~\ref{alg:nashq-main-algo} converges to $\check Q^i_{\ckpi}$ %
under the following assumption, which is classical in the literature on NashQ-learning, see e.g.~\cite{hu2003nash,yang2018mean}. We use it for the proof, although it seems that in practice the algorithm works well even when this assumption does not hold.
\begin{assumption}\label{assm-NashQ}
{\bf (a)}    Every state $\check{s}\in\check{S}$ and action $\check a^i\in\check A^i$, $i=1,\dots,m$, are visited infinitely often.\\
{\bf (b)}    $\alpha_t$ satisfies the following two conditions for all $t, \check{s}, \check a^1,\dots, \check a^m$:
    {\bf 1.} $0\le\alpha_t(\check{s}, \check a^1,\dots, \check a^m)<1$, $\sum_{t=0}^{\infty}\alpha_t(\check{s}, \check a^1,\dots, \check a^m)=\infty$, $\sum_{t=0}^{\infty}\alpha_t^2(\check{s}, \check a^1,\dots, \check a^m)<\infty$, the latter two hold uniformly and with probability 1.
    {\bf 2.} $\alpha_t(\check{s}, \check a^1,\dots, \check a^m)=0$, if $(\check{s}, \check a^1,\dots, \check a^m)\ne (\check{s}_t, \check a^1_t,\dots, \check a^m_t)$.\\
{\bf (c)}    One of the following two conditions holds:
    {\bf 1.} Every stage game $(\check{Q}^1_t(\check{s}'),\dots, \check{Q}^m_t(\check{s}'))$ for all $t$ and $\check{s}$, has a global optimal point, and players' payoff in this equilibrium are used to update their Q-functions.
    {\bf 2.} Every stage game $(\check{Q}^1_t(\check{s}'),\dots,\check{Q}^m_t(\check{s}'))$ for all $t$ and $\check{s}$, has a saddle point, and players' payoff in this equilibrium are used to update their Q-functions.
\end{assumption}

    Here, a \defi{global optimal point} is a joint policy of the stage game such that each player receives her highest reward following this policy. A \defi{saddle point} is a Nash equilibrium policy of the stage game such that each player would receive a higher reward if at least one of the other players takes a policy different from the Nash equilibrium policy.

\begin{theorem}[NashQ-learning convergence]\label{thm:nashq-convergence}
    Under Assm.~\ref{assm-NashQ}, $\check{Q}_t=(\check{Q}^1_t,\dots, \check{Q}^m_t)$, updated by~(\ref{eq:NashQ-updates})  
    converges to the Nash equilibrium Q-functions %
    $\check{Q}_{\check{\bm{\pi}}_*}=(\check{Q}^1_{\check{\bm{\pi}}_*},\dots, \check{Q}^m_{\check{\bm{\pi}}_*})$.
\end{theorem}

We omit the proof of Theorem \ref{thm:nashq-convergence} as it is essentially the same as in \citep{hu2003nash}. We then focus on the difference between the approximated Nash Q-function,\linebreak $\check Q^i_t(\psa)$ and the true Nash Q-function, $\bar{Q}^i_{\opi}(\sa)$, in the infinite space $\bar S\times\bar{A}^i\times\cdots\times\bar{A}^m$. For this proof, we use the following assumption, which is a multi-player version of the assumptions in~\citep{carmona2023model}.
\begin{assumption}\label{assm-mf-lip}
{\bf (a)}    For each $i$, {$\bar{r}^i$} is bounded and Lipschitz continuous w.r.t. $(\bar{s}_t, \bar{a}^i_t)$ with constant $L_{\bar{r}^i}$. 
    {$\bar{F}$} is Lipschitz continuous w.r.t. $(\bar{s},\bar{a}^1,\dots,\bar{a}^m)$ with constant $L_{\bar{F}}$ in expectation. \\
{\bf (b)}    $\bar{v}^i_{\bar{\bm{\pi}}}$ is Lipschitz continuous w.r.t. $\bar{s}$ with constant $L_{\bar{v}_{\bar{\bm{\pi}}}}$. 
\end{assumption}
Assm.~\ref{assm-mf-lip}~{\bf (a)} can be achieved with suitable conditions on the game. The boundedness of the reward function, together with the discount factor $0<\gamma<1$, can also lead to the boundedness of the payoff function $\bar{v}^i_{\bar{\bm{\pi}}_*}$. {For classical MDPs, Lipschitz continuity of the value function can be derived from assumptions on the model as in~\citep{motte2022mean}.}

To alleviate the notation, we let:  $\pj(\sa) = (\psa).$

\begin{theorem}[Discrete problem analysis]\label{thm:discrete-Nash-Q}
    Let $\epsilon>0$. Suppose Assm.~\ref{assm-mf-lip} holds and there is a unique pure policy $\opi^p$ such that $\opi^p$ is a global optimal point for the stage game $\bar{Q}^i_{\opi^p}(\bar{s})$ for each $i=1,2,\dotsm, m$ and $\bar{s}\in\bar{S}$. Then, if $t$ is large enough, for each $i$, $\bar s\in\bar{S}$, we have
    $
        |\check Q^i_t(\pj(\sa))-\bar{Q}^i_{\opi^p}(\sa)|\le\epsilon',
    $
    where 
    $
        \epsilon'=\epsilon+C_1\epsilon_{A}+C_2\epsilon_S,
    $
    with 
    $\epsilon_{S}$ and $\epsilon_A$ defined above, respectively, $C_1 = \frac{1}{1-\gamma}(L_{\bar{r}^i}+\gamma L_{\bar{v}^i_{\opi}}L_{\bar F}m)$ and $C_2 = \frac{\gamma}{1-\gamma}L_{\bar{v}^i_{\opi}}+L_{\bar{r}^i}+\gamma L_{\bar{v}^i_{\opi}}L_{\bar F}$.
\end{theorem}
Note that the first $\epsilon$ in the bound $\epsilon'$ can be arbitrarily chosen small, provided that $t$ is large enough. The second and third terms are controlled by $\epsilon_A$ and $\epsilon_S$ and can be small if we choose a finer simplex approximation. The proof is provided in Suppl.~\ref{sec:sensitivity-nashQ-learning}. %

\section{Deep RL for MFTG} 
\label{sec:deepRL-algo}

While the above extension of the NashQ learning algorithm has the advantage of being fully analyzable and enjoying convergence guarantees, it is not scalable to large state and action spaces. Indeed, it requires discretizing the simplexes of distributions on states and actions. The number of points increases exponentially with the number of states and actions, making the algorithm intractable for very fine discretizations. Furthermore, each step relies on solving a stage game, and computing a Nash equilibrium is a difficult task for large games, even if they are static.

For this reason, we now present a deep RL algorithm whose main advantages are that it does not require discretizing the simplexes and does not require solving any stage game. The state and action distributions are represented as vectors (containing the probability mass functions) and passed as inputs to neural networks for the policies and the value functions. At the level of the central player for coalition $i$, an action is an element $\bar{a}^i\in\bar{A}^i$. The input $\bar{s}$ is a simplex that represents the distribution of the population over the finite-state space. Although it corresponds to a mixed policy at the individual agent level, it represents a single action for the central player. We focus on learning deterministic central policies, which are functions that map a mean-field state $\bar{s}$ to a mean-field action $\bar{a}^i$. To this end, we use a variant of the deep deterministic policy gradient algorithm (DDPG)~\citep{lillicrap2016continuousDDPG}, as shown in Algo.~\ref{alg:ddpg-mftg-main}. Our algorithm differs substantially from the DDPG algorithm, as the behaviors of the two players are coupled. Each player interacts with a dynamic environment that is also influenced by the other player.  Unlike the tabular Nash Q-learning algorithm, it is generally difficult to have a rigorous proof of convergence due to the complexity of deep neural networks. Although the theoretical convergence of some algorithms has been studied, such as deep Q-learning~\citep{fan2020theoretical}, deterministic policy gradient \citep{xiong2022dpg}, and actor-critic algorithms with multilayer neural networks~\citep{tian2024ac}, to the best of our knowledge, the convergence of DDPG under assumptions that could be applied to our setting has not been established. Also, in the case of MFTGs, we would need to analyze whether the solution converges to a Nash equilibrium, which is more complex than solving an MDP.  Therefore, we leave the theoretical analysis for future work and focus on the numerical analysis. We use several numerical metrics to measure the performance of DDPG-MFTG Algo.~\ref{alg:ddpg-mftg-main}, as detailed in the next section.

\begin{algorithm}[]
\caption{DDPG for MFTG}
\label{alg:ddpg-mftg-main}
\begin{algorithmic}[1]
\STATE \textbf{Inputs:} A number of episodes $N$; a length $T$ for each episode; a minibatch size $N_{\text{batch}}$; \linebreak a learning rate $\tau$.
\STATE \textbf{Outputs:} Policy functions for each central player represented by $\pi^i_{\omega_i}$.
\STATE Initialize parameters $\theta_i$ and $\omega_i$ for critic networks $Q^i_{\theta_i}$ and actor networks $\pi^i_{\omega_i}$,\linebreak $i=1,...,m$
\STATE Initialize $\theta'_i\gets\theta_i$ and $\omega'_i\gets\omega_i$ for target networks ${Q^i}'_{\theta'_i}$ and ${\pi^i}'_{\omega'_i}$, \linebreak $i=1,...,m$
\STATE Initialize replay buffer $R_{\text{buffer}}$
\FOR{$k=0,1,...,N-1$}
  \STATE Initialize distribution $\bar s_0$ 
  \FOR{$t=0,1,\dots,T-1$}
  \STATE Select actions $\bar{a}^i_t=\pi^i_{\omega_i}(\bar{s}_t)+\epsilon_t$, where $\epsilon_t$ is the exploration noise, for $i=1,...,m$
  \STATE Execute $\bar{a}^i_t$, observe reward $\bar{r}^i(\bar{s}_t,\bar{a}^i_t)$, for $i=1,...,m$
  \STATE Observe $\bar{s}_{t+1}$
  \STATE Store transition $(\bar{s}_t,\bar{a}^1_t,...,\bar{a}^m_t,\bar{r}_t^1,...,\bar{r}_t^m,\bar{s}_{t+1})$ in $R_{\text{buffer}}$
  \STATE Sample a random minibatch of $N_{\text{batch}}$ transitions $(\bar{s}_j,\bar{a}^1_j,...,\bar{a}^m_j,\bar{r}_j^1,...,\bar{r}_j^m,\bar{s}_{j+1})$ \linebreak from $R_{\text{buffer}}$
  \STATE Set $y^i_j=\bar{r}_j^i+\gamma {Q^i}'_{\theta'_i}(\bar{s}_{j+1},{\pi^i}'_{\omega'_i}(\bar{s}_{j+1}))$  for $i=1,...,m$, $j=1,...,N_{\text{batch}}$
  \STATE Update the critic networks by minimizing the loss: $L^i(\theta_i)=\frac{1}{N_{\text{batch}}}\sum_{j}(y^i_j-Q^i_{\theta_i}(\bar{s}_j,\bar{a}^i_j))^2$, for $i=1,...,m$
  \STATE Update the actor policies using the sampled policy gradients $\nabla_{\omega_i}v^i$, for $i=1,...,m$:
  $$\nabla_{\omega_i}v^i(\omega_i)\approx\frac{1}{N_{\text{batch}}}\sum_j\nabla_{\bar{a}^i}Q^i_{\theta_i}(\bar{s}_j,\pi^i_{\omega_i}(\bar{s}_j))\nabla_{\omega_i}\pi^i_{\omega_i}(\bar{s}_j)$$
  \STATE Update target networks:  $\theta'_i\gets\tau\theta_i+(1-\tau)\theta'_i$, $\omega'_i\gets\tau\omega_i+(1-\tau)(\omega'_i)$, \linebreak for $i=1,...,m$.
  \ENDFOR
  \ENDFOR
\end{algorithmic}
\end{algorithm}

\section{Numerical experiments}
\label{sec:num-expe}
{\bf Metrics.}
To assess the convergence of our algorithms, we use several metrics. First, we check the testing rewards of each central player (i.e., the total reward for each coalition, averaged over the testing set of initial distributions). But this is not sufficient to show that the policies form a Nash equilibrium of the MFTG. For this, we compute the exploitability. {This requires training a best response (BR) policy for each player independently,} which is also done with deep RL, using the DDPG method. Our experiments for hyperparameter sweeping in Suppl.~\ref{sec:sweeps} show that the DDPG provides a reasonable approximation for the best-response policy. Lastly, we also check the evolution of the distributions to ensure that they align with our expectations for the Nash equilibrium. The pseudo-codes for evaluating a policy profile and computing the exploitability are respectively provided in Algs.~\ref{algo:policy-eval} and~\ref{algo:exploitability-comp general} in Suppl.~\ref{app:metrics-pseudocodes}. 
\\
{\bf Training and testing sets.}
The training set consists of randomly generated tuples of distributions, and each element of the tuple represents the initial distribution of a player.
The testing set consists of a finite number of tuples of distributions that are not in the training set. Details of the training and testing sets are described on a case-by-case basis.
\\
{\bf Baseline.} To the best of our knowledge, there are no RL algorithms that can be applied to the type of MFTG problems we study here. In the absence of standard baselines, we will use two types of baselines for each of our algorithms. For small-scale examples, we discretize the mean-field state and action spaces and employ DNashQ-MFTG. Here, we use an algorithm where each coalition runs an independent mean-field type Q-learning (after suitable discretization of the simplexes) as a baseline. We call this method Independent Learning-Mean Field Type Game (IL-MFTG, for short, explained in Appendix~\ref{explain: IL-MFTG}). For larger-scale examples with many states, the baseline is an ablated DDPG method in which each central player can only see her own (mean-field) state; i.e., the states of the other players are masked. For both our algorithms and the baselines, the exploitability is computed using our original class of policies, see Algo.~\ref{algo:exploitability-comp general}. \\
{\bf Games. } We present here 3 examples. One more is presented in Suppl.~\ref{app:ex2-distrib-plan}. 
Table~\ref{table:improvements} in Suppl. summarizes the average improvements obtained by our method (at least $30\%$ in each game). %

\begin{figure}[tbh]
    \centering
        \includegraphics[width=0.28\linewidth]{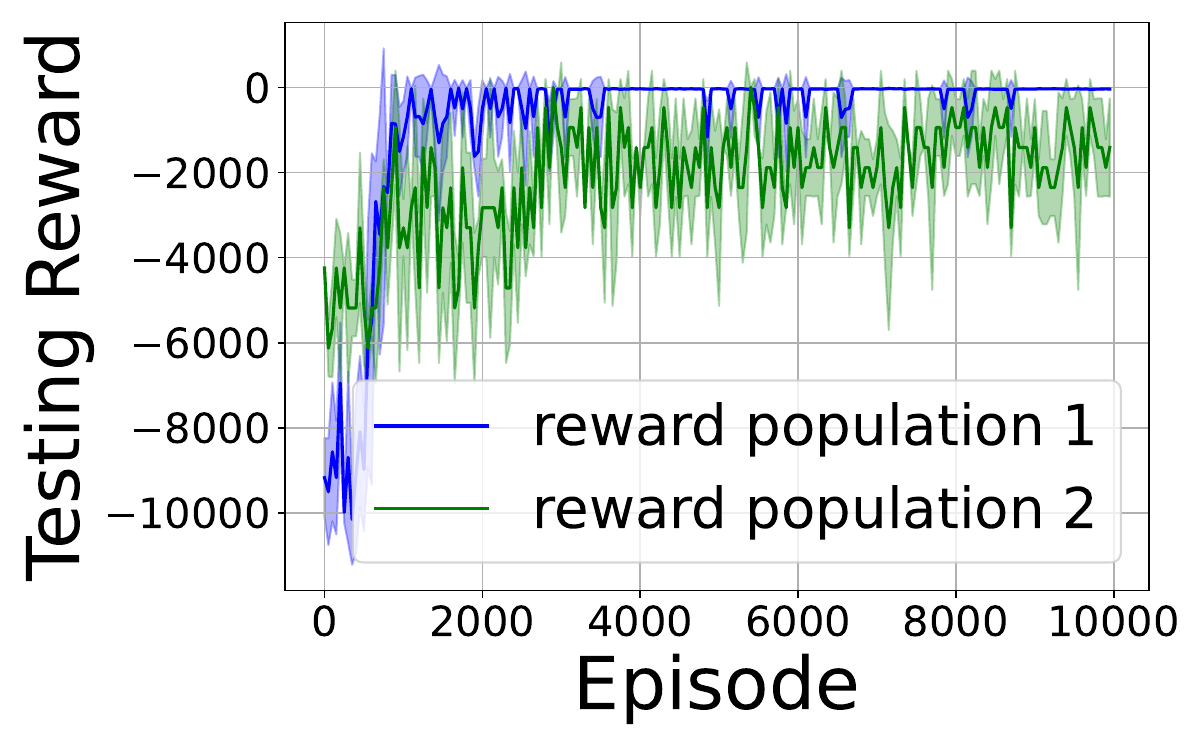}%
        \includegraphics[width=0.28\linewidth]{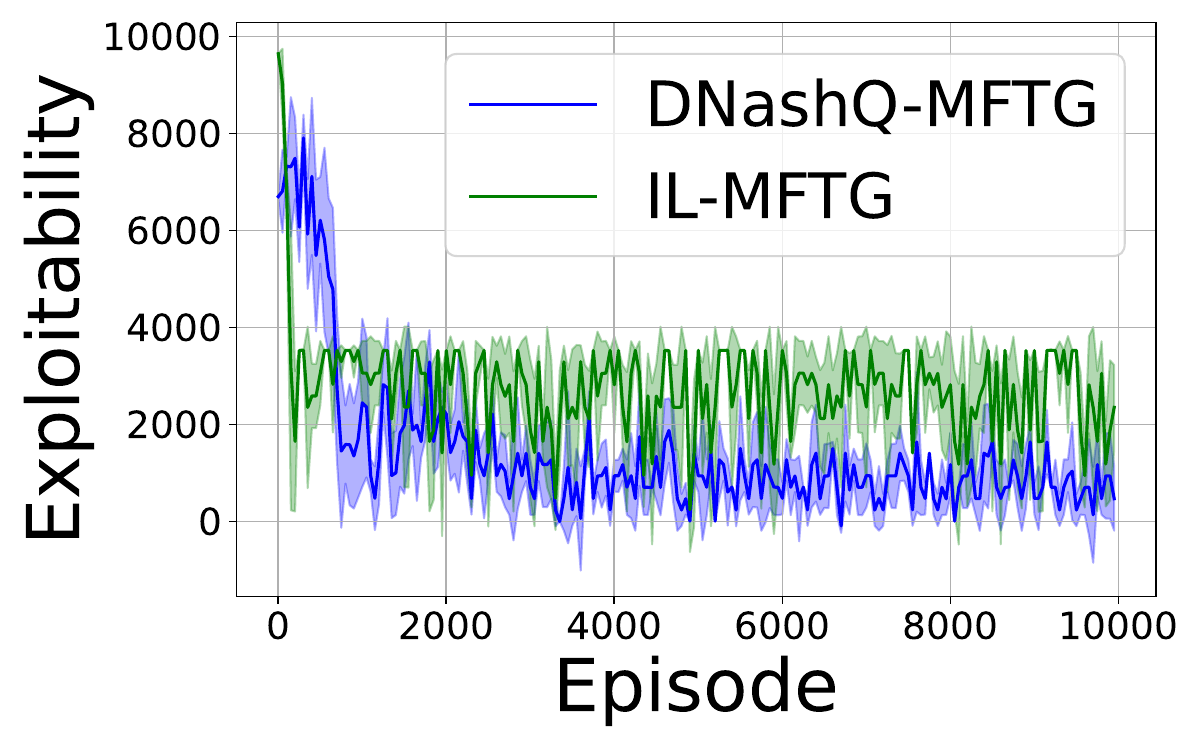} 
        \includegraphics[width=0.42\linewidth]{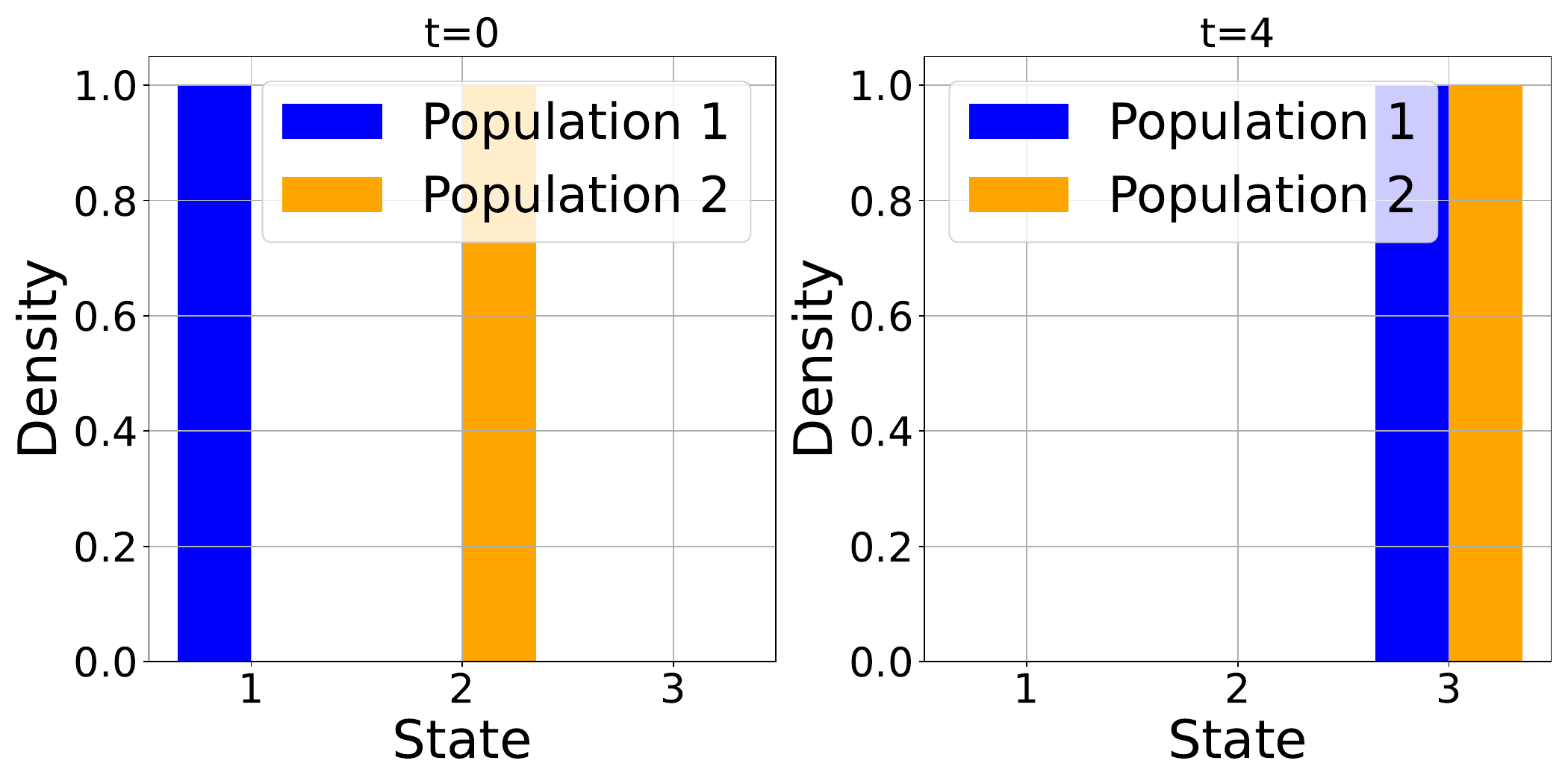} 
        \caption{{\bf Ex.~1:} Left and middle: {averaged} testing rewards and exploitabilities resp. {(mean $\pm$ stddev)}. Right: {one realization of} population evolution at $t=0$ and $4$
        for one testing distribution.}
        \label{fig:predator-prey-1D-main}
\end{figure}

\paragraph{Example 1: 1D Population Matching Grid Game}
There are $m=2$ populations. The agent's state space is a 3-state 1D grid world. The possible actions are moving left, staying, and moving right, with individual noise perturbing the movements. The rewards encourage Coalition 1 to stay where it is initialized, but also to consider avoiding Coalition 2, and encourage Coalition 2 to match Coalition 1. For the model details and the training and testing distributions, see Suppl.~\ref{app:example1-targetmoving}. We implement {\bf DNashQ-MFTG} to solve this game. The numerical results are presented in Fig.~\ref{fig:predator-prey-1D-main}. We make the following observations.
    \textbf{Testing reward curves: } Fig~\ref{fig:predator-prey-1D-main} (left) shows the testing rewards. In this game setting, the Nash equilibrium is for Coalition 1 to maintain its current position and consider the impact of Coalition 2 simultaneously, while Coalition 2 aligns with Coalition 1 perfectly.  The testing reward for Coalition 1 increases over the first 2000 episodes. The testing reward for Coalition 2 increases during the first 3000 episodes and fluctuates below 0 due to the noise in the environment dynamics. 
    \textbf{Exploitability curves:} Fig.~\ref{fig:predator-prey-1D-main} (middle) shows the averaged exploitabilities over the testing sets and players. The game reaches an approximate Nash equilibrium around 4000 episodes, with slight fluctuations thereafter. However, the independent learner remains high exploitability. The exploitability oscillates due to the noise in the environment dynamics.
    \textbf{Distribution plots:} Fig.~\ref{fig:predator-prey-1D-main} (right) illustrates the distribution evolution during the game. After training, Coalition 1 mainly stays where it is, while Coalition 2 tries to match with Coalition 1. See Suppl. \ref{app:example1-targetmoving} for details.

\paragraph{Example 2: Four-room with crowd aversion}

\begin{figure}[!htbp]%
    \centering
    \begin{minipage}{0.27\linewidth}
    \centering
    \includegraphics[width=1\linewidth]{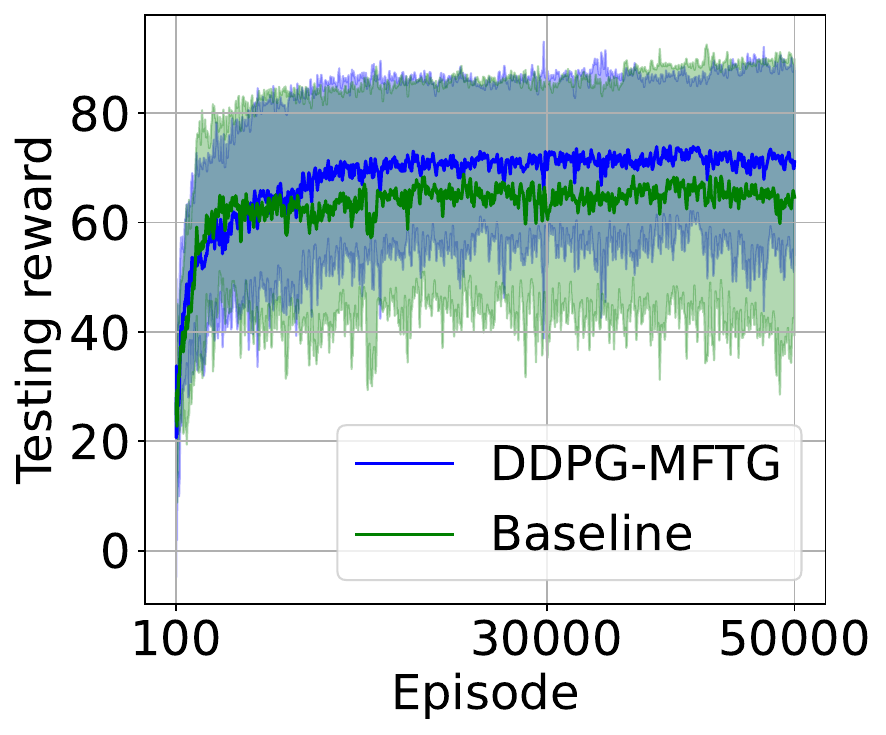} 
    \includegraphics[width=1\linewidth]{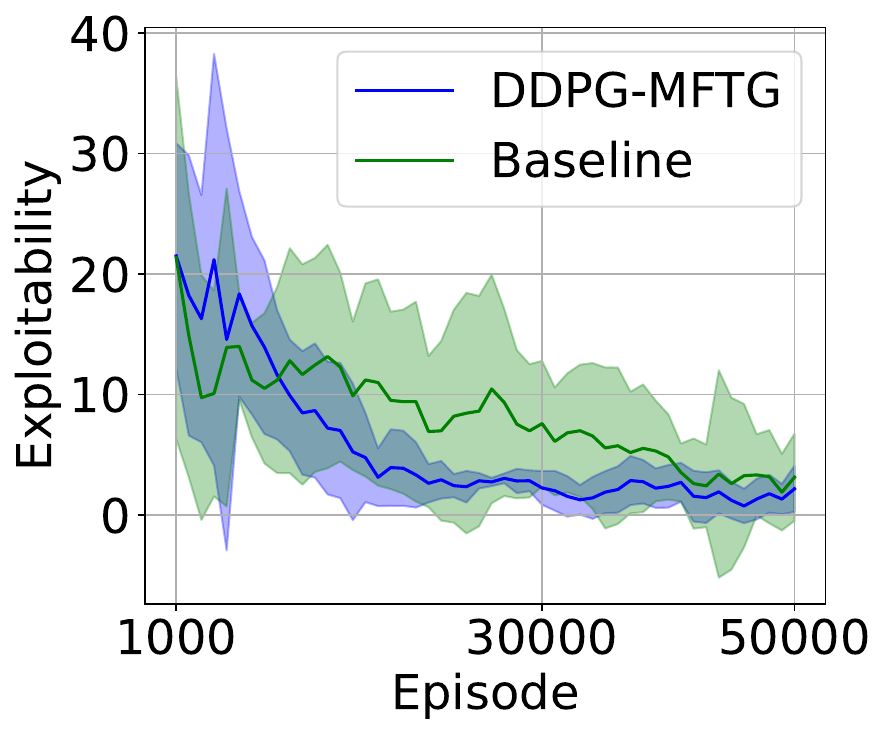} 
    \end{minipage}
    \hspace{-0.2cm}
    \begin{minipage}{0.72\linewidth}
    \centering
    \includegraphics[width=1\linewidth]{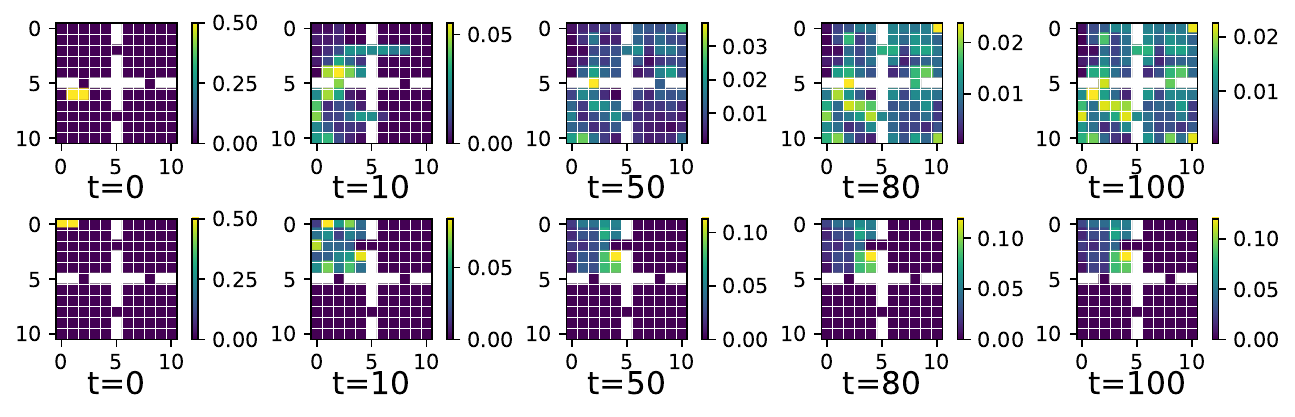} 
    \includegraphics[width=1\linewidth]{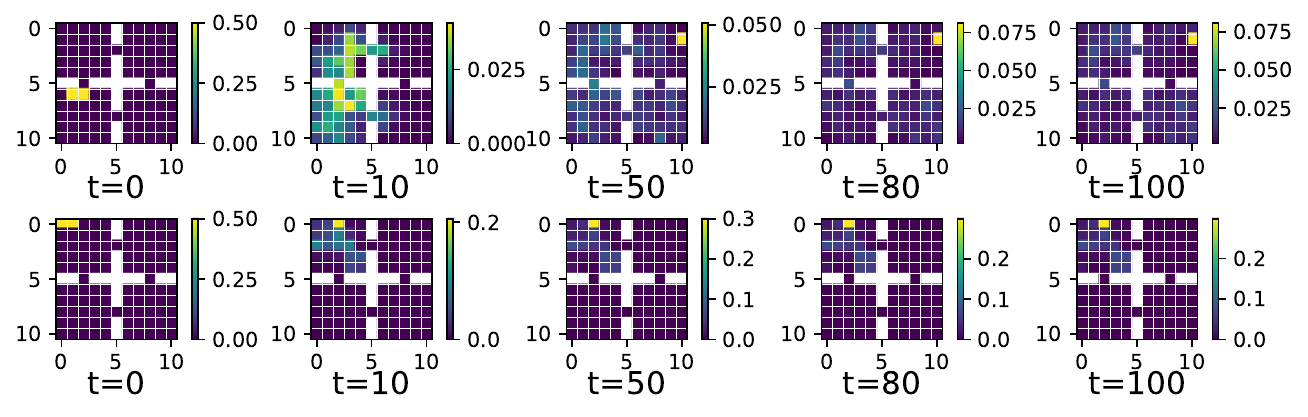} 
    \end{minipage}
    \caption{\textbf{Ex.~2:} 
    Left, top, and bottom: {averaged} testing rewards and exploitabilities resp. {(mean $\pm$ stddev)}. Right, the top two rows: distribution evolution of the two populations using our method. The bottom two rows on the right: distribution evolution using the baseline. {Color bars indicate density values.}
}
    \label{fig:four_room_main}
\end{figure}

There are $m=2$ populations. The agent's state space is a 2D grid world composed of $4$ rooms of size $5\times 5$ connected by $4$ doors, as shown in Fig.~\ref{fig:four_room_main} (right). The policies' inputs are thus of dimension $2 \times 4 \times 5 \times 5 = 200$. The reward function encourages the two populations to spread as much as possible (to maximize the entropy of the distribution) while avoiding each other. Furthermore, Coalition 2 has a penalty for moving to rooms other than the one in which she started. See Suppl.~\ref{app:ex3-4rooms} for details of the reward and the training and testing distributions. 
We implement {\bf DDPG-MFTG} to solve this game. The numerical results are presented in Fig.~\ref{fig:four_room_main}. We make the following observations.
    \textbf{Testing reward curves:} Fig.~\ref{fig:four_room_main} (left, top) shows the testing rewards.
    \textbf{Exploitability curves:} Fig.~\ref{fig:four_room_main} (left, bottom) shows the average exploitabilities over the testing set and players. The DDPG-MFTG algorithm performs better. 
    \textbf{Distribution plots:} Figs.~\ref{fig:four_room_main} (right) illustrate the distribution evolution during the game for a (pair of) initial distributions and for the policy obtained by the DDPG-MFTG algorithm and the baseline. We see that the populations spread well in any case, but with DDPG-MFTG, Coalition 1 can see where Coalition 2 is and then decides to avoid that room. This explains the better performance of the DDPG-MFTG algorithm.
    
\begin{figure}[!htbp]%
    \centering
    
    \includegraphics[width=0.29\linewidth]{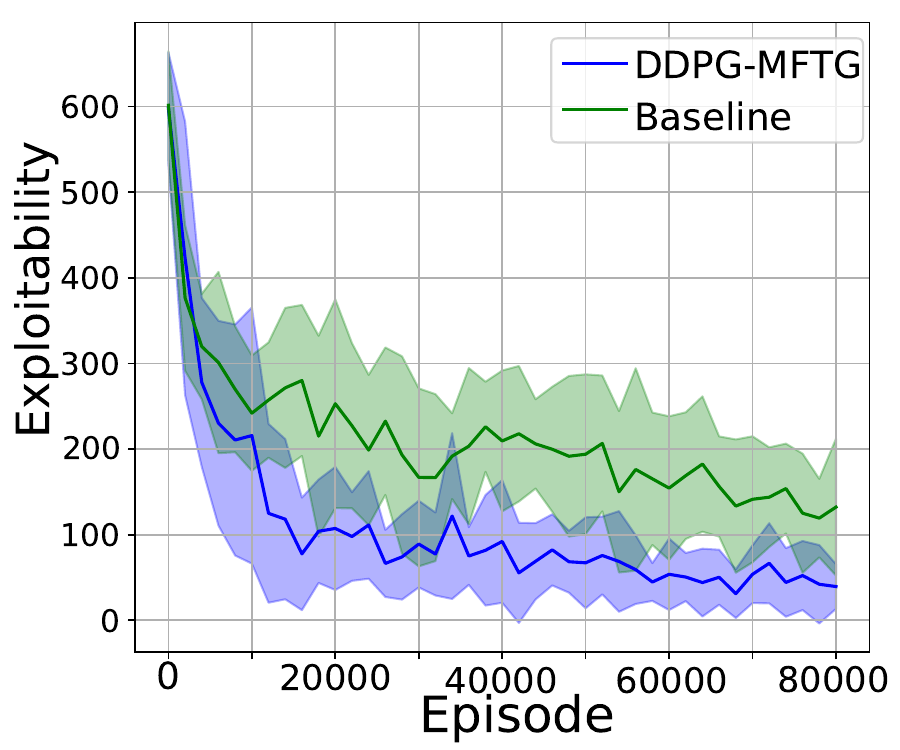} 
    \includegraphics[width=0.69\linewidth]{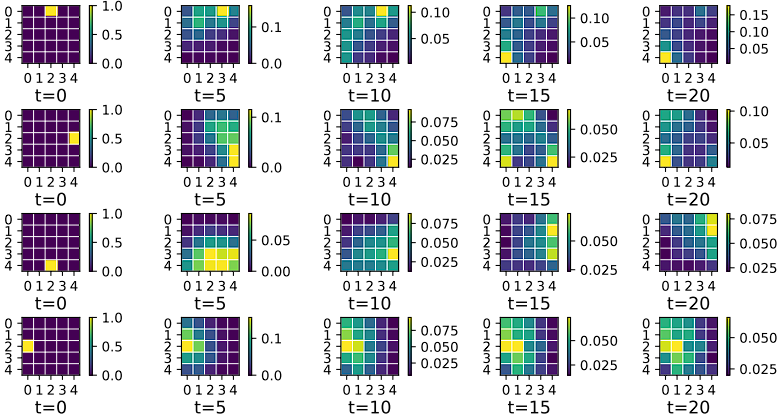} 
    \caption{\textbf{Ex.~3:}  
    Left: {averaged} exploitabilities {(mean $\pm$ stddev)}. Right: populations' evolution, one coalition per row and one time per column: \(t = 0, 5, 10, 15, 20\). {Color bars indicate density values.}
    \label{fig: predator-prey 2D 4 groups main}
   }
\end{figure}

\paragraph{Example 3: Predator-prey 2D with 4 groups} We now present an example with more coalitions. There are $m=4$ populations. The player's state space is a $5 \times 5$-state 2D grid world with walls on the boundaries (no periodicity). The reward functions represent the idea that Coalition 1 is a predator of Coalition 2. Coalition 2 avoids Coalition 1 and chases Coalition 3, which avoids Coalition 2 while chasing Coalition 4. Coalition 4 tries to avoid Coalition 3. There is also a cost for moving. See Suppl.~\ref{app:ex4-predator-prey} for details of the reward and the training and testing distributions. 
We implement {\bf DDPG-MFTG} to solve this game. The numerical results are presented in Fig.~\ref{fig: predator-prey 2D 4 groups main}. We make the following observations.
    {The \textbf{testing reward curves} (Fig.~\ref{fig: predator-pray 2D with 4 groups testing reward only} in Suppl.) do not show a clear increase for the same reason as the previous example.} 
    \textbf{Exploitability curves:} {Fig.~\ref{fig: predator-prey 2D 4 groups main} (left)} shows the averaged exploitabilities over the testing set and players. Initially, the baseline and DDPG-MFTG have similar exploitability for the first several thousand episodes. However, after that period, the baseline maintains higher exploitability than DDPG-MFTG. The exploitability of DDPG-MFTG decreases to zero faster, although it fluctuates between 0 and 100. 
    
    \textbf{Distribution plots:} {Fig.~\ref{fig: predator-prey 2D 4 groups main} (right)} shows the distribution evolution during testing. Coalition 1 chases Coalition 2. Coalition 2 tries to catch Coalition 3 while avoiding Coalition 1. Coalition 3 tries to catch Coalition 4 while escaping from Coalition 2. Coalition 4 simply escapes from Coalition 3. The testing rewards are shown in Suppl.~\ref{app:ex4-predator-prey}.

\section{Conclusion}
\label{sec:conclusion}

{\bf Summary.}
In this work, we made both theoretical and numerical contributions. First, we proved that the Nash equilibrium for a mean-field type game provides an approximate Nash equilibrium for a game between coalitions of finitely many agents, and we obtained a rate of convergence.
We then proposed the first (to our knowledge) value-based RL methods for MFTGs: a tabular RL and a deep RL algorithm. We applied them to several MFTGs. Our proposed methods provide a way to approximately compute the Nash equilibrium of a finite number of players, which is hard to solve numerically.  
We proved the convergence of the tabular algorithm, and through extensive experiments, we illustrated the scalability of the deep RL method.

{{\bf Related works. } \cite{carmona2020policyCDC,zaman2024independent,zaman2024robust} studied RL for MFTGs of LQ form only, with specific methods when the policy is deterministic and linear, while our algorithms are for generic MFTGs with discrete spaces. \citep{motte2022mean,carmona2023model} focused on single MFMDPs while we consider a game between MFMDPs. \cite{SubramanianMahajan-2018-RLstatioMFG,guo2019learning,elie2020convergence,cui2021approximately} propose RL for MFGs but are limited to population-independent policies. \cite{perrin2022generalization} studied population-dependent policies, but only for MFGs, in which players are infinitesimal; their method cannot solve MFTG because each player has a macroscopic impact on the other groups. }

{\bf Limitations and future directions.} We did not provide proof of convergence for the deep RL algorithm due to the difficulties related to analyzing deep neural networks { and because we aim for Nash equilibria rather than just MDPs}. Furthermore, we would like to apply our algorithms to more realistic examples and investigate the differences further in comparison to the baseline. We are also interested in applying other deep RL algorithms and seeing their performance in MFTGs of increasing complexity. 

{\bf Reproducibility statement.} We have included all relevant details to ensure reproducibility and provided pseudo-code for all algorithms, including the evaluation of our method's performance using the exploitability metric. Suppl.~\ref{sec:defails-expe} gives {all the detailed definitions of the environments, provides extra numerical results, and also gives all the details about the implementation, including neural network architectures and hyperparameter choices for training.} Suppl.~\ref{sec:sweeps} shows sweeps over hyperparameters to illustrate the sensitivity of our algorithms. 

\subsubsection*{Broader Impact Statement}
\label{sec:broaderImpact}
The research question addressed in this paper does not have any negative impact on the real world.
\subsubsection*{Acknowledgements}
\label{sec:acknowledgement}
We thank Chijie An for  fruitful discussions, the anonymous reviewers for their valuable comments, and Shanghai Frontiers Science Center of Artificial Intelligence and Deep Learning at NYU Shanghai for its support.
\appendix

\bibliography{mainbib}
\bibliographystyle{plainnat}

\newpage
\appendix

\section{Proof of Approximate Nash Property}
\label{app:approx-Nash-finite}

We prove Theorem~\ref{thm:approx-Nash}.

\begin{proof}
For each $i\in [m]$, we first define the distance between two distributions $\mu^i_t$, $\tilde\mu^i_t\in \Delta(\states^i)$ to be 
$$d(\mu^i_t, \tilde\mu^i_t)=||\mu^i_t-\tilde\mu^i_t||_{1}=\sum_{x\in\states^i}|\mu^i_t(x)-\tilde\mu^i_t(x)|$$
For $\mu_t,\tilde\mu_t\in\Delta(\states^1) \times \dots \times \Delta(\states^m)$, we also define 
$$d(\mu_t,\tilde\mu_t)=\max_{i}d(\mu^i_t,\tilde\mu^i_t)$$ 
We first derive a bound for $\EE ||\mu^{i,\bar{N}}_0-\mu^i_0||_{1}$. The idea is inspired by the Lemma 7 in \citep{guan2024zero}. Since $x^{ij}_0$ are i.i.d. from $\mu^i_0$, for all $x\in \states^i$,
\begin{align}
    \EE ||\mu^{i,\bar{N}}_0-\mu^i_0||^2_2&=\EE \left[\sum_{x\in\states^i}\left(\frac{1}{N_i}\sum_{j=1}^{N_i}\delta_{x^{ij}_{0}}(x)-\mu^i_0(x)\right)^2\right]
    \notag
    \\
    &=\EE \left[\sum_{x\in\states^i}\frac{1}{N_i^2}\left(\sum_{j=1}^{N_i}\left(\delta_{x^{ij}_{0}}(x)-\mu^i_0(x)\right)\right)^2\right]
    \notag
    \\
    &=\sum_{x\in\states^i}\frac{1}{N_i^2}\EE\left[\left(\sum_{j=1}^{N_i}\left(\delta_{x^{ij}_{0}}(x)-\mu^i_0(x)\right)\right)^2\right]
    \notag
    \\
    &=\sum_{x\in\states^i}\frac{1}{N_i^2}\text{Var}\left(\sum_{j=1}^{N_i}\delta_{x^{ij}_{0}}(x)\right)
    \notag
    \\
    &=\frac{1}{N_i^2}\sum_{x\in\states^i}\sum_{j=1}^{N_i}\text{Var}\left(\delta_{x^{ij}_{0}}(x)\right)\quad \text{as } x^{ij}_{0} \text{ are i.i.d.}
    \notag
    \\
    &=\frac{1}{N_i^2}\sum_{j=1}^{N_i}\sum_{x\in\states^i}\left(\EE\left[\delta^2_{x^{ij}_{0}}(x)\right]-\left(\mu^i_0(x)\right)^2\right)
    \notag
    \\
    &=\frac{1}{N_i^2}\sum_{j=1}^{N_i}\sum_{x\in\states^i}\left(\mu^i_0(x)-\left(\mu^i_0(x)\right)^2\right) \quad\text{as } \EE\left[\delta^2_{x^{ij}_{0}}(x)\right]=\mu^i_0(x)
    \notag
    \\
    &\le\frac{1}{N_i^2}\sum_{j=1}^{N_i}\sum_{x\in\states^i}\mu^i_0(x)=\frac{1}{N_i}
        \label{bd1}
\end{align}
So we have:
$$
    \EE ||\mu^{i,\bar{N}}_0-\mu^i_0||_{1}\le\sqrt{|\states^i|}\EE ||\mu^{i,\bar{N}}_0-\mu^i_0||_{2}\le\sqrt{\frac{|\states^i|}{N_i}}
$$
the second inequality above is due to the Jensen's inequality. Thus, for each $i\in[m]$, as $N_i\rightarrow +\infty$, we have 
$$\EE d(\mu^{\bar{N}}_0, \mu_0)\rightarrow 0\ \text{a.e.}$$
Next, we consider the distance between the joint state-action distribution of population $i$ at time $t$ and its empirical distribution. We denote the joint state-action distribution of population $i$ at time $t$ to be $$\nu^i_t(x,a)=\mu^i_t(x)\pi^i_t(a|x,\mu_t)$$ and the empirical state-action distribution of population $i$ at time $t$ to be $$\nu^{i,\bar{N}}_t = \frac{1}{N_i} \sum_{j=1}^{N_i} \delta_{x^{ij}_t, a^{ij}_t}$$ then, we have
\begin{align*}
    &\EE\sum_{x,a}|\nu^i_t(x,a) - \nu^{i,\bar{N}}_t(x,a)|
    \\
    &= \EE\sum_{x,a}|\mu^i_t(x) \pi^i_t(a|x,\mu_t) - \mu^{i,\bar{N}}_t(x) \pi^i_t(a|x,\mu_t)
    \\
    &\qquad +\mu^{i,\bar{N}}_t(x) \pi^i_t(a|x,\mu_t)-\mu^{i,\bar{N}}_t(x) \pi^i_t(a|x,\mu^{\bar{N}}_t)
    \\
    &\qquad +\mu^{i,\bar{N}}_t(x) \pi^i_t(a|x,\mu^{\bar{N}}_t) - \nu^{i,\bar{N}}_t(x,a)| 
    \\
    &\le\EE\sum_{x,a}|\pi^i_t(a|x,\mu_t)(\mu^i_t(x)-\mu^{i,\bar{N}}_t(x))|\\ 
    &\quad+ \EE\sum_{x,a}|\mu^{i,\bar{N}}_t(x) (\pi^i_t(a|x,\mu_t) - \pi^i_t(a|x,\mu^{\bar{N}}_t))|
    \\
    &\quad + \EE\sum_{x,a}\left|\frac{1}{N_i} \sum_{j=1}^{N_i} \delta_{x^{ij}_t}(x) \left(\pi^i_t(a|x,\mu^{\bar{N}}_t)-\frac{\frac{1}{N_i} \sum_{j=1}^{N_i} \delta_{x^{ij}_t, a^{ij}_t}(x,a)}{\frac{1}{N_i} \sum_{j=1}^{N_i} \delta_{x^{ij}_t}(x)}\right)\right|
    \\
    &\le\EE\sum_{x,a}|\pi^i_t(a|x,\mu_t)||(\mu^i_t(x)-\mu^{i,\bar{N}}_t(x))|\\ 
    &\quad+ \EE\sum_{x,a}|\mu^{i,\bar{N}}_t(x)|| (\pi^i_t(a|x,\mu_t) - \pi^i_t(a|x,\mu^{\bar{N}}_t))|
    \\
    &\quad + \EE\sum_{x,a}\left|\frac{1}{N_i} \sum_{j=1}^{N_i} \delta_{x^{ij}_t}(x) \left(\pi^i_t(a|x,\mu^{\bar{N}}_t)-\frac{\frac{1}{N_i} \sum_{j=1}^{N_i} \delta_{x^{ij}_t, a^{ij}_t}(x,a)}{\frac{1}{N_i} \sum_{j=1}^{N_i} \delta_{x^{ij}_t}(x)}\right)\right|
    \\
    &\le\EE\sum_{x}|\mu^i_t(x)-\mu^{i,\bar{N}}_t(x)|\\ 
    &\quad+ \EE\sum_{x} |\mu^{i,\bar{N}}_t(x)|L_{\pi}d(\mu_t, \mu^{\bar{N}}_t)
    \\
    &\quad + \sum_{x,a}\EE\left|\frac{1}{N_i} \sum_{j=1}^{N_i} \delta_{x^{ij}_t}(x) \left(\pi^i_t(a|x,\mu^{\bar{N}}_t)-\frac{\frac{1}{N_i} \sum_{j=1}^{N_i} \delta_{x^{ij}_t, a^{ij}_t}(x,a)}{\frac{1}{N_i} \sum_{j=1}^{N_i} \delta_{x^{ij}_t}(x)}\right)\right|
    \\
    &\le (1+L_\pi)\EE d(\mu_t, \mu^{\bar{N}}_t)\\
    &\quad+\sum_{x,a}\EE\left|\frac{1}{N_i} \sum_{j=1}^{N_i} \delta_{x^{ij}_t}(x) \left(\pi^i_t(a|x,\mu^{\bar{N}}_t)-\frac{\frac{1}{N_i} \sum_{j=1}^{N_i} \delta_{x^{ij}_t, a^{ij}_t}(x,a)}{\frac{1}{N_i} \sum_{j=1}^{N_i} \delta_{x^{ij}_t}(x)}\right)\right|
\end{align*}  
Given $\{x^{ij}_t\}_{j=1}^{N_i}$, let $N^t_i(x)=\sum^{N_i}_{j=1}\delta_{x^{ij}_t}(x)=N_i\mu^{i,\bar{N}}_t(x)$. We can decompose $\states^i$ into $\states^i=\states^i_{+}\cup \states^i_0$, where $\states^i_+=\{x\in\states^i:N^t_i(x)>0\}$ and $\states^i_0=\{x\in\states^i:N^t_i(x)=0\}$. 
For $x\in\states^i_0$, we have
$\mu^{i,\bar{N}}_t(x)=0$ and
$\nu^{i,\bar{N}}_t(x,a)=0$, so
$$\EE\left|\mu^{i,\bar{N}}_t(x) \pi^i_t(a|x,\mu^{\bar{N}}_t) - \nu^{i,\bar{N}}_t(x,a)\right|=0$$
For a fixed $x\in\states^i_{+}$, since $a^{ij}_t$ are i.i.d. with distribution $\pi^i(\cdot|x,\mu^{\bar{N}}_t)$, we have 
$$
    \EE_{a^{ij}_t} \left[\frac{\sum_{j=1}^{N_i} \delta_{x^{ij}_t, a^{ij}_t}(x,a)}{\sum_{j=1}^{N_i} \delta_{x^{ij}_t}(x)}\right]=\pi^i_t(a|x,\mu^{\bar{N}}_t).
$$
Thus, similarly to (\ref{bd1}), for $x\in\states^i_{+}$ we have
\begin{align*}
    &\EE_{a_t^{ij}}\left|\left|\pi^i_t(\cdot|x,\mu^{\bar{N}}_t)-\frac{\frac{1}{N_i} \sum_{j=1}^{N_i} \delta_{x^{ij}_t, a^{ij}_t}(x,\cdot)}{\frac{1}{N_i} \sum_{j=1}^{N_i} \delta_{x^{ij}_t}(x)}\right|\right|_2^2
    \\
    &=\EE_{a_t^{ij}}\left[\sum_{a\in A^i}\left(\pi^i_t(a|x,\mu^{\bar{N}}_t)-\frac{1}{N_i(x)}\sum_{j=1}^{N_i} \delta_{x^{ij}_t, a^{ij}_t}(x,a)\right)^2\right]
    \\
    &\le\left[\frac{1}{N^t_i(x)}\right],
\end{align*}
and
$$\EE_{a_t^{ij}}\left|\left|\pi^i_t(\cdot|x,\mu^{\bar{N}}_t)-\frac{\frac{1}{N_i} \sum_{j=1}^{N_i} \delta_{x^{ij}_t, a^{ij}_t}(x,\cdot)}{\frac{1}{N_i} \sum_{j=1}^{N_i} \delta_{x^{ij}_t}(x)}\right|\right|_{1}\le\frac{\sqrt{|A^i|}}{\sqrt{N^t_i(x)}}$$
Thus,
\begin{align*}
    &\sum_{x,a}\EE\left|\frac{1}{N_i} \sum_{j=1}^{N_i} \delta_{x^{ij}_t}(x) \left(\pi^i_t(a|x,\mu^{\bar{N}}_t)-\frac{\frac{1}{N_i} \sum_{j=1}^{N_i} \delta_{x^{ij}_t, a^{ij}_t}(x,a)}{\frac{1}{N_i} \sum_{j=1}^{N_i} \delta_{x^{ij}_t}(x)}\right)\right|
    \\
    &\le\sum_{x}\EE\left[\mu_t^{i,N_1\dots N_m}(x)\frac{\sqrt{|A^i|}}{\sqrt{N^t_i(x)}}\right]
    \\
    &=\sum_{x}\EE\sqrt{\frac{\mu_t^{i,N_1\dots N_m}(x)|A^i|}{N_i}}\le\frac{|S^i|\sqrt{|A^i|}}{\sqrt{N_i}}
\end{align*}
Therefore, we have 
$$
    \EE\sum_{x,a}|\nu^i_t(x,a) - \nu^{i,\bar{N}}_t(x,a)|\le(1+L_\pi)\EE d(\mu_t, \mu^{\bar{N}}_t)+\frac{|S^i|\sqrt{|A^i|}}{\sqrt{N_i}}$$
On the other hand, for any $t\ge 1$, we have
$$
    \mu^i_{t+1}(x')=\sum_{x,a}p(x'|x,a,\mu_t)\nu^i_t(x,a)
$$
and
$$
    \mu^{i, \bar{N}}_{t+1}(x')=\sum_{x,a}p(x'|x,a,\mu^{\bar{N}}_{t})\nu^{i,\bar{N}}_t(x,a).
$$

Moreover,
\begin{align*}
&\quad\EE\|\mu^i_{t+1}-\mu^{i, \bar{N}}_{t+1}\|_1\\
&=\EE\sum_{x'}|\mu^i_{t+1}(x')-\mu^{i, \bar{N}}_{t+1}(x')|\\
&=\EE\sum_{x'}|\sum_{x,a}p(x'|x,a,\mu_t)\nu^i_t(x,a)-\sum_{x,a}p(x'|x,a,\mu^{\bar{N}}_{t})\nu^{i,\bar{N}}_t(x,a)|
\\
&\le\EE\sum_{x'}|\sum_{x,a}p(x'|x,a,\mu_t)\nu^i_t(x,a)-\sum_{x,a}p(x'|x,a,\mu_t)\nu^{i,\bar{N}}_t(x,a)|
\\
&\quad+\EE\sum_{x'}|\sum_{x,a}p(x'|x,a,\mu_t)\nu^{i,\bar{N}}_t(x,a)-\sum_{x,a}p(x'|x,a,\mu^{\bar{N}}_{t})\nu^{i,\bar{N}}_t(x,a)|
\\
&\le\sum_{x,a}\EE|\nu^i_t(s,a)-\nu^{i,\bar{N}}_t(s,a)|
\\
&\quad+\EE\sum_{x'}\sum_{x,a}|(p(x'|x,a,\mu_t)-p(x'|x,a,\mu^{\bar{N}}_{t}))\nu^{i,\bar{N}}_t(x,a)|
\\
&\le\sum_{x,a}\EE|\nu^i_t(s,a)-\nu^{i,\bar{N}}_t(s,a)|+\EE\sum_{x,a}L_p d(\mu^i_t, \mu^{i,\bar{N}}_t)\nu^{i,\bar{N}}_t(x,a)
\\
&\le(1+L_{\pi}+L_p)\EE d(\mu_t, \mu^{\bar{N}}_t)+|\states^i|\sqrt{|A^i|}\frac{1}{\sqrt{N_i}}
\end{align*}

Thus, for $t\ge1$
\begin{equation}\label{step1}
\EE d(\mu_{t+1},\mu^{\bar{N}}_{t+1})\le(1+L_{\pi}+L_p)\EE d(\mu_t, \mu^{\bar{N}}_t)+\frac{|S|\sqrt{|A|}}{\sqrt{N}}
\end{equation}
where $\frac{|S|\sqrt{|A|}}{\sqrt{N}}=\max_i\{\frac{|S^i|\sqrt{|A^i|}}{\sqrt{N_i}}\}_{i=1}^m$. Therefore,
$$\EE d(\mu_{t},\mu^{\bar{N}}_{t})\le(1+L_{\pi}+L_p)^t\EE d(\mu_0, \mu^{\bar{N}}_0)+M(t)\frac{|S|\sqrt{|A|}}{\sqrt{N}}$$
where $M(t)=\frac{(1+L_{\pi}+L_p)^t-1}{L_{\pi}+L_p}$.

We can also rewrite the reward functions using $\nu^i_t$ and $\nu^{i,\bar{N}}_t$ as:
\begin{align*}
    J^{i}(\pi^1,\dots,\pi^m) 
    &= \EE\left[ \sum_{t \ge 0} \gamma^t r^i(x^{i}_t, a^{i}_t, \mu_t)\right]\\
    &= \sum_{t \ge 0} \gamma^t \sum_x \mu^i_t(x) \sum_a \pi^i_t(a|x,\mu_t)r^i(x, a, \mu_t)\\ 
    &= \sum_{t \ge 0} \gamma^t \sum_{x,a} \nu^i_t(x,a) r^i(x, a, \mu_t)
\end{align*} 
and 
\begin{align*}
   J^{i,\bar{N}}(\pi^1,\dots,\pi^m) 
    &= \EE\Big[ \frac{1}{N_i} \sum_{j=1}^{N_i}  \sum_{t \ge 0} \gamma^t r^i(x^{ij}_t, a^{ij}_t, \mu^{\bar{N}}_t)\Big]\\
    &= \sum_{t \ge 0} \gamma^t \sum_{x,a} \EE \Big[\nu^{i,\bar{N}}_t(x,a) r^i(x, a, \mu^{\bar{N}}_t)\Big]. 
\end{align*}
    
Given a joint policy $(\pi^1,\dots,\pi^m) \in \Pi^1\times\dots\times\Pi^m$, we have
\begin{align*}
&\quad |J^{i,\bar{N}}(\pi^1,\dots,\pi^m) -J^{i}(\pi^1,\dots,\pi^m)|\\
& =|\sum_{t \ge 0} \gamma^t \sum_{x,a} \EE \Big[\nu^{i,\bar{N}}_t(x,a) r^i(x, a, \mu^{\bar{N}}_t)\Big]-\sum_{t \ge 0} \gamma^t \sum_{x,a} \nu^i_t(x,a) r^i(x, a, \mu_t)|
\\
&\le|\sum_{t \ge 0} \gamma^t \sum_{x,a} \EE \Big[\nu^{i,\bar{N}}_t(x,a) r^i(x, a, \mu^{\bar{N}}_t)\Big]-\sum_{t \ge 0} \gamma^t \sum_{x,a} \EE \Big[\nu^{i,\bar{N}}_t(x,a) r^i(x, a, \mu_t)\Big]|
\\
&\quad+|\sum_{t \ge 0} \gamma^t \sum_{x,a} \EE \Big[\nu^{i,\bar{N}}_t(x,a) r^i(x, a, \mu_t)\Big]-\sum_{t \ge 0} \gamma^t \sum_{x,a} \nu^i_t(x,a) r^i(x, a, \mu_t)|
\\
&\le\Big|\sum_{t \ge 0} \gamma^t \EE\sum_{x,a} \Big[\nu^{i,\bar{N}}_t(x,a) \Big(r^i(x, a, \mu^{\bar{N}}_t)-r^i(x, a, \mu_t)\Big)\Big]\Big|
\\
&\quad+\sum_{t \ge 0} \gamma^t \sum_{x,a}C_r \EE \Big|\nu^{i,\bar{N}}_t(x,a)-\nu^i_t(x,a)\Big|
\\
&\le\sum_{t \ge 0} \gamma^tL_r\EE d(\mu^{\bar{N}}_t,\mu_t)+\sum_{t \ge 0} \gamma^tC_r(1+L_{\pi})\EE d(\mu^{\bar{N}}_t,\mu_t)
+\sum_{t \ge 0} \gamma^tC_r|\states^i|\sqrt{|A^i|}\frac{1}{\sqrt{N_i}}
\\
&\le\sum_{t \ge 0} \gamma^t(L_r+C_r(1+L_{\pi}))\EE d(\mu^{\bar{N}}_t,\mu_t)+\sum_{t \ge 0} \gamma^tC_r|\states^i|\sqrt{|A^i|}\frac{1}{\sqrt{N_i}}\\
&\le\sum_{t \ge 0}(L_r+C_r(1+L_{\pi}))\gamma^t\left(1+L_{\pi}+L_p\right)^t\EE d(\mu^{\bar{N}}_0,\mu_0)\\
&\quad+\sum_{t \ge 0}(L_r+C_r(1+L_{\pi}))\gamma^tM(t)\frac{|S|\sqrt{|A|}}{\sqrt{N}}+ \sum_{t \ge 0} \gamma^tC_r\frac{|\states|\sqrt{|A|}}{\sqrt{N}}
\end{align*}
When the discount factor $\gamma$ satisfies 
\begin{equation}
\gamma(1+L_{\pi}+L_p)<1
\end{equation} 
we have
\[\sum_{t \ge 0}(L_r+C_r(1+L_{\pi}))\gamma^t\left(1+L_{\pi}+L_p\right)^t<\infty\] 
\[\sum_{t \ge 0}(L_r+C_r(1+L_{\pi}))\gamma^tM(t)<\infty, \quad \sum_{t \ge 0} \gamma^tC_r<\infty\]
Thus,
\begin{equation}\label{step2}
|J^{i,\bar{N}}(\pi^1,\dots,\pi^m) -J^{i}(\pi^1,\dots,\pi^m)|\le M\frac{|\states|\sqrt{|A|}}{\sqrt{N}}
\end{equation}
where 
\begin{align*}
    M&=\sum_{t \ge 0}(L_r+C_r(1+L_{\pi}))\gamma^t\left(1+L_{\pi}+L_p\right)^t\\
    &\quad+\sum_{t \ge 0}(L_r+C_r(1+L_{\pi}))\gamma^tM(t)+\sum_{t \ge 0} \gamma^tC_r
\end{align*}
is finite.

Let $(\pi_*^1,\dots,\pi_*^m) \in \Pi^1\times\dots\times\Pi^m$ be a Nash equilibrium for the mean-field type game and $\tilde\pi^i$ be the policy for an agent in coalition $i$ of the finite-population $m$-coalition game such that 
\[J^{i,\bar{N}}(\tilde\pi^i; \pi_*^{-i})=\max_{\pi^i\in\Pi^i}J^{i,\bar{N}}(\pi^i; \pi_*^{-i}),\]
we have
\begin{align*}
J^{i,\bar{N}}(\tilde\pi^i; \pi_*^{-i})-J^{i,\bar{N}}(\pi_*^i; \pi_*^{-i})
&=J^{i,\bar{N}}(\tilde\pi^i; \pi_*^{-i})-J^{i}(\tilde\pi^i; \pi_*^{-i})\\
&\quad+J^{i}(\tilde\pi^i; \pi_*^{-i})-J^{i}(\pi_*^i; \pi_*^{-i})\\
&\quad+J^{i}(\pi_*^i; \pi_*^{-i})-J^{i,\bar{N}}(\pi_*^i; \pi_*^{-i})
\\
&\le|J^{i,\bar{N}}(\tilde\pi^i; \pi_*^{-i})-J^{i}(\tilde\pi^i; \pi_*^{-i})|\\
&\quad+|J^{i}(\pi_*^i; \pi_*^{-i})-J^{i,\bar{N}}(\pi_*^i; \pi_*^{-i})|
\\
&\le\frac{2M|\states|\sqrt{|A|}}{\sqrt{N}}
\end{align*}
The last two inequalities are due to the definition of $\pi_*^i$ and (\ref{step2}).
\end{proof}

\section{Connection between MFTG and stage-game Nash equilibria} 
\label{app:equiv-Nash-stage}

We prove Proposition~\ref{prop:Nash-equiv-stage}.

\begin{proof}
    {\bf Proof of $\Leftarrow$:} If (ii) is true, without loss of generality, we consider player $i$. we have for $\bar{s}\in\bar{S}$,
    \begin{align*}
        \bar{v}^i_{\bar{\bpi}_*}(\bar{s})
        &\ge\bar{\pi}_*^1(\bar{s})\cdots\bar{\pi}^{i-1}_*(\bar{s})\bar{\pi}^{i}(\bar{s})\bar{\pi}_*^{i+1}(\bar{s})\cdots\bar{\pi}_*^m(\bar{s})\bar{Q}^i_{\bar{\bpi}}(\bar{s})\\
         &=\bar{r}^i(\bar{s},\bar{\pi}^i(\bar{s}))+\gamma\int_{\bar{S}}\int_{\bar{\bm\actions}}\bar{p}(\mathrm d\bar{s}'|\bar{s},\bar{a}^1,\dots, \bar{a}^m)\bar{\pi}_*^1(\mathrm{d}\bar{a}^1|\bar{s})\cdots\bar{\pi}^i(\mathrm{d}\bar{a}^i|\bar{s})\cdots\bar{\pi}_*^m(\mathrm{d}\bar{a}^m|\bar{s})\bar{v}_{\bar\bpi_*^i}(\bar{s}')
    \end{align*}
    By iteration and substituting $\bar{v}_{\bar\bpi_*^i}(\bar{s}')$ with the above inequality, we have
    \begin{equation*}
        \bar{v}^i_{\bar{\bpi}_*}(\bar{s})\ge\bar{v}^i_{\bar{\bpi}'}(\bar{s})
    \end{equation*}
    for all $\bar{\pi}^i\in\bar{\Pi}^i$, where $\bar{\bpi}'=(\bar{\pi}_*^1,\dots,\bar{\pi}^i,\dots\bar{\pi}_*^m)$. Since $i$ is arbitrary, by the definition of Nash equilibrium, we have $(\bar\pi_*^1,\dots, \bar\pi_*^m)$ is a Nash equilibrium for the MFTG.

    {\bf Proof of $\Rightarrow$:} If (i) is true, then $\bar{\pi}^i_*$ is also the optimal policy for the MDP($\bar{\bpi}_*^{-i}$). For each $\bar{s}$, $\bar{\pi}^i_*(\bar{s})$ maximizes 
\begin{equation}
\bar{r}_{\bar\bpi^{-i}}(\bar{s},\bar{a}^i)+\gamma\int_{\bar{S}}\bar{p}_{\bar\bpi^{-i}}(\mathrm{d}\bar{s}'|\bar{s},\bar{a}^i)\bar{v}_{\bar\bpi_*^i}(\bar{s}')
\end{equation}
So $\bar{\pi}^i_*(\bar{s})$ is the best response of player $i$ in stage game $(\bar{Q}^1_{\bar{\bpi}_*}(\bar{s}),\dots,\bar{Q}^m_{\bar{\bpi}_*}(\bar{s}))$. The result also applies to other players, so $(\bar{\pi}_*^1(\bar{s}),\dots,\bar{\pi}_*^m(\bar{s}))$ is a Nash equilibrium in the stage game $(\bar{Q}^1_{\bar{\bpi}_*}(\bar{s}),\dots,\bar{Q}^m_{\bar{\bpi}_*}(\bar{s}))$. 
\end{proof}

\section{Analysis of Discretized NashQ Learning}
\label{sec:sensitivity-nashQ-learning}

We now prove Theorem~\ref{thm:discrete-Nash-Q}.

\begin{proof}
Let $\ckpi^p$ be a unique pure policy for the discretized MFTG such that for each $i$ and $\check{s}\in\check{S}$, the payoff function $v^i_{\ckpi^p}(\check{s})$ is a global optimal point for the stage game $\check{Q}^i_{\ckpi^p}(\check{s})$. \begin{equation}\label{q}
\begin{aligned}
    &|\check Q^i_t(\psa)-\bar{Q}^i_{\opi^p}(\sa)|\\
    &\le |\check Q^i_{t}(\psa)-\check{Q}^i_{\ckpi^p}(\psa)|\\
    &\ +|\check{Q}^i_{\ckpi^p}(\psa)-\bar{Q}^i_{\opi^p}(\psa)|\\
    &\ +|\bar{Q}^i_{\opi^p}(\psa)-\bar{Q}^i_{\opi^p}(\sa)|
\end{aligned}
\end{equation}
From Theorem \ref{thm:nashq-convergence}, when $t$ is large enough, we have
\begin{equation}\label{nashq_bd1}
    |\check Q^i_{t}(\psa)-\check{Q}^i_{\ckpi^p}(\psa)|<\epsilon.
\end{equation}
We now consider the second term on the RHS of (\ref{q}). Using the notation 
$$
    (\psa)=(\csa).
$$
and
$$\check{F}(\csa)=\text{Proj}(\bar{F}(\csa))$$
then we have
\begin{equation}
\begin{aligned}
    &\quad|\check Q^i_{\ckpi^p}(\csa)-\bar Q^i_{\opi^p}(\csa)|\\
    &\le\gamma\mathbb{E}[v^i_{\ckpi^p}(\check{F}(\csa))-v^i_{\opi^p}(\bar{F}(\csa))]\\
    &\le\gamma\mathbb{E}[v^i_{\ckpi^p}(\check{F}(\csa))-v^i_{\opi^p}(\check{F}(\csa)]\\
    &\quad +\gamma\mathbb{E}[v^i_{\opi^p}(\check{F}(\csa))-v^i_{\opi^p}(\bar{F}(\csa)]\\
    &\le\gamma\mathbb{E}[v^i_{\ckpi^p}(\check{F}(\csa))-v^i_{\opi^p}(\check{F}(\csa)]+\gamma L_{\bar{v}_{\opi}}\epsilon_{S}\\
    &\le\gamma\mathbb{E}[|\Nash\check{Q}^i_{\ckpi^p}(\check{F}(\csa))-\Nash\bar{Q}^i_{\opi^p}(\check{F}(\csa))|]+\gamma L_{\bar{v}_{\opi}}\epsilon_{S}
\end{aligned}
\end{equation}
where we used the assumption that $\bar{v}^i_{\bar{\bm{\pi}}_*}$ is Lipschitz continuous w.r.t. $\bar{s}$ with constant $L_{\bar{v}_{\opi}}$. Namely,
$$|\bar{v}^i_{\bar{\bm{\pi}}_*}(\bar{s})-\bar{v}^i_{\bar{\bm{\pi}}_*}(\bar{s'})|\le L_{\bar{v}_*}d_{\bar{S}}(\bar{s},\bar{s'})$$
Let $\check{F}(\csa)=\check{s}'$, and $(\oa)$, $(\ckoa)$ such that 
\begin{align*}
    &\Nash\check{Q}^i_{\ckpi^p}(\check{F}(\csa))=\check{Q}^i_{\ckpi^p}(\check{s}',\ckoa)\\
    &\Nash\bar{Q}^i_{\opi^p}(\check{F}(\csa))=\bar{Q}^i_{\opi^p}(\check{s}',\oa)
\end{align*}
consider the term
\begin{equation}
\begin{aligned}
    &\quad\check{Q}^i_{\ckpi^p}(\check{s}',\ckoa)-\bar{Q}^i_{\opi^p}(\check{s}',\oa)\\
    &=\check{Q}^i_{\ckpi^p}(\check{s}',\ckoa)-\check{Q}^i_{\ckpi^p}(\check{s}',\poa)\\
    &\quad+\check{Q}^i_{\ckpi^p}(\check{s}',\poa)-\bar{Q}^i_{\opi^p}(\check{s}',\poa)\\
    &\quad+\bar{Q}^i_{\opi^p}(\check{s}',\poa)-\bar{Q}^i_{\opi^p}(\check{s}',\oa)\\
    &\ge-||\check{Q}^i_{\ckpi^p}-\bar{Q}^i_{\opi^p}||_{\infty}+\bar{r}^i(\check{s}',\pj_{\check{A}^i}(\bar{a}^i_*))-\bar{r}^i(\check{s}',\bar{a}^i_*)\\
    &\quad+\gamma\mathbb{E}v^i_{\opi^p}(\bar{F}(\check{s}',\poa))-\gamma\mathbb{E}v^i_{\opi^p}(\bar{F}(\check{s}',\oa))\\
    &\ge-||\check{Q}^i_{\ckpi^p}-\bar{Q}^i_{\opi^p}||_{\infty}-L_{\bar{r}^i}d(\bar{a}^i_*,\pj_{\check{A}^i}(\bar{a}^i_*))-\gamma L_{\bar{v}^i_{\opi}}L_{\bar{F}}\sum_{i=1}^md(\bar{a}^i_*,\pj_{\check{A}^i}(\bar{a}^i_*))
\end{aligned}
\end{equation}
the last inequality is due to the Lipschitz continuous assumptions on $\bar{r}^i$ and $\bar{F}$. Namely,
$$|\bar{r}^i(\bar{s},\bar{a}^i)-\bar{r}^i(\bar{s'},\bar{a'}^i)|\le L_{\bar{r}^i}\Big(d_{\bar{S}}(\bar{s},\bar{s'})+d_{\bar{A^i}}(\bar{a}^i,\bar{a'}^i)\Big)$$
and
$$\EE|\bar{F}(\bar{s},\bar{a}^1,\dots,\bar{a}^m)-\bar{F}(\bar{s'},\bar{a'}^1,\dots,\bar{a'}^m)|\le L_{\bar{F}}\Big(d_{\bar{S}}(\bar{s},\bar{s'})+\sum_{i\in[m]}d_{\bar{A^i}}(\bar{a}^i,\bar{a'}^i)\Big)$$
On the other hand,
\begin{equation}
\begin{aligned}
    &\check{Q}^i_{\ckpi^p}(\check{s}',\ckoa)-\bar{Q}^i_{\opi^p}(\check{s}',\oa)\\
    &=\check{Q}^i_{\ckpi^p}(\check{s}',\ckoa)-\bar{Q}^i_{\opi^p}(\check{s}',\ckoa)+\bar{Q}^i_{\opi^p}(\check{s}',\ckoa)-\bar{Q}^i_{\opi^p}(\check{s}',\oa)\\
    &\le||\check{Q}^i_{\ckpi^p}-\bar{Q}^i_{\opi^p}||_{\infty}
\end{aligned}
\end{equation}
Thus, we have 
\begin{equation}
\begin{aligned}
    &|\check{Q}^i_{\ckpi^p}(\check{s}',\ckoa)-\bar{Q}^i_{\opi^p}(\check{s}',\oa)|\\
    &\le\gamma(||\check{Q}^i_{\ckpi^p}-\bar{Q}^i_{\opi^p}||_{\infty}+L_{\bar{r}^i}\epsilon_A+\gamma L_{\bar{v}^i_{\opi}}L_{\bar{F}}m\epsilon_A)+\gamma L_{\bar{v}_{\opi}}\epsilon_{S}
\end{aligned}
\end{equation}
Therefore, we have
\begin{equation}\label{nashq_bd2}
||\check{Q}^i_{\ckpi^p}-\bar{Q}^i_{\opi^p}||_{\infty}\le\frac{\gamma}{1-\gamma}\left(L_{\bar{r}^i}\epsilon_A+\gamma L_{\bar{v}^i_{\opi}}L_{\bar{F}}m\epsilon_A+L_{\bar{v}_{\opi}}\epsilon_{S}\right)
\end{equation}
For the last term on the RHS of (\ref{q}), we have
\begin{equation}\label{nashq_bd3}
\begin{aligned}
    &|\bar{Q}^i_{\opi^p}(\psa)-\bar{Q}^i_{\opi^p}(\sa)|\\
    &\le |\bar{r}^i(\pj_{\check{S}}(\bar{s}),\pj_{\check{A}^i}(\bar{a}^i))-\bar{r}^i(\bar{s},\bar{a}^i)|\\
    &\quad+\gamma\mathbb{E}[\bar{v}^i_{\opi^p}(\bar{F}(\psa))-\bar{v}^i_{\opi^p}(\bar{F}(\sa))]\\
    &\le L_{\bar{r}^i}(d_{\bar{S}}(\pj_{\check S}(\bar s),\bar{s})+d_{\bar{A}^i}(\pj_{\check A^i}(\bar{a}^i),\bar{a}^i))\\
    &\quad+\gamma L_{\bar{v}^i_{\opi}}\mathbb{E}(\bar{F}(\psa)-\bar{F}(\sa))\\
    &\le L_{\bar r^i}(\epsilon_S+\epsilon_{A})+\gamma L_{\bar{v}^i_{\opi}}L_{\bar{F}}(\epsilon_S+m\epsilon_{A})
\end{aligned}
\end{equation}
Finally, we get the result by combining inequalities (\ref{nashq_bd1}), (\ref{nashq_bd2}), and (\ref{nashq_bd3}) together.
\end{proof}

\clearpage 
\section{Pseudo-code for the Discretized Nash Q-learning}
\label{app:nashq-algo}
\begin{algorithm}[H]
\caption{Discretized Nash Q-learning for Mean Field Type Game (\textbf{DNashQ-MFTG})} 
\label{alg:nashq-main-algo}
\begin{algorithmic}[1]
    \STATE \textbf{Inputs: } A series of learning rates $\alpha_t\in(0,1)$, $t \ge 0$, and exploration levels $\epsilon_t$, $t \ge 0$ \\
    \STATE \textbf{Outputs:} Nash Q-functions $\check{Q}_{N}^{i}$ for $i=1,\dots,m$
    \STATE  Initialization: $\check{Q}^i_{0,0}(\check{s}, \check a^1,\dots, \check a^m)=0$ for all $\check{s}\in\check{S}$ and $\check a^i\in\check{A}^i$; 
    \FOR{$k=0,1,\dots,N-1$}
    \STATE Initialize state $\check{s}_0$
    \FOR{$t=0,\dots, T-1$} \label{line:while-loop-forever}
            \STATE Generate a random number $\zeta_{t}\sim \mathcal{U}[0,1]$
            \IF{$\zeta_{t} \ge \epsilon_t $}
            \STATE Solve the stage game $\check{Q}^i_{k,t}(\check{s}_{t})$ and get strategy profile $(\check{\pi}^{i,1}_{*},\dots,\check{\pi}^{i,m}_{*})$ for $i=1,\dots,m$
            \STATE Sample 
                $\check{a}_{t}^i \sim \check{\pi}^{i,i}_*$ for $i=1,\dots, m$
            \ELSE
            \STATE Sample action $\check{a}^i_t$ uniformly from $\check{A}^{i}$ for $i=1,\dots, m$
            \ENDIF
            
            \STATE Observe $r^1_t$,\dots, $r^m_t$, $\check{a}^1_t$,\dots, $\check{a}^m_t$, and $\check{s}_{t+1}=\pj_{\check S}(\bar{F}(\check{s}_{t},\check{a}^1_t$,\dots, $\check{a}^m_t))$
            \STATE Solve the stage game $\check{Q}^i_{k,t}(\check{s}_{t+1})$ and get strategy profile $(\check{\pi}^{'i,1}_{*},\dots,\check{\pi}^{'i,m}_{*})$ for $i=1,\dots,m$
            \STATE Compute $\Nash\check{Q}^i_{k,t}(\check{s}_{t+1})=\check{\pi}^{'i,1}_{*}\dots \check{\pi}^{'i,m}_{*}\check{Q}_{k,t}^{i}(\check{s}_{t+1})$ 
            \STATE Copy $\check{Q}^i_{k,t+1}=\check{Q}^i_{k,t}$ for $i=1,\dots,m$ and update $\check{Q}^i_{k,t+1}$ by: \\
            \quad \quad $\check{Q}^i_{k,t+1}(\check{s}_{t},\check a^1,\dots, \check a^m)=(1-\alpha_t)\check{Q}^i_{k,t}(\check{s}_{t},\check a^1,\dots, \check a^m)+\alpha_t(r^i_t+\beta \Nash\check{Q}^i_{k,t}(\check{s}_{t+1}))$
            \ENDFOR
        \STATE Copy $\check{Q}^i_{k+1,0}=\check{Q}^i_{k,T-1}$ for $i=1. \dots.m$
    \ENDFOR
\end{algorithmic}
\end{algorithm}

\clearpage
\section{Pseudo-codes for the evaluation metrics}
\label{app:metrics-pseudocodes}

In this section, we present pseudo-codes used for evaluation. 
\begin{itemize}
    \item Algorithm~\ref{alg:nash Q inference} shows how to do the inference of DNash-MFTG given the Q-functions of agents.
    \item Algorithm~\ref{algo:policy-eval} explains the way to evaluate policies.
    \item Algorithm~\ref{algo:exploitability-comp general} presents the general structure of computing exploitability.
    \item Algorithm~\ref{algo:exploitability-comp detailed} presents a detailed version of computing the exploitability.
\end{itemize}
\begin{algorithm}[H]
\caption{DNashQ-MFTG inference} 
\label{alg:nash Q inference}
\begin{algorithmic}[1]
    \STATE \textbf{Inputs:} Nash Q-functions $\check{Q}^i_{N}$ for $i=1, \dots,m$; number of steps $T$
    \STATE \textbf{Outputs:} $\upsilon^{i}=(\check{s}^{i}_0,\check{a}^{i}_0,r^i_0,\dots, \check{s}^{i}_{T-1},\check{a}^{i}_{T-1},r^i_{T-1})$ for $i=1,\dots,m$
    \STATE  Initialize $\check{s}_0$ and trajectory $\upsilon^{i}$
    \FOR{$t=0,\dots,T-1$} \label{line:while-loop-forever_2}
        \STATE Solve the stage game $\check{Q}^i_{N}(\check{s}_t)$ and get strategy profile $(\check{\pi}^{i,1}_{*},\dots,\check{\pi}^{i,m}_{*})$ for $i=1,\dots,m$\\
        \STATE Sample 
            $\check{a}_t^i \sim \check{\pi}^{i,i}_*$ for $i=1,\dots, m$
        
        \STATE Observe $r^1_t$,\dots, $r^m_t$ and $\check{s}_{t+1}=\pj_{\check S}(\bar{F}(\check{s}_t,\check{a}^1_t$,\dots, $\check{a}^m_t))$
        \STATE Store $(\check{s}^{i}_t,\check{a}^{i}_t,r^i_t)$ to $\upsilon^{i}$
    \ENDFOR
    \RETURN{Trajectory $\upsilon^{i}$}
\end{algorithmic}
\end{algorithm}

\begin{algorithm}[H]
   \caption{Policies evaluation}
   \label{algo:policy-eval}
\begin{algorithmic}[1]
   \STATE {\bfseries Inputs:} Policy profile $\bar\pi = (\bar\pi^1,\dots,\bar\pi^m)$, testing set of initial distributions $\cD_{\text{test}}$
   \STATE {\bfseries Outputs:} Values $J^i(\bar\pi)$
   \STATE Initialize $V^i = 0$, $i=1,\dots,m$
   \FOR{$\mu_0 \in \cD_{\text{test}}$}
   \STATE Run an episode starting from initial distribution $\mu_0$ and using policies $\bar\pi$
   \STATE Let $V^i_{\mu_0}$ be the total reward, $i=1,\dots,m$
   \STATE Let $V^i = V^i + V^i_{\mu_0}$, $i=1,\dots,m$
   \ENDFOR
   \STATE Let $J^i = \frac{1}{|\cD_{\text{test}}|} V^i$
   \STATE Return $J^i$, $i=1,\dots,m$
\end{algorithmic}
\end{algorithm}

\begin{algorithm}[H]
   \caption{Exploitability computation}
   \label{algo:exploitability-comp general}
\begin{algorithmic}[1]
   \STATE {\bfseries Inputs:} Policy profile $\bar\pi = (\bar\pi^1,\dots,\bar\pi^m)$, training set of initial distributions $\cD_{\text{train}}$,\linebreak testing set of initial distributions $\cD_{\text{test}}$
   \STATE {\bfseries Outputs:} Exploitabilities $E^i(\bar\pi)$, $i=1,\dots,m$
   \FOR{$i=1,\dots,m$}
        \STATE Compute BR $\bar\pi^{i*} = \argmax_{\tilde{\bar\pi}^i} J^i(\tilde{\bar\pi}^i; \bar\pi^{-i})$ using RL with testing set  $\cD_{\text{test}}$
        \STATE Compute $M^i = J^i(\bar\pi^{i*};\bar\pi^{-i})$ using Algo.~\ref{algo:policy-eval} with policy profile $(\bar\pi^{i*};\bar\pi^{-i})$ and $\cD_{\text{test}}$
        \STATE Compute $V^i = J^i(\bar\pi^{i};\bar\pi^{-i})$ using Algo.~\ref{algo:policy-eval} with policy profile $(\bar\pi^{i};\bar\pi^{-i})$ and $\cD_{\text{test}}$
        \STATE Let $E^i = M^i - V^i$
   \ENDFOR
   \STATE Return $E^i$, $i=1,\dots,m$
\end{algorithmic}
\end{algorithm}

\begin{algorithm}[H]
   \caption{Exploitability computation}
   \label{algo:exploitability-comp detailed}
\begin{algorithmic}[1]
   \STATE {\bfseries Inputs:} Policy profile $\bar\pi = (\bar\pi^1,\dots,\bar\pi^m)$, testing set of initial distributions $\cD_{\text{test}}$,
   \STATE {\bfseries Outputs:} Exploitabilities $E^i(\bar\pi)$, $i=1,\dots,m$
   \STATE Initialize $M^i = 0$, $E^i=0$, $i=1,\dots,m$
   \FOR{$i=1,\dots,m$}
   \FOR{$\mu_0$ in $\cD_{\text{test}}$}
        \STATE Initialize replay buffer and optimizers
        \FOR{$j=1,\dots, N$}
        \STATE Compute BR $\bar\pi^{i*}_j = \argmax_{\tilde{\bar\pi}^i} J^i(\tilde{\bar\pi}^i; \bar\pi^{-i})$ using RL with the initial distribution $\mu_0$ 
        \STATE Compute $M^i_j = J^i(\bar\pi^{i*}_j;\bar\pi^{-i})$ using Algo.~\ref{algo:policy-eval} with policy profile $(\bar\pi^{i*}_j;\bar\pi^{-i})$ and $\mu_0$
        \STATE $M^i=M^i+M^i_j$
        \ENDFOR
        \STATE $M^i=M^i/N$
        \STATE Compute $V^i = J^i(\bar\pi^{i};\bar\pi^{-i})$ using Algo.~\ref{algo:policy-eval} with policy profile $(\bar\pi^{i};\bar\pi^{-i})$ and $\mu_0$
        \STATE $E^i = E^i+ M^i - V^i$
        \ENDFOR
    \STATE $E^i =\frac{1}{|\cD_{\text{test}}|}E^i$
   \ENDFOR
   \STATE Return $E^i$, $i=1,\dots,m$
\end{algorithmic}
\end{algorithm}

\newpage
\section{Details on numerical experiments} 
\label{sec:defails-expe}

\subsection{IL-MFTG}
\label{explain: IL-MFTG}
Here, \textbf{IL-MFTG} stands for \textbf{I}ndependent \textbf{L}earning - \textbf{M}ean \textbf{F}ield \textbf{T}ype \textbf{G}ame, where each coalition independently performs standard Q-learning (see~\cite{watkins1989learning}) without access to the states of other coalitions, following a suitable discretization of both the state and action spaces to enable learning in the mean field type game setting. We propose IL-MFTG as a baseline for comparison with DNashQ-MFTG (Algorithm~\ref{alg:nashq-main-algo}).

\subsection{Example 1: 1D Target Moving Grid Game} 
\label{app:example1-targetmoving}

\paragraph{\bf Model.}
The model is as follows:
\begin{itemize}
    \item {\bf Number of populations:} $m=2$.
    \item {\bf State space:} $\states^{i} = \states = \{1,2,3,\dots, G\}$ for $i=1,2$, which represents locations. 
    \item {\bf Action space:} $\actions^{i} = \{0,-1,1\}$ for $i=1,2$, represents the agent will stay, move left, or move right, respectively
    \item {\bf Individual dynamics:} $x^i_{t+1} = x^i_t + a^i_t + \xi^i_t$, where $(\xi^i_t)_{n \ge 1}$ is a sequence of i.i.d. random variables and sampled from a predefined distribution as noises. We use periodic boundary conditions, meaning that agents who move left (resp. right) while in the $0$ (resp. $G$) state end up on the other side, at the $G$ (resp. $0$) state.
    \item  
     {\bf Mean-field transitions:} The element in the $k$-th row, $\ell$-th column in the $G \times G$ transition matrix $\bar{P}^{i}(\bar{s}^{i}_t,\bar{a}^i_t)$ is equal to $p^{i}(\bar{s}^{i}_{t+1}=k|\bar{s}^{i}_t=\ell,\bar{a}^i_t,\xi^i_t)$
    \item {\bf Rewards:} Population 1 receives a high penalty when it moves, while Population 2 tries to match with Population 1's current position. We use the following rewards:
    $$
    \bar{r}^1(\bar{s},\bar{a}^{1}_{t}) = -c_1 (
    \|\bar{a}^{1}_{\text{stay}} - \bar{a}^{1}_{t}\|_{2})- c_2(\bar{s}^{1}\times \bar{s}^{2}) , \quad \bar{r}^2(\bar{s}) = -c_1
    (\|\bar{s}^{1} - \bar{s}^{2}\|_{2})
    $$ where $c_1=1000$ and $c_2=10$. As a consequence, we expect that, at the Nash equilibrium, Coalition 1 stays where it is but also tries to avoid Coalition 2, while Coalition 2 matches Coalition 1 perfectly.
\end{itemize}

\paragraph{\bf Training and testing sets.}
In this example, we use $G=3$ points in the 1D grid. (Scaling up to larger spaces would require a huge amount of memory due to the required discretization of the state space. This motivates the deep RL algorithm we use in the next examples.) We use the following sets of initial distributions for training and testing.
\begin{itemize}
    \item Training distributions: We employ a random sampling technique to generate the training distribution at the beginning of each training episode. Specifically, we first sample each element in the state matrix from a uniform distribution over the interval $[0,1)$ and then divide each element by the total sum of the matrix to normalize it.
    \item Testing distributions: we use the following pairs: 
    
    $\cD_{\text{test}} = \{
        \big((1.0,0.0,0.0),(0.0,0.0,1.0)\big),$ 
        $\big((0.0,0.0,1.0),(1.0,0.0,0.0)\big),$ \\
        $\big((0.0,1.0,0.0),(0.0,1.0,0.0)\big) \}$ 
\end{itemize}
\paragraph{\bf Parameters and Hyper-parameters}

In the tabular case, we use the following hyperparameters for both inner Q-learning and outer Nash Q-learning: \begin{itemize}
    \item learning rate $\alpha_{t}=\frac{1}{n_t(\bar{s}_t,\bar{a}^{1},\bar{a}^{2})}$, where $n_t(\bar{s}_t,\bar{a}^{1},\bar{a}^{2})$ is the number of times that tuple $( \bar{s}_t,\bar{a}^{1},\bar{a}^{2})$ has visited.
    \item  $\epsilon_t=\epsilon_{end}+(\epsilon_{start}-\epsilon_{end})\exp({-\frac{t}{T}})$, where $T$ is the total training episode, $\epsilon_{end}=0.01$, and $\epsilon_{start}=0.99$.
    \item $\xi_t \sim \{0.99,0.005,0.005\}$
\end{itemize} 
\paragraph{Evaluation}
We evaluate the policy of each player by computing exploitability in Algo.~\ref{algo:exploitability-comp detailed}. We employ tabular Q-learning to solve an MDP and generate the best response.
\paragraph{Baseline}
The baseline for DNashQ-MFTG is different from other examples. Each coalition learns the game independently through Q-learning using the same discretization as our DNashQ-MFTG. For the exploitability computation, we still perform standard Q-learning with full observation of mean-field states to generate the best response.

\begin{figure}[!htbp]
    \centering
    
    \begin{minipage}{0.9\linewidth}
    \includegraphics[width=1\linewidth]{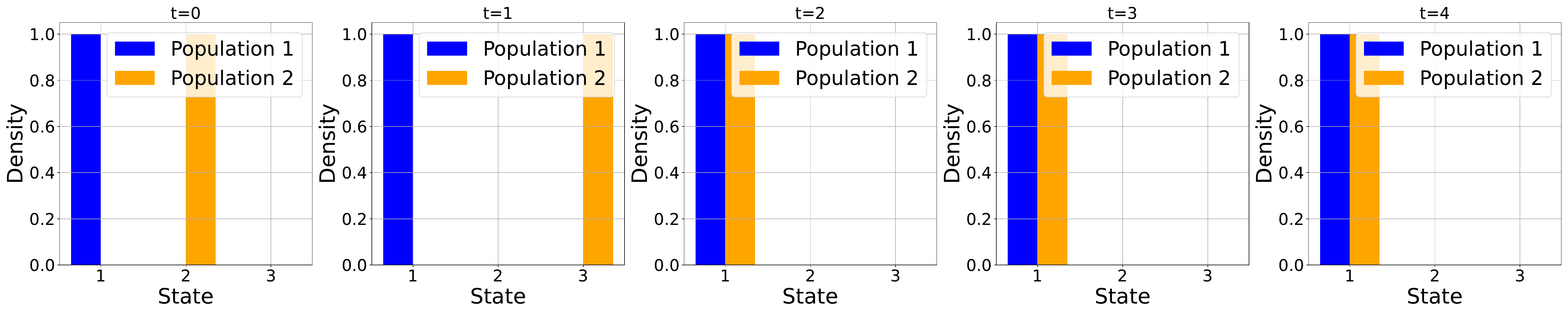} 
    \end{minipage}
    \begin{minipage}{0.9\linewidth}
    \includegraphics[width=1\linewidth]{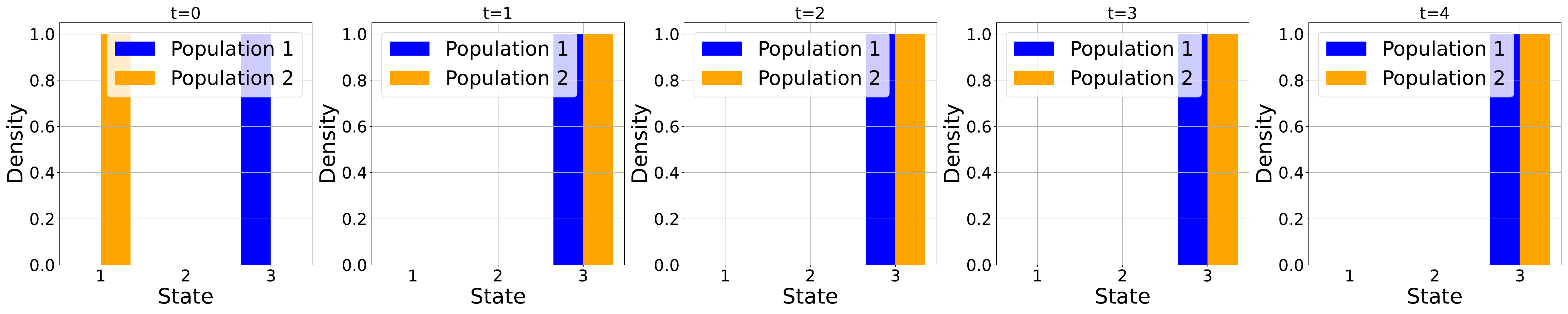} 
    \end{minipage}
    \begin{minipage}{0.9\linewidth}
    \includegraphics[width=1\linewidth]{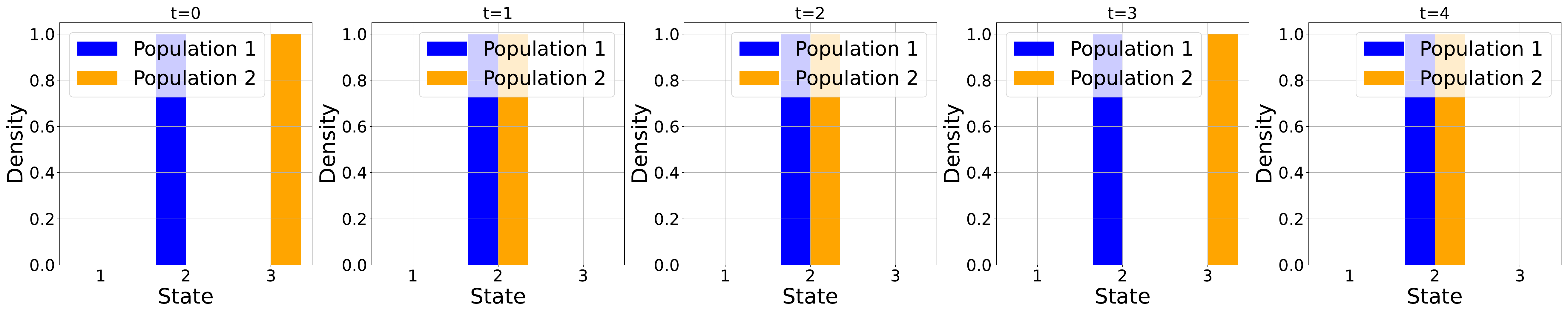} 
    \end{minipage}
    
    \caption{
   1D Target Moving Grid Game: Population evolution of testing distribution at $t=0,1,2,3,4$. From top to bottom are the evolutions of testing distributions 1, 2, and 3.
    }
    \label{fig: predator-prey 1D testing 1 - more}
\end{figure}

We show more examples of distribution evolution in Fig.~\ref{fig: predator-prey 1D testing 1 - more}.

\subsection{Example 2: Four-room with crowd aversion}
\label{app:ex3-4rooms}

\paragraph{\bf Model.}
We consider a 2-dimensional grid world with four rooms and obstacles. Each room has only one door that connects to the next room and has $5\times 5$ states. 
\begin{itemize}
    \item {\bf Number of populations:} $m=2$.
    \item {\bf State space:} $ S=\{0,\dots,N^1_x\}\times\{0,\dots,N^2_x\}$, where $N^1_x=N^2_x=10$. 
    \item {\bf Action space:} $ A=\{(-1,0), (1,0), (0,0), (0,1), (0,-1)\}$, which represents move left, move, right, stay, move up, and move down, respectively.
    \item {\bf Transitions:} At time $n$, the agent at position $s_n=(x,y)$ chooses an action $a_n$, the next state is computed according to 
    \begin{equation}
    s_{n+1}=
    \begin{cases}
    s_n + a_n +\epsilon_{n+1}, & \text{if } s_n + a_n +\epsilon_{n+1} \text{ is not in a forbidden state}\\
    s_n, & \text{otherwise}
    \end{cases}
    \end{equation}
    where $\{\epsilon_n\}_n$ is a sequence of i.i.d. random variables taking values in $A$, representing the random disturbance.
    
    The mean-field distribution $\bar{s}^i_{t}(x,y)$ is computed according to
    \begin{align*}
    \bar{s}^i_{t+1}(x,y)&=\bar{s}^i_{t}(x,y)\bar{a}^i((0,0)|(x,y))+\bar{s}^i_{t}(x,y-1)\bar{a}^i((0,1)|(x,y-1))\\
    &\quad+\bar{s}^i_{t}(x,y+1)\bar{a}^i((0,-1)|(x,y+1))+\bar{s}^i_{t}(x+1,y)\bar{a}^i((-1,0)|(x+1,y))\\
    &\quad+\bar{s}^i_{t}(x-1,y)\bar{a}^i((1,0)|(x-1,y))\\
    \end{align*}
    where $\bar{s}^i_t(a,b)$ is the density of Coalition i at the location $(a,b)$ at time step $t$.
    \item {\bf One-step reward function:} $$\bar r^1(\bar{s}^1_t,\bar{s}^2_t)=-\bar{s}^1_t \cdot \log(\bar{s}^1_t+\bar{s}^2_t)/\log(100)$$
$$\bar r^2(\bar{s}^1_t,\bar{s}^2_t)=-\bar{s}^2_t \cdot \log(\bar{s}^1_t+\bar{s}^2_t)/\log(100)-30\times\Big(\bar{s}^2_t(2,5)+\bar{s}^2_t(8,5)+\bar{s}^2_t(5,2)+\bar{s}^2_t(5,8)\Big)$$
where $\cdot$ is the inner product.
    \item {\bf Time horizon:} $N_T=40$.
\end{itemize}

\begin{figure}[!htbp]
    \centering
    \begin{minipage}{0.47\linewidth}
    \includegraphics[width=1\linewidth]{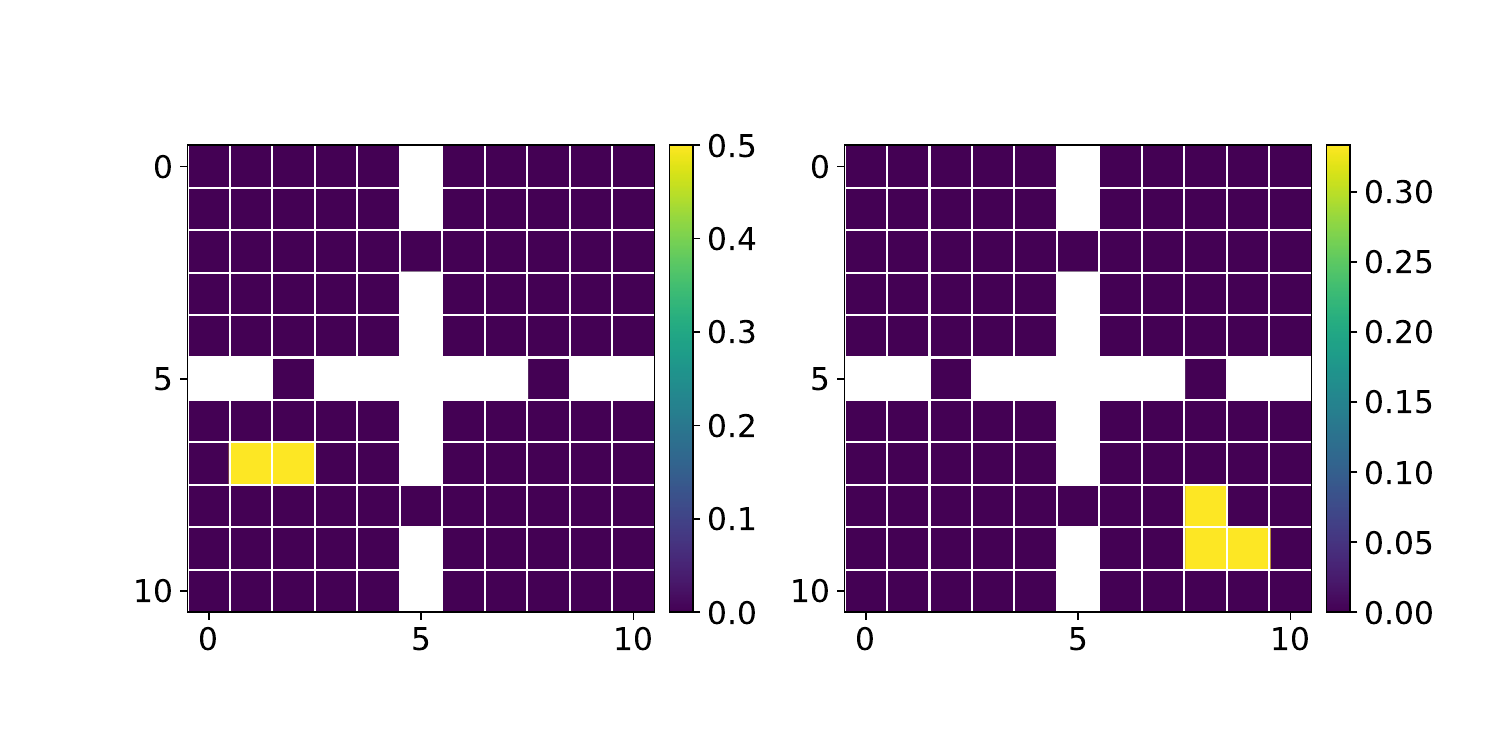} 
    \end{minipage}
    \begin{minipage}{0.47\linewidth}
    \includegraphics[width=1\linewidth]{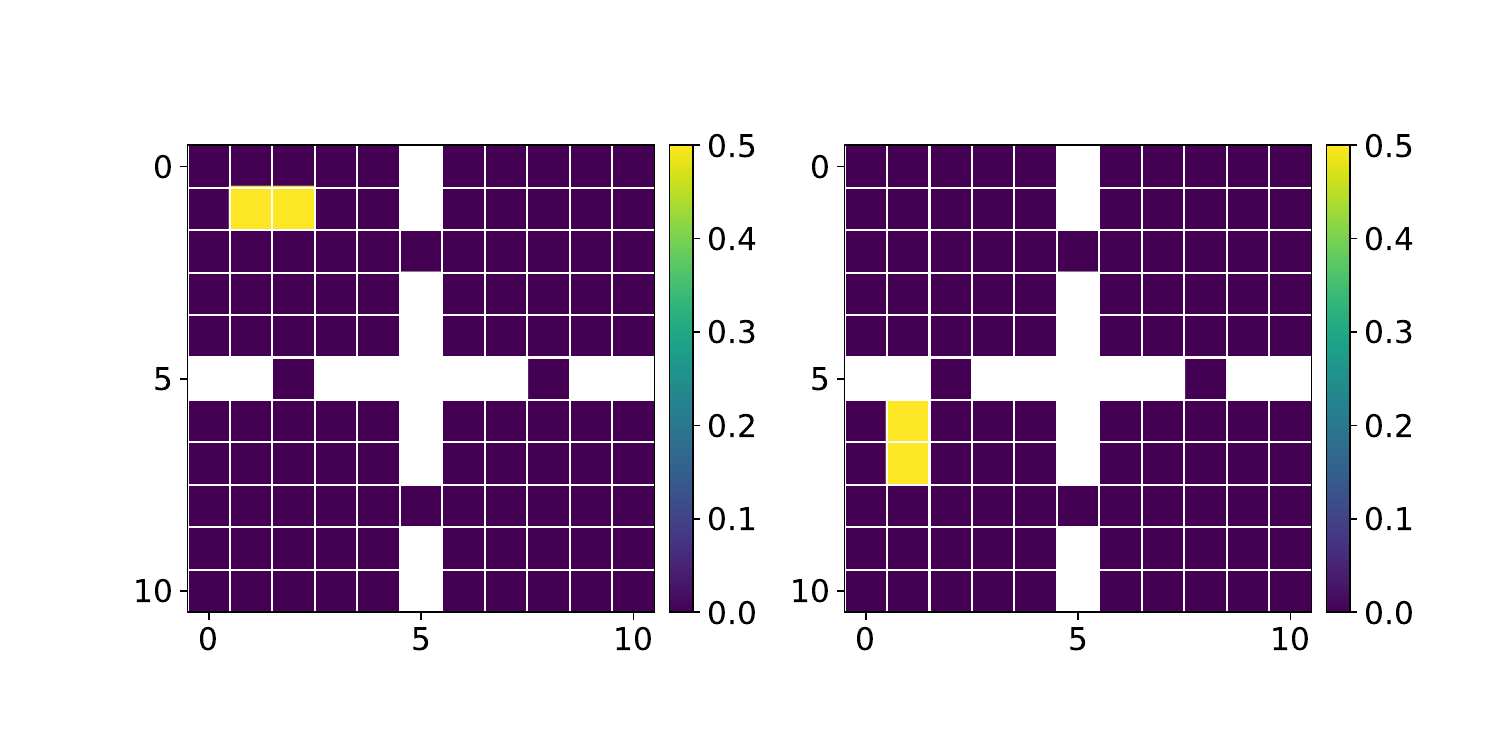}
    \end{minipage} 
    \begin{minipage}{0.47\linewidth}
    \includegraphics[width=1\linewidth]{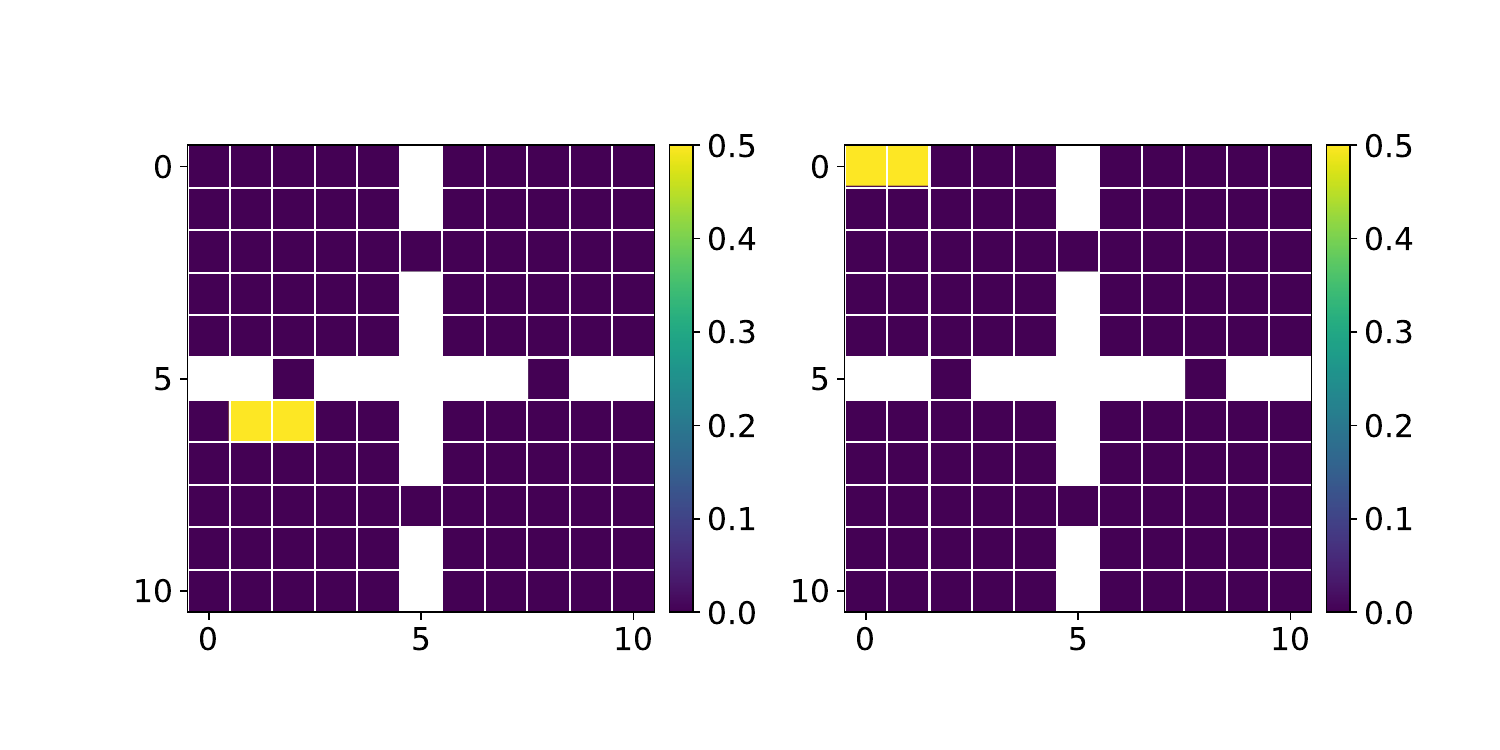}
    \end{minipage}
\caption{3 pairs of testing distributions. For each pair, the left one is the initial distribution of Player 1, and the right one is the initial distribution of Player 2.}
\label{fig:test_four_room}
\end{figure}

\paragraph{\bf Training and testing sets} For the training set, each player chooses locations among the four rooms with the sum of probability density equal to $1$ as the initial distribution. We used three pairs of distributions with different random seeds as the testing set. Each of them is a uniform distribution among selected locations. The testing distributions are illustrated in Fig.~\ref{fig:test_four_room}.

\paragraph{Neural network architecture and hyper-parameters}
In the actor network, each state vector is initially flattened and fed into a fully connected network with a Tanh activation function, resulting in a 200-dimensional output for each. These outputs are then concatenated and processed through a two-layer fully connected network, each with 200 hidden neurons, utilizing ReLU and Tanh activation functions. The final output dimension is $|S|\times|A|$. The output is then normalized using the softmax function. The critic network follows a similar architecture. During the training, we use the Adam optimizer with the actor network learning rate equal to $5 \times 10^{-5}$ and the critic network learning rate equal to 0.0001. The standard deviation used in the Ornstein–Uhlenbeck process is 0.08. We also use target networks to stabilize the training and the update rate is 0.005. The replay buffer is of size 100000, and the batch size is 32. The model is trained using one GPU with 256GB of memory, and it takes at most seven days to finish 50000 episodes.

\begin{figure}[!htbp]
    \centering
    \begin{minipage}{1\linewidth}
    \includegraphics[width=1\linewidth]{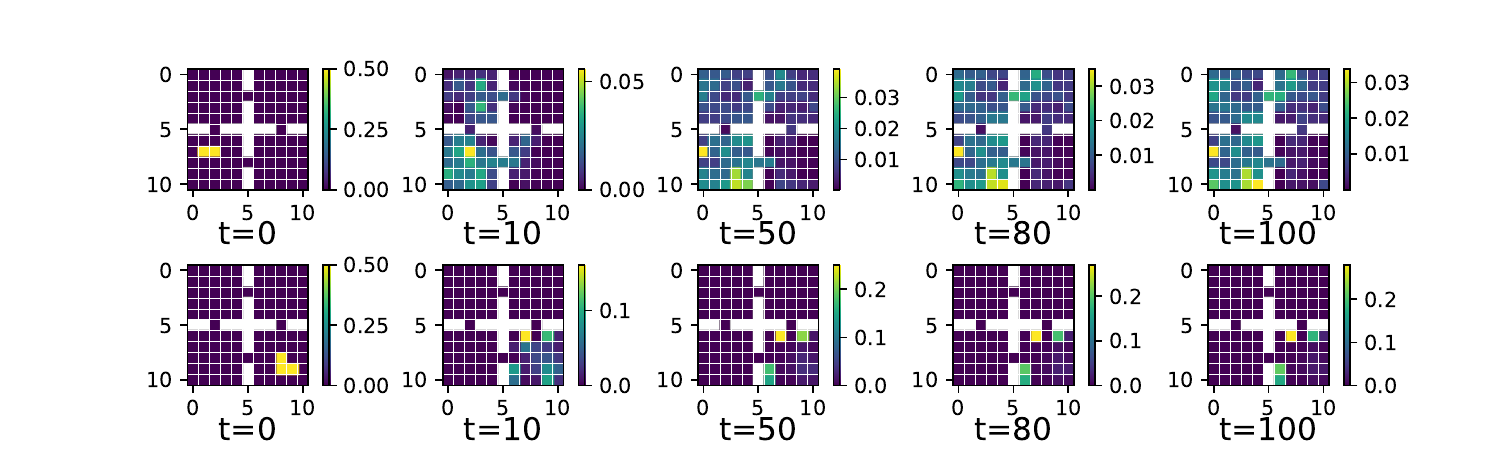}
    \end{minipage}
    \begin{minipage}{1\linewidth}
    \includegraphics[width=1\linewidth]{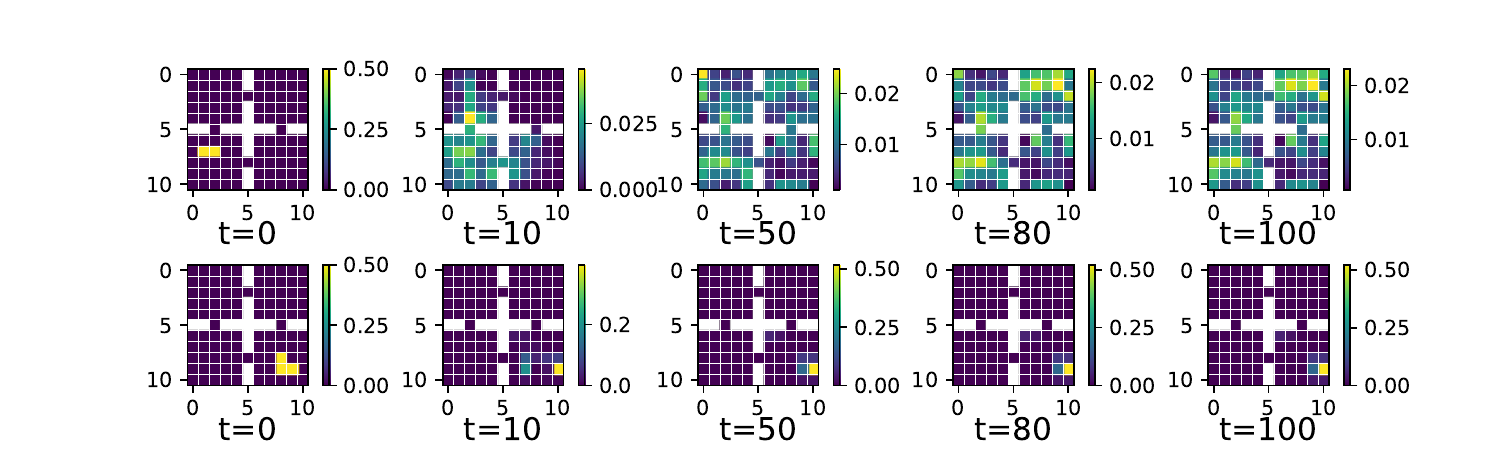}
    \end{minipage} 
\caption{\textbf{Ex.~2:} 
    populations evolution 2. The top two rows show the distribution evolution of the two players. The bottom two rows show the corresponding distribution evolution of the baseline model. }
\label{fig:four_room_distrib_evo_2}
\end{figure}

\subsection{Example 3: Predator-prey 2D with 4 groups}
\label{app:ex4-predator-prey}
\paragraph{Model.}

\begin{figure}[!htbp]
    \centering
    \begin{minipage}
        {\linewidth}
    \includegraphics[width=1\linewidth]{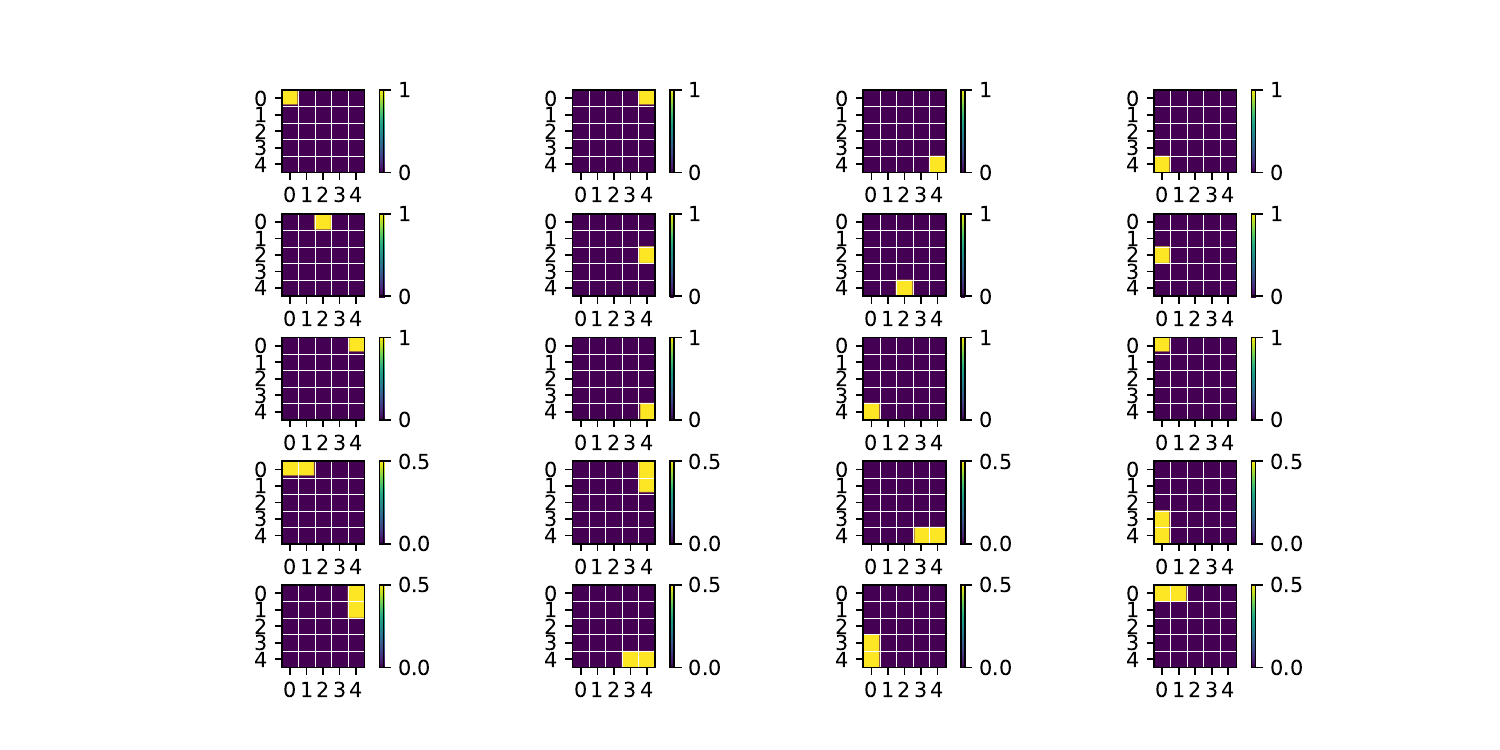}

    \caption{
   5 sets of testing distributions for predator-prey 2D with 4 groups. Each row shows one set of testing distributions for 4 coalitions. For each row, from left to right, are Coalitions 1 to 4.
    }
    \label{fig:testing distribution of predator-prey 2D}
    \end{minipage}
\end{figure}

In this $5 \times 5$ dimensional grid world, the transition dynamics and the action space are the same as in Example 2. In this game, we have one coalition acting as the predator and another coalition as the prey. Their reward function can be formulated as follows:
\begin{equation*}
    \bar{r}^1(\bar{s}_t,\bar{a}^1)=c_1 r_{\text{move}}(\bar{s}^1,\bar{a}^1)+c_2  \bar{s}^1\cdot\bar{s}^{2}
\end{equation*}
\begin{equation*}
    \bar{r}^4(\bar{s}_t,\bar{a}^4)=c_1 r_{\text{move}}(\bar{s}^4_t,\bar{a}^4)-c_2 \bar{s}^{3}\cdot\bar{s}^{4}
\end{equation*}
The remaining two coalitions act as predator and prey at the same time, with rewards:\begin{equation*}
    \bar{r}^2(\bar{s}_t,\bar{a}^2)=c_1 r_{\text{move}}(\bar{s}^2,\bar{a}^2)+c_2 (\bar{s}^2\cdot\bar{s}^{3}-\bar{s}^{1}\cdot\bar{s}^{2})
\end{equation*}
\begin{equation*}
    \bar{r}^3(\bar{s}_t,\bar{a}^3)=c_1 r_{\text{move}}(\bar{s}^3,\bar{a}^3)+c_2 (\bar{s}^3\cdot\bar{s}^{4}-\bar{s}^{2}\cdot\bar{s}^{3}),
\end{equation*}
where $c_1=c_2=100$. Each episode has a time horizon $T=21$ and $\gamma=0.99$.
\paragraph{Training and testing set}
For the training set, we sample each element in the grid world from a uniform distribution over the interval $[0,1)$ and then divide each element by the total sum of the matrix to normalize it. The testing set is shown in Fig.~\ref{fig:testing distribution of predator-prey 2D}.
\paragraph{Neural network architecture and hyperparameters} 
The architectures of the actor and critic networks are the same as those used in the discrete planning 2D (Suppl.~\ref{app:ex2-distrib-plan}). We use the Adam optimizer, with learning rates set to 0.0005 for the actor network and 0.001 for the critic network. The Ornstein-Uhlenbeck noise standard deviation is set to 0.8. Target networks are updated at a rate of 0.0025. The replay buffer has a capacity of 50,000 and a batch size of 64. This experiment was run on a GPU with 64GB of memory, taking two days to complete 80,000 episodes of training.

\begin{figure}[!htbp]%
    \centering
    \includegraphics[width=0.4\linewidth]{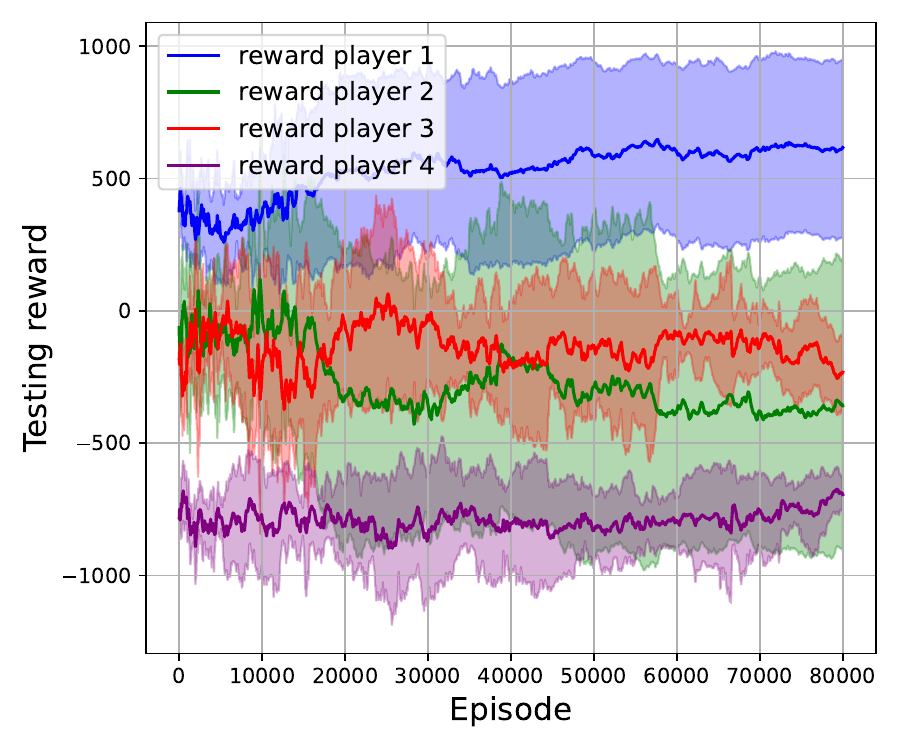} 
    \caption{
    \textbf{Ex.~3:} testing rewards. %
    }
    \label{fig: predator-pray 2D with 4 groups testing reward only}
\end{figure}

\paragraph{\bf Numerical results.}
 We conducted this experiment over 5 runs, with each run corresponding to a specific testing distribution from the testing set. For each run, we averaged the exploitability of all players to determine the run's exploitability. We then calculated the mean and standard deviation of exploitability across the 5 runs. Additionally, for the testing reward, we calculated the mean and standard deviation for each player over the 5 runs. Fig.~\ref{fig: predator-pray 2D with 4 groups testing reward only} shows the testing rewards.
\subsection{Example 4: Distribution planning in 2D}
\label{app:ex2-distrib-plan}
\begin{figure}[!htbp]
    \centering
    \begin{minipage}{.35\linewidth}
    \includegraphics[width=1\linewidth]{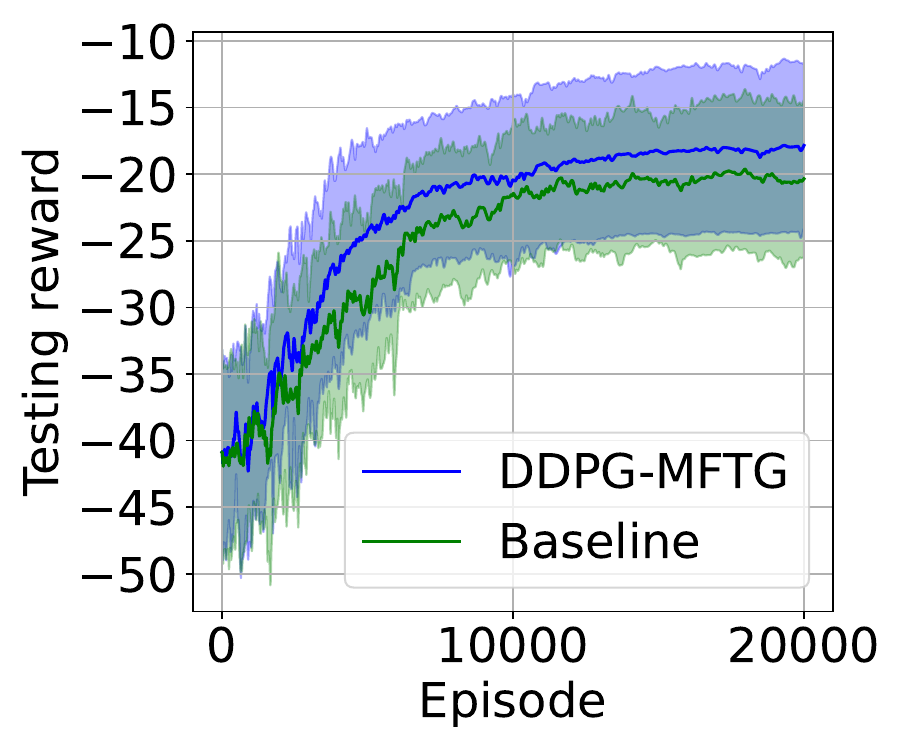} 
    \end{minipage}
    \,
    \begin{minipage}{.35\linewidth}
    \includegraphics[width=1\linewidth]{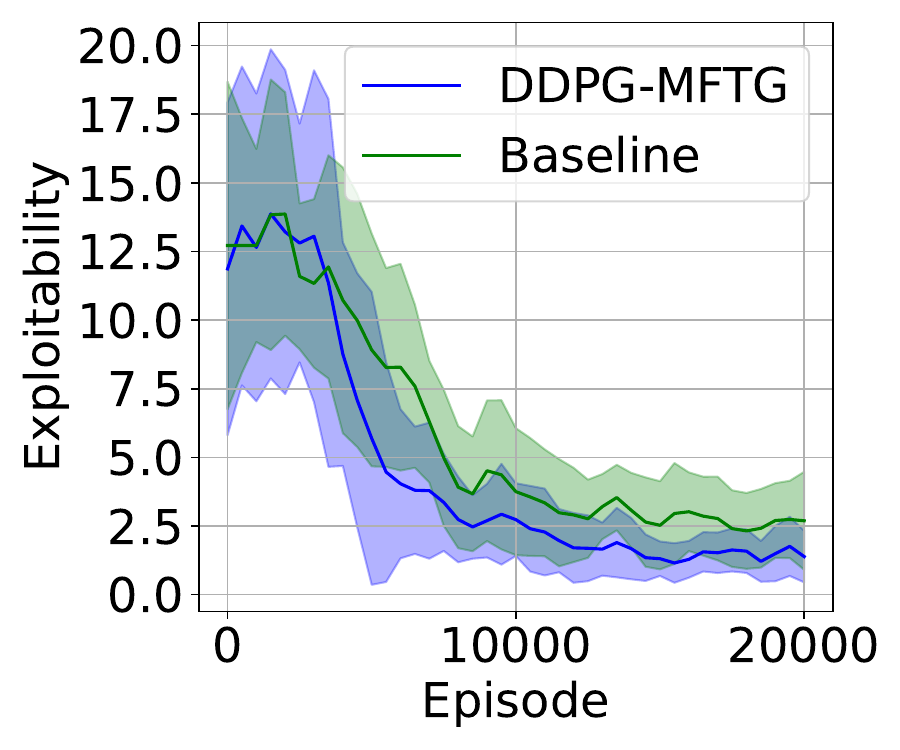} 
    \end{minipage} 

    \caption{
    Left:  Testing rewards. Right: exploitabilities.
    }
    \label{fig:2D-Plan-rewards-exploitability}
\end{figure}

\begin{figure}[!htbp]
    \centering
    \begin{minipage}{.8\linewidth}
    \includegraphics[width=1\linewidth]{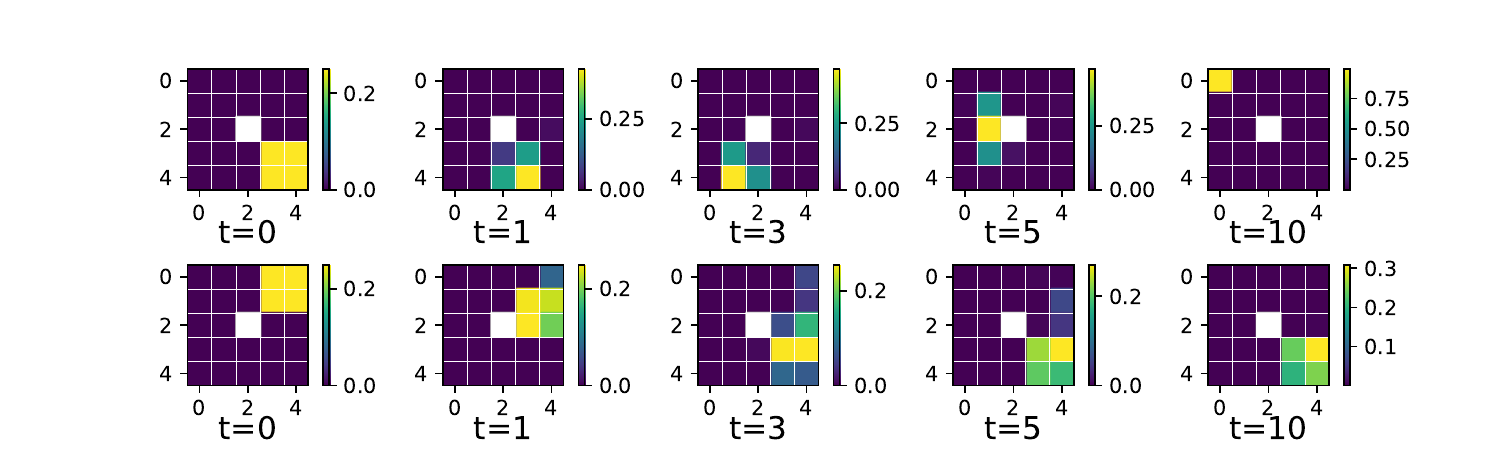} 
    \end{minipage}

    \caption{
    Distribution planning in 2D: The top row and the bottom row, respectively, show the distribution evolution of players 1 and 2 using the policy learned by DDPG-MFTG.
    }
    \label{fig:2D-Plan-distributions}
\end{figure}

\begin{figure}[H]
    \centering
    \includegraphics[width=0.5\linewidth]{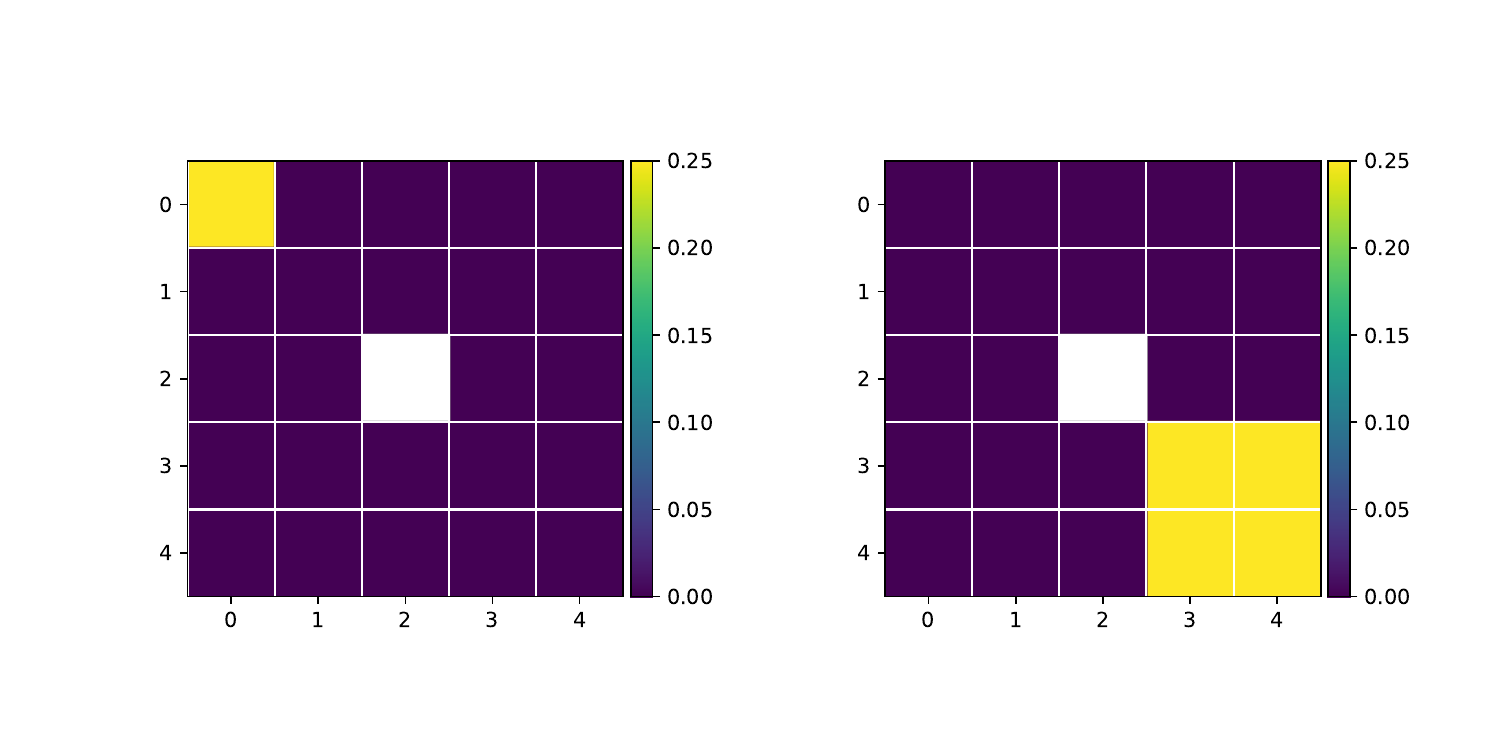} 

\caption{Target distributions for player 1 (left) and player 2 (right).}
\label{fig:2D-Plan-target}
\end{figure}

There are $m=2$ populations. The agent's state space is a $5 \times 5$ state 2D grid world, with the center as a forbidden state. The possible actions are to move up, down, left, right, or stay, and there is no individual noise perturbing the movements. The rewards encourage each population to match a target distribution (hence the name ``planning''): Population 1 and 2 move respectively towards the top left and bottom right corners, with a uniform distribution over fixed locations (see Fig.~\ref{fig:2D-Plan-target}). We describe the model details and the training and testing distributions below. We implement {\bf DDPG-MFTG} to solve this game. The numerical results are presented in Figs.~\ref{fig:2D-Plan-rewards-exploitability} and~\ref{fig:2D-Plan-distributions}. We make the following observations.
    \textbf{Testing reward curves: } Fig.~\ref{fig:2D-Plan-rewards-exploitability} (left) shows the testing rewards. In this game setting, the Nash equilibrium for each coalition is to move to its target position without interacting with the other coalition. We observe that the testing rewards increase and then stabilize with minimal oscillation. The reward curve of the baseline stays below that of the one using DDPG-MTFG.
    \textbf{Exploitability curves:} Fig.~\ref{fig:2D-Plan-target} (right) shows the averaged exploitabilities over the testing set and players. We observe that the exploitability stabilizes near zero after around 15000 episodes, indicating that players reach an approximate Nash equilibrium. The baseline shows higher exploitability than the {\bf DDPG-MFTG} algorithm.
    \textbf{Distribution plots:} Fig.~\ref{fig:2D-Plan-distributions} illustrates the distribution evolution during the game. With the policy learned using {\bf DDPG-MFTG}, each player deterministically moves to the target position in several steps and avoids overlapping with the other player during movement.

\paragraph{Model.}
\begin{itemize}
    \item {\bf Number of populations:} $m=2$.
    \item {\bf State space:} $ S=\{0,\dots,N^1_x\}\times\{0,\dots,N^2_x\}$, where we set $N^1_x=N^2_x=4$. 
    \item {\bf Action space:} $ A=\{(-1,0), (1,0), (0,0), (0,1), (0,-1)\}$, which represents move left, move right, stay, move up, and move down, respectively.
    \item {\bf Transitions:} At time $n$, the agent at position $s_n=(x,y)$ chooses an action $a_n$, the next state is computed according to 
    \begin{equation}
    s_{n+1}=
    \begin{cases}
    s_n + a_n, & \text{if } s_n + a_n \text{ is not in a forbidden state}\\
    s_n, & \text{otherwise}
    \end{cases}
    \end{equation}
    
    The mean-field distribution $\bar{s}^i_{t}(x,y)$ is computed according to
    \begin{align*}
    \bar{s}^i_{t+1}(x,y)&=\bar{s}^i_{t}(x,y)\bar{a}^i((0,0)|(x,y))+\bar{s}^i_{t}(x,y-1)\bar{a}^i((0,1)|(x,y-1))\\
    &\quad+\bar{s}^i_{t}(x,y+1)\bar{a}^i((0,-1)|(x,y+1))+\bar{s}^i_{t}(x+1,y)\bar{a}^i((-1,0)|(x+1,y))\\
    &\quad+\bar{s}^i_{t}(x-1,y)\bar{a}^i((1,0)|(x-1,y))\\
    \end{align*}
    where $\bar{s}^i_t(a,b)$ is the density of Population i at the location $(a,b)$ at time step $t$.
    \item {\bf One-step reward function:} Each central player $i$ aims to make the population match a target distribution $m_i$ while maximizes the reward. For each player $i$, the reward of each step is $$\bar{r}^i(\bar{s}^1_t, \bar{s}^2_t,\bar{a}^i)=c_1r_{\text{move}}(\bar{s}^i,\bar{a}^i)+c_2r(\bar{s}^i,m_i)+c_3r(\bar{s}^1,\bar{s}^2),$$ where $r_{\text{move}}(\bar{s}^i,\bar{a}^i)=-\bar{s}^i\cdot||\bar{a}^i||$ is the cost for moving, $r(\bar{s}^i,m_i)=-\text{dist}(\bar{s}^i,m_i)$ is the distance to a target distribution, $r(\bar{s}^1,\bar{s}^2)=-\bar{s}^1\cdot\bar{s}^2$ is the inner product of the two population distributions. $c_i$ is the coefficient, for $i=1, 2, 3$. Here, $c_1=1$, $c_2=2$, and $c_3=5$.
    \item {\bf Time horizon:} $N_T=10$.
\end{itemize}

\begin{figure}[H]
    \centering
    \begin{minipage}{0.47\linewidth}
    \includegraphics[width=1\linewidth]{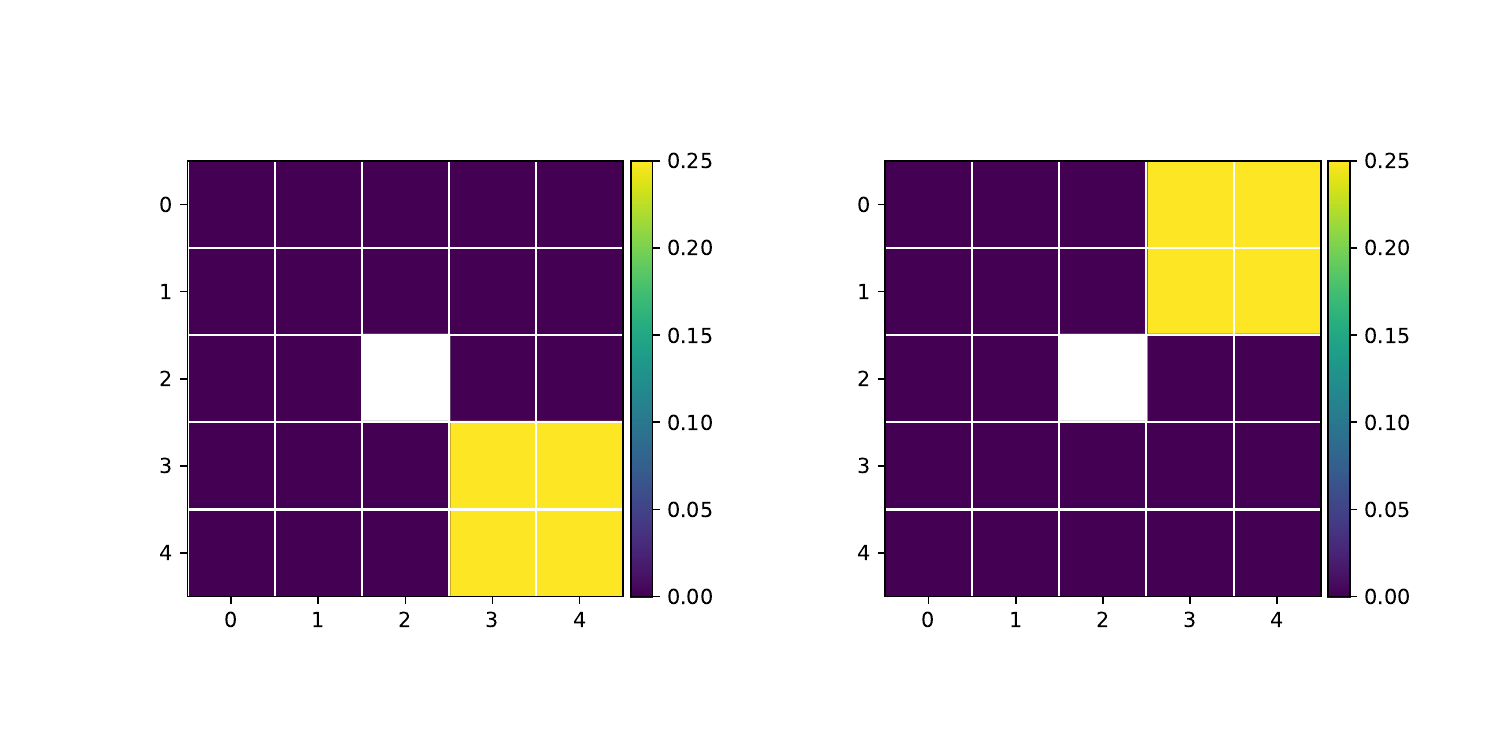} 
    \end{minipage}
    \begin{minipage}{0.47\linewidth}
    \includegraphics[width=1\linewidth]{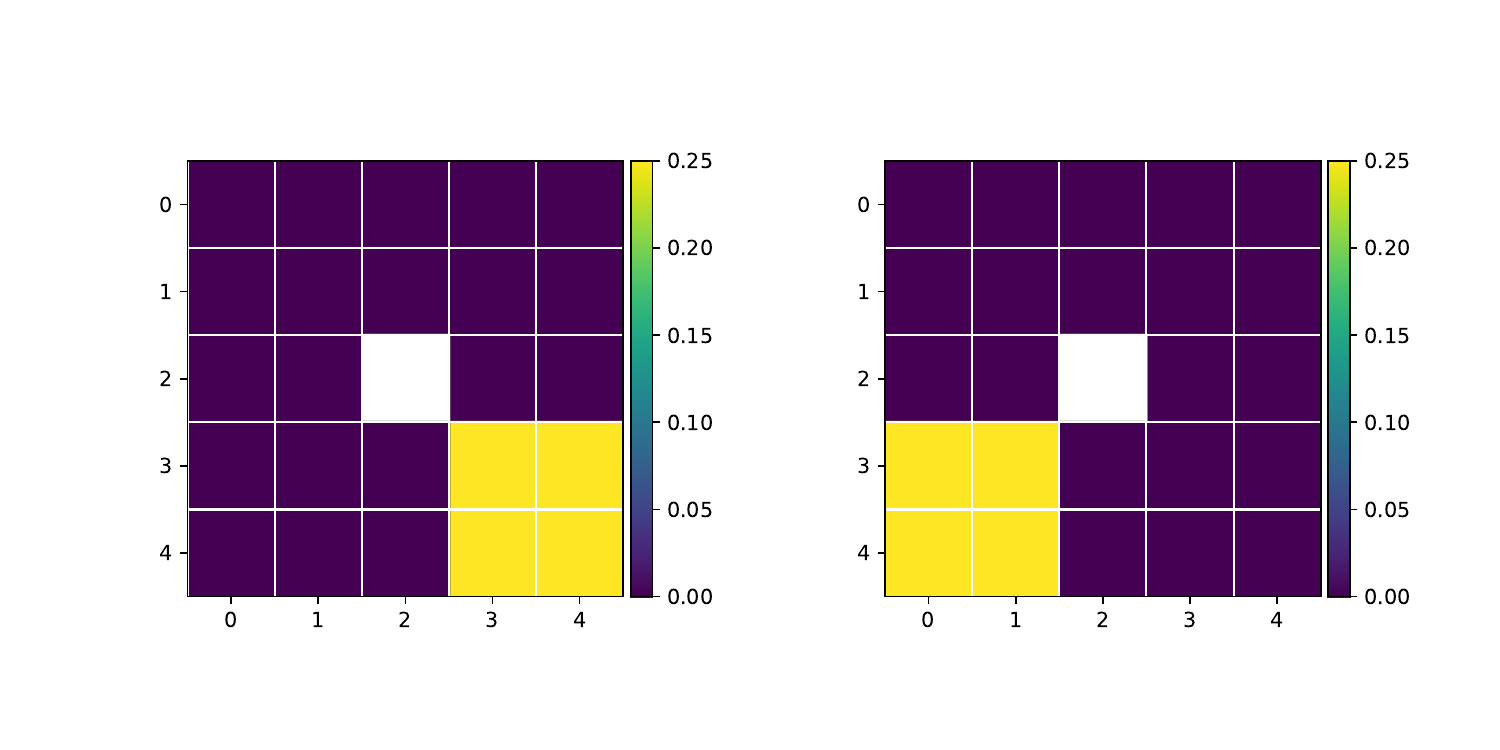}
    \end{minipage} 
     \begin{minipage}{0.47\linewidth}
    \includegraphics[width=1\linewidth]{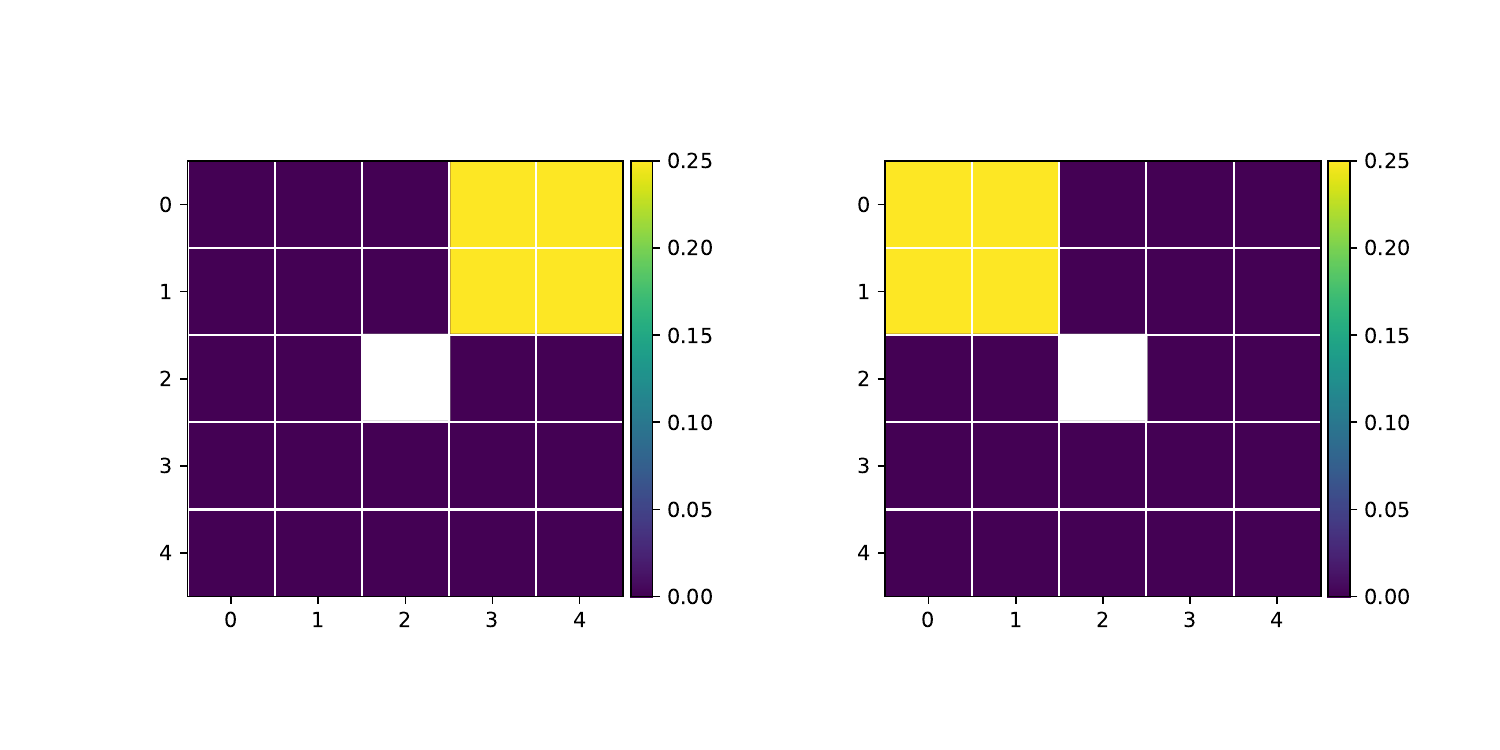} 
    \end{minipage}
    \begin{minipage}{0.47\linewidth}
    \includegraphics[width=1\linewidth]{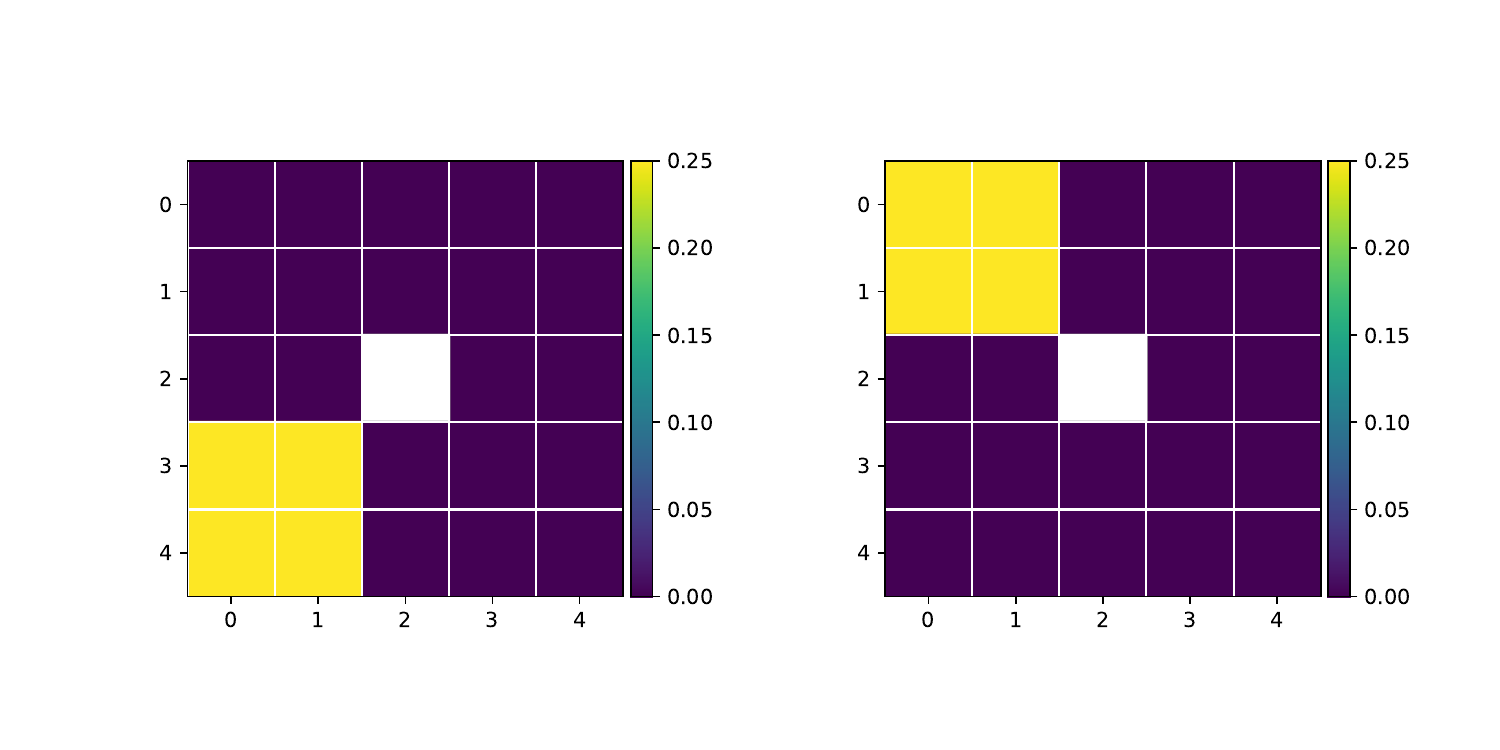}
    \end{minipage} 
\caption{4 pairs of testing distributions. For each pair, the left one is the initial distribution of player 1, and the right one is the initial distribution of player 2.}
\label{fig:test_2D}
\end{figure}

\paragraph{\bf Training and testing sets.}
The training set consists of a randomly sampled location with a probability density $1$ representing the initial state. See Fig.~\ref{fig:test_2D} for testing distribution.

\paragraph{Neural network architecture and hyperparameters}

In the actor network, each state vector is initially flattened and fed into a fully connected network with a ReLU activation function, resulting in a 200-dimensional output for each. These outputs are then concatenated and processed through a two-layer fully connected network with 200 hidden neurons, utilizing ReLU and Tanh activation functions. The final output dimension is $|S|\times|A|$. The output is then normalized using the softmax function. The critic network follows a similar architecture, where we use the ReLU in the last layer. During the training, we use the Adam optimizer with the actor-network learning rate equal to $5 \times 10^{-5}$ and the critic-network learning rate equal to 0.0001. Both learning rates are reduced by half after around 6000 and 12000 episodes.  The standard deviation used in the Ornstein–Uhlenbeck process is 0.08 and is also reduced by half after around 6000 and 12000 episodes. We also use target networks to stabilize the training, and the update rate is 0.005. The replay buffer is of size 50000, and the batch size is 128. The model is trained using one GPU with 256GB of memory, and it takes at most two days to finish 20000 episodes.

\subsection{Summary of improvements}
\label{sec:improvements}

In Table~\ref{table:improvements}, we summarize the improvements brought about by our method compared to the corresponding baseline in each example. The quantities are:
\begin{itemize}
    \item {\bf Baseline Exploitability:} The baseline's mean value (as described in the paper).
    \item {\bf Our Exploitability:} Our method's mean value (as described in the paper).
    \item {\bf Improvement:} The percentage improvement is calculated as:
    $$
      \text{Improvement (percentage)} = \frac{\text{Baseline} - \text{Ours}}{\text{Baseline}} \times 100.
    $$
\end{itemize}

\begin{table}[h!]
\centering
\begin{tabular}{|c|c|c|c|c|}
\hline
                 & \textbf{Example 1} & \textbf{Example 2} & \textbf{Example 3} & \textbf{Example 4} \\
\hline
\textbf{Baseline Exploitability} & 2355.35 & 3.13 & 131.43 & 2.69 \\
\hline
\textbf{Our Exploitability} & 471.40 & 2.16 & 38.75 & 1.39 \\
\hline
\textbf{Improvement} & 79.98\% & 31.0\% & 70.52\% & 48.3\\ 
\hline
\end{tabular}
\caption{Comparison of baseline and our exploitability metrics across the 4 examples described in the text, along with percentage improvement.}
\label{table:improvements}
\end{table}

\clearpage
\section{Hyperparameters sweep}
\label{sec:sweeps}

We explore various batch sizes, actor learning rates, and standard deviations of Ornstein-Uhlenbeck noise (OU noise) across all numerical experiments. Heuristically, we set $\alpha_{\text{critic}} = 2 \times \alpha_{\text{actor}}$ and $\tau = 5 \times \alpha_{\text{actor}}$. Each hyperparameter group is evaluated during one player's exploitability computation stage, and the results are presented as follows:
\subsection{Predator-prey 2D with 4 groups}
\begin{figure}[!htbp]
    \centering
    
    \begin{minipage}{0.23\linewidth}
    \includegraphics[width=1\linewidth]{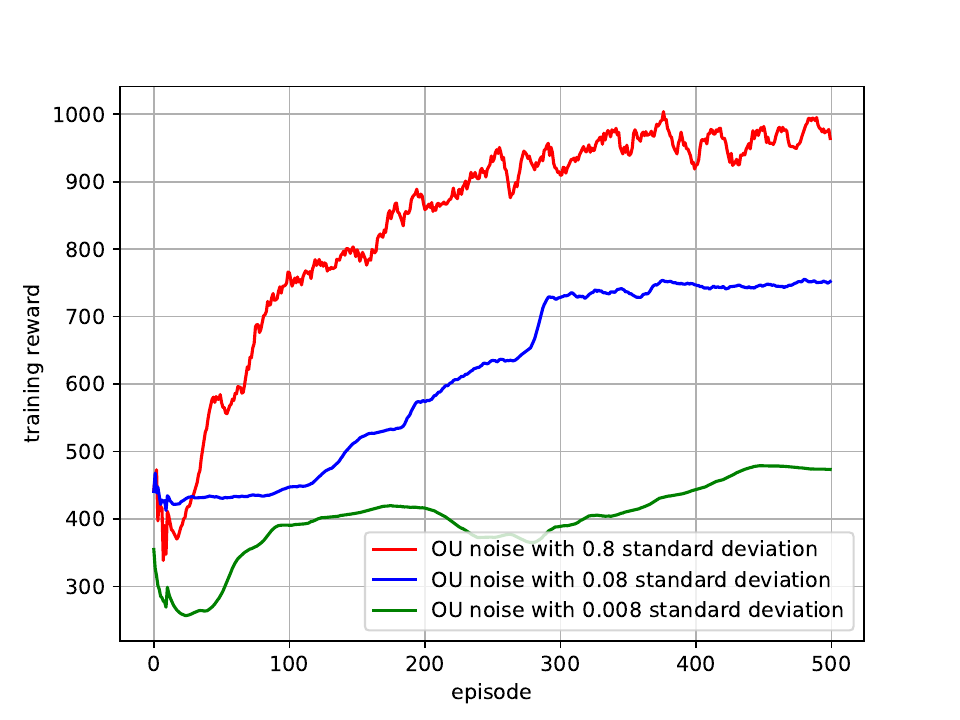} 
    \end{minipage}
    \begin{minipage}{0.23\linewidth}
    \includegraphics[width=1\linewidth]{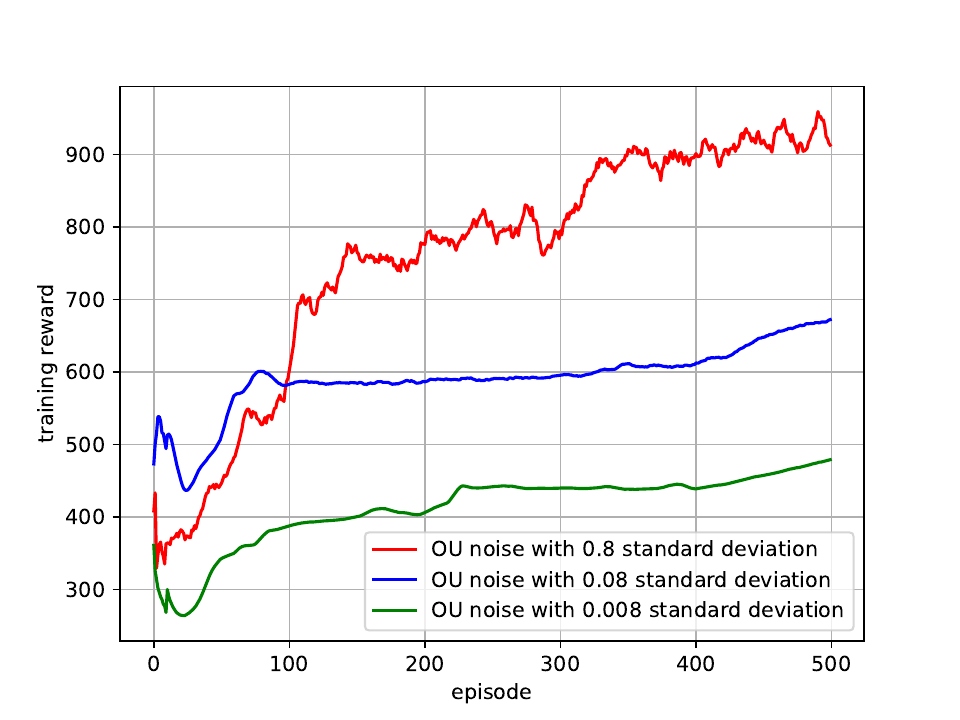} 
    \end{minipage}
    \begin{minipage}{0.23\linewidth}
    \includegraphics[width=1\linewidth]{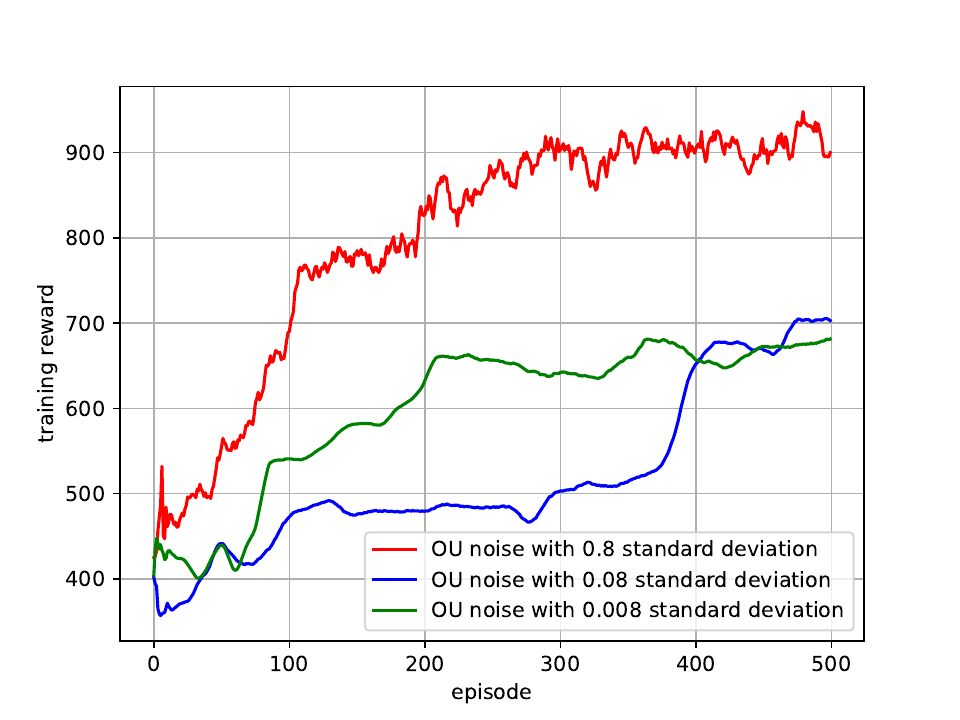} 
    \end{minipage} 
    \begin{minipage}{0.23\linewidth}
    \includegraphics[width=1\linewidth]{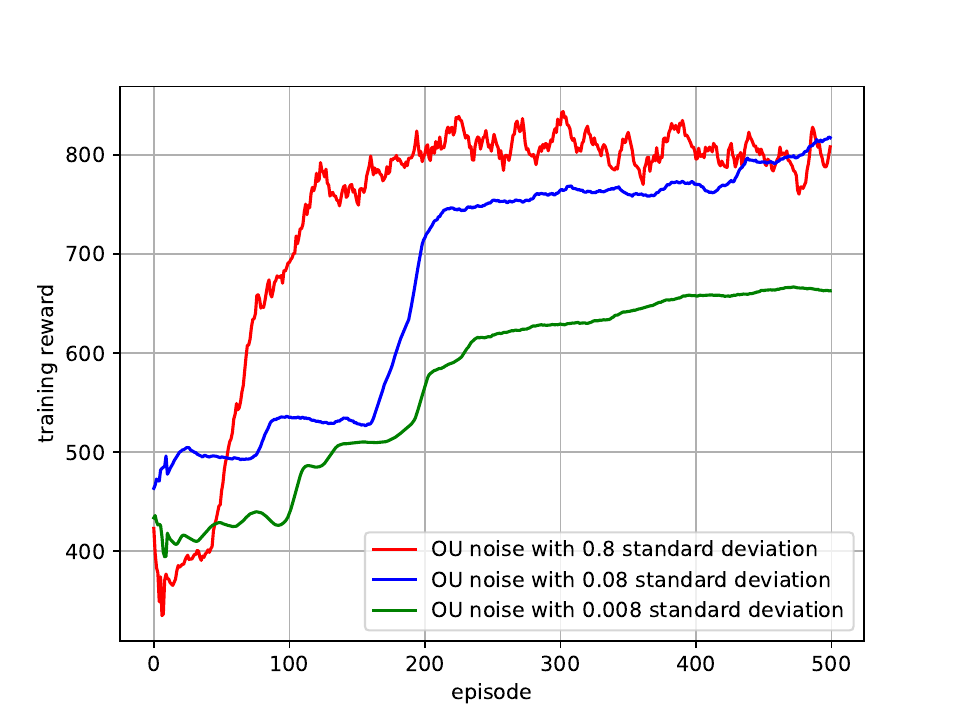} 
    \end{minipage} 
    \caption{
    Exploitability computation training reward with $\alpha_{\text{actor}}=5\times10^{-5}$. Batch size from left to right: 16, 32, 64, 128.
    }
    \label{fig:hyper predator-prey 2D with 4 groups with lr=5e-5}
\end{figure}
\begin{figure}[!htbp]
    \centering
    
    \begin{minipage}{0.23\linewidth}
    \includegraphics[width=1\linewidth]{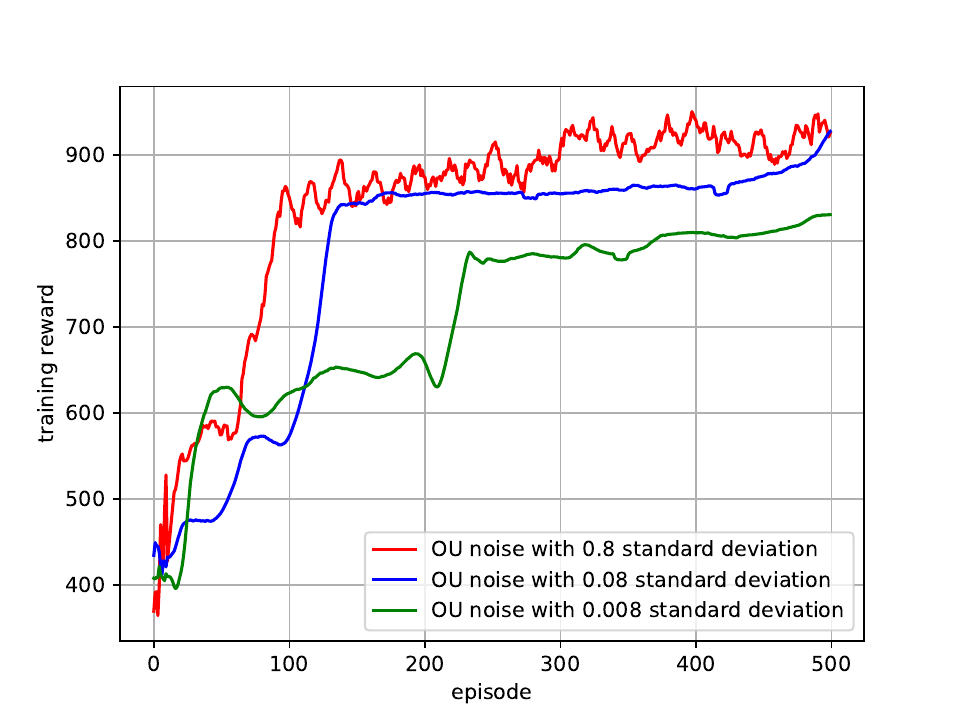} 
    \end{minipage}
    \begin{minipage}{0.23\linewidth}
    \includegraphics[width=1\linewidth]{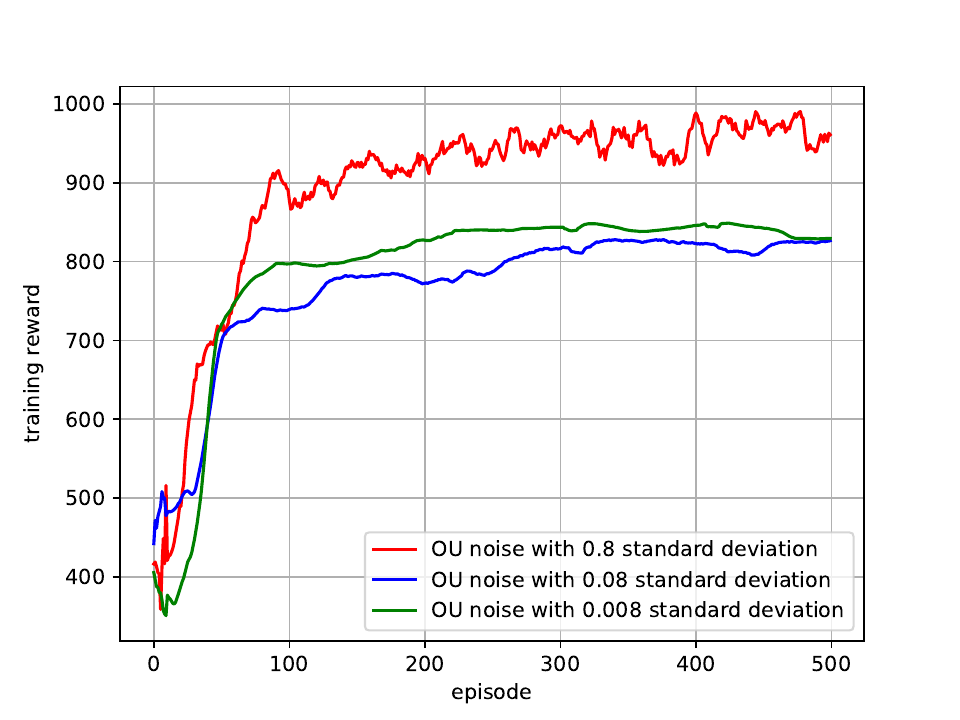} 
    \end{minipage}
    \begin{minipage}{0.23\linewidth}
    \includegraphics[width=1\linewidth]{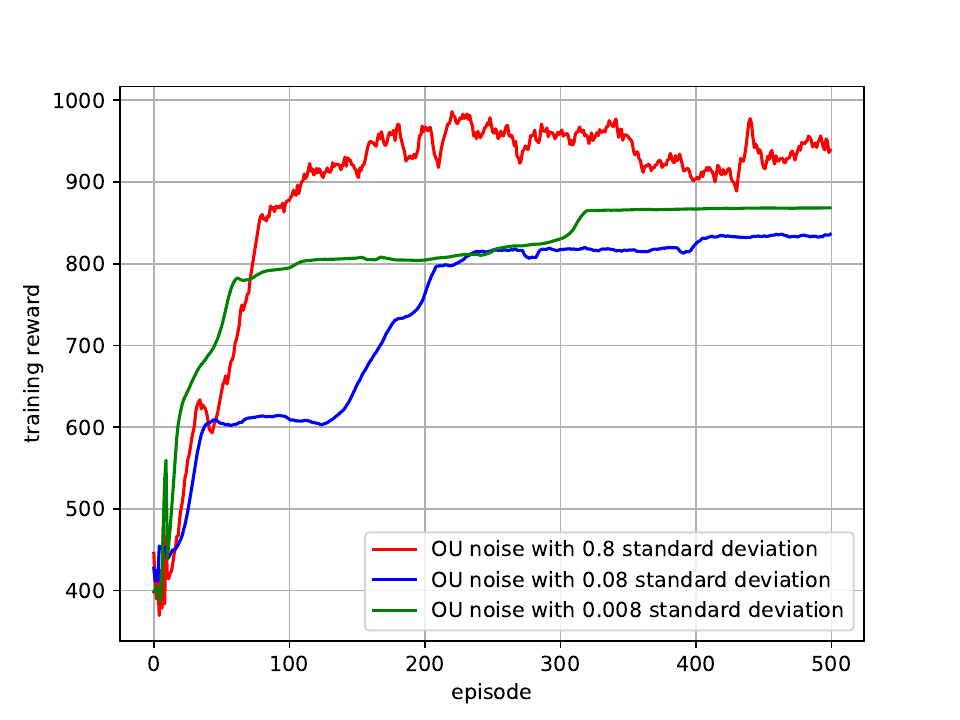} 
    \end{minipage} 
    \begin{minipage}{0.23\linewidth}
    \includegraphics[width=1\linewidth]{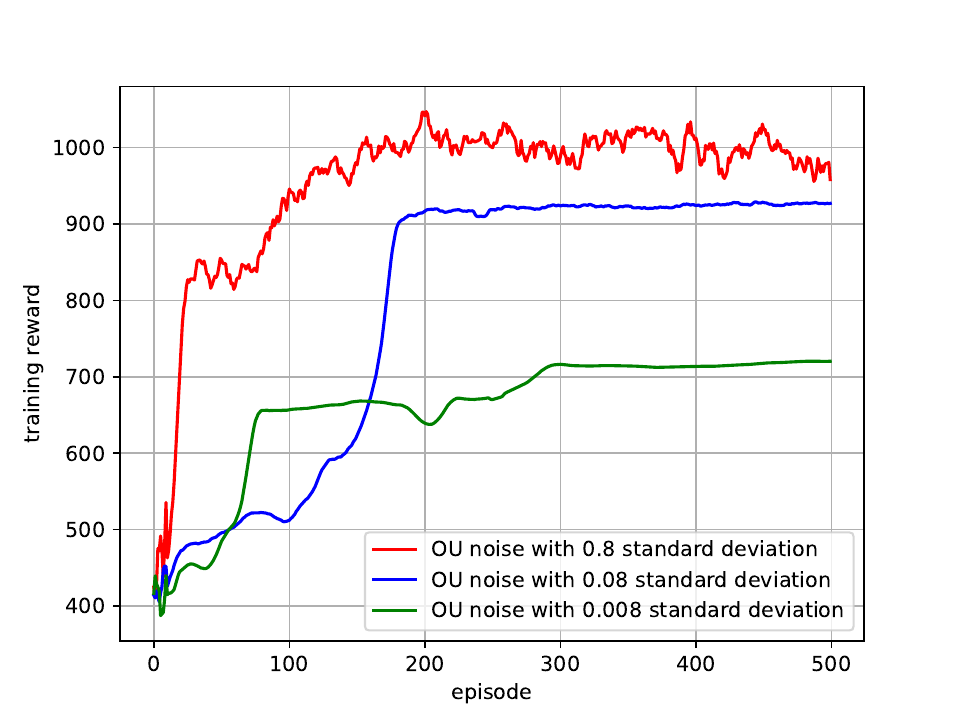} 
    \end{minipage} 
    \caption{
    Exploitability computation training reward with $\alpha_{\text{actor}}=0.0005$. Batch size from left to right: 16, 32, 64, 128.
    }
    \label{fig:hyper predator-prey 2D with 4 groups with lr 5e-4}
\end{figure}
\begin{figure}[!htbp]
    \centering
    
    \begin{minipage}{0.23\linewidth}
    \includegraphics[width=1\linewidth]{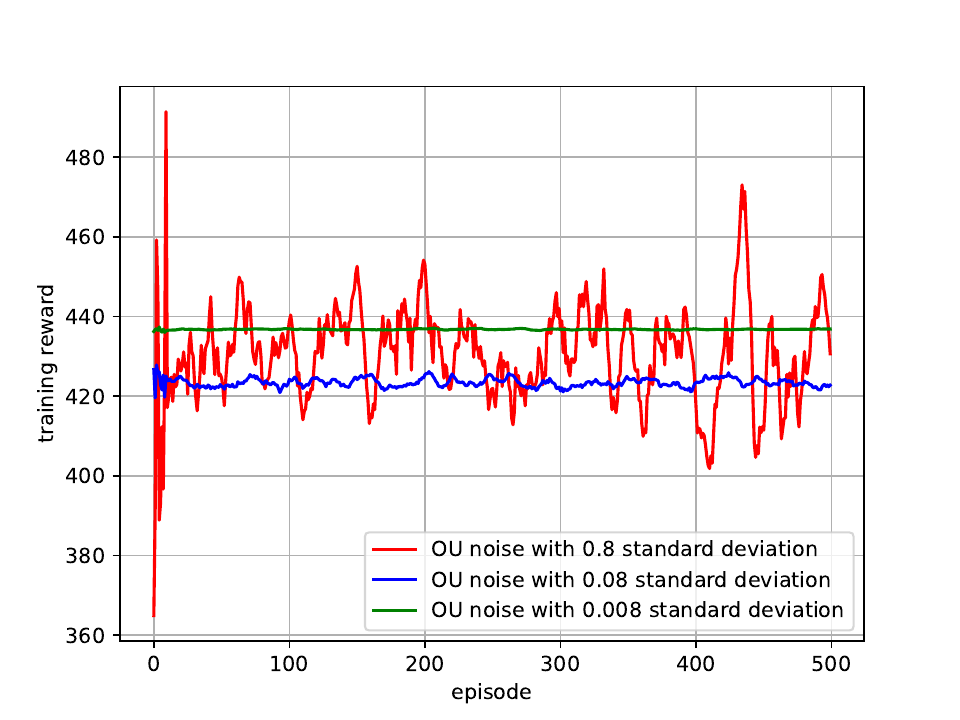} 
    \end{minipage}
    \begin{minipage}{0.23\linewidth}
    \includegraphics[width=1\linewidth]{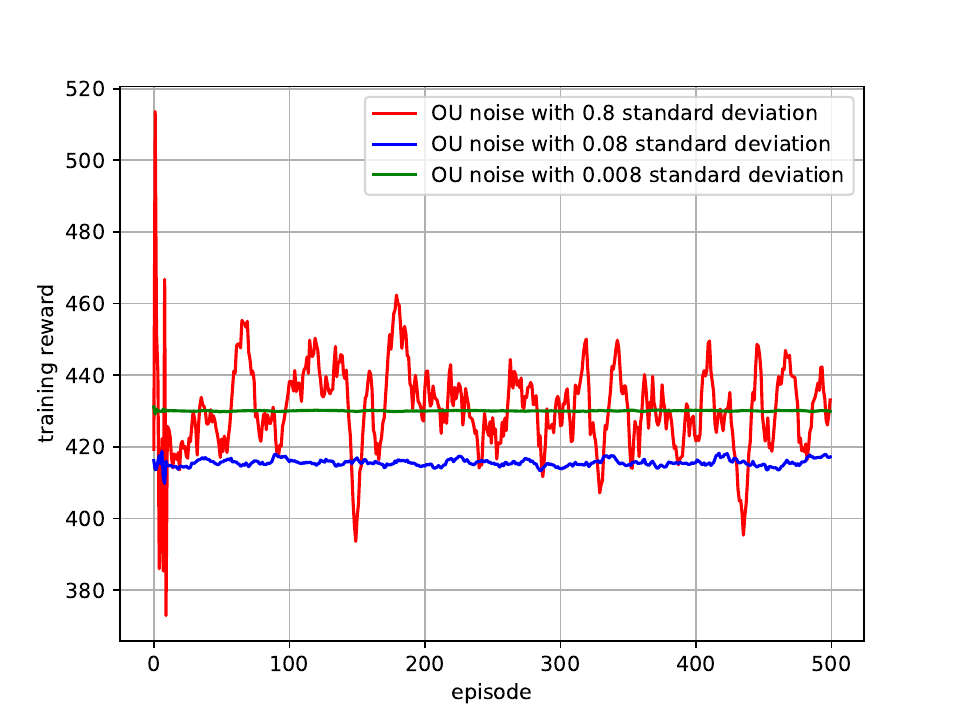} 
    \end{minipage}
    \begin{minipage}{0.23\linewidth}
    \includegraphics[width=1\linewidth]{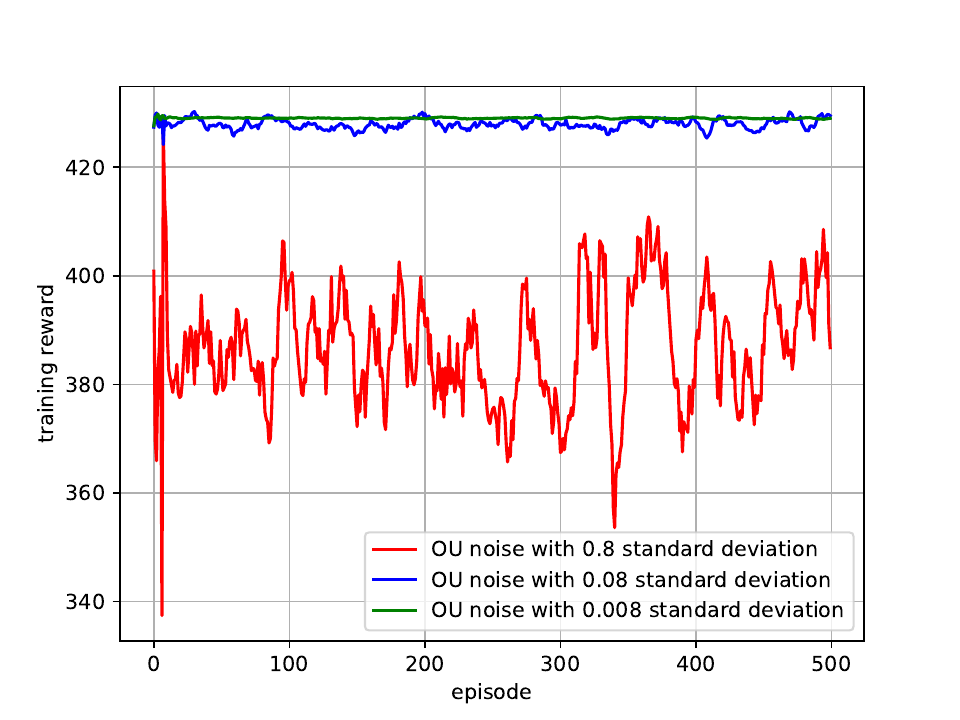} 
    \end{minipage} 
    \begin{minipage}{0.23\linewidth}
    \includegraphics[width=1\linewidth]{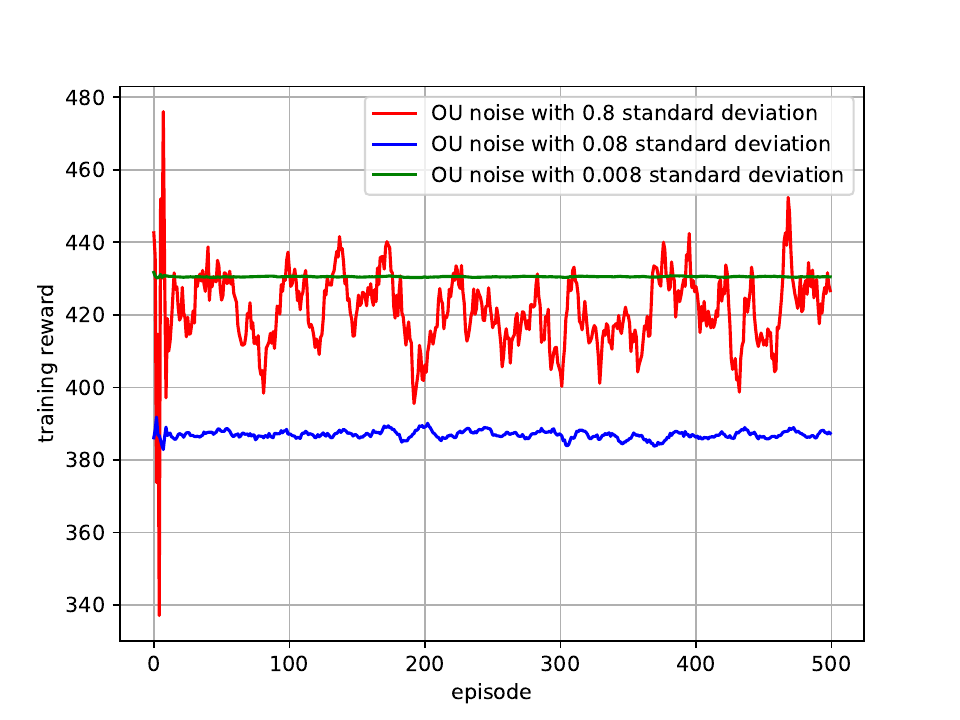} 
    \end{minipage} 
    \caption{
    Exploitability computation training reward with $\alpha_{\text{actor}}=0.005$. Batch size from left to right: 16, 32, 64, 128.
    }
    \label{fig:hyper predator-prey 2D with 4 groups with learning rate 5e-3}
\end{figure}

\begin{figure}[!htbp]
    \centering
    
    \begin{minipage}{0.23\linewidth}
    \includegraphics[width=1\linewidth]{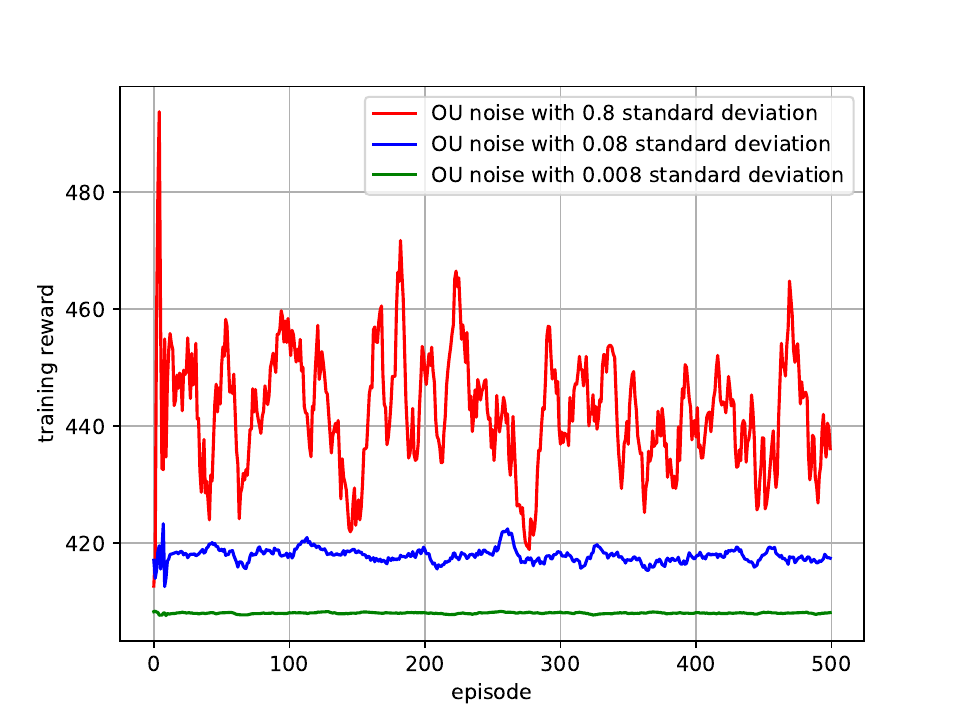} 
    \end{minipage}
    \begin{minipage}{0.23\linewidth}
    \includegraphics[width=1\linewidth]{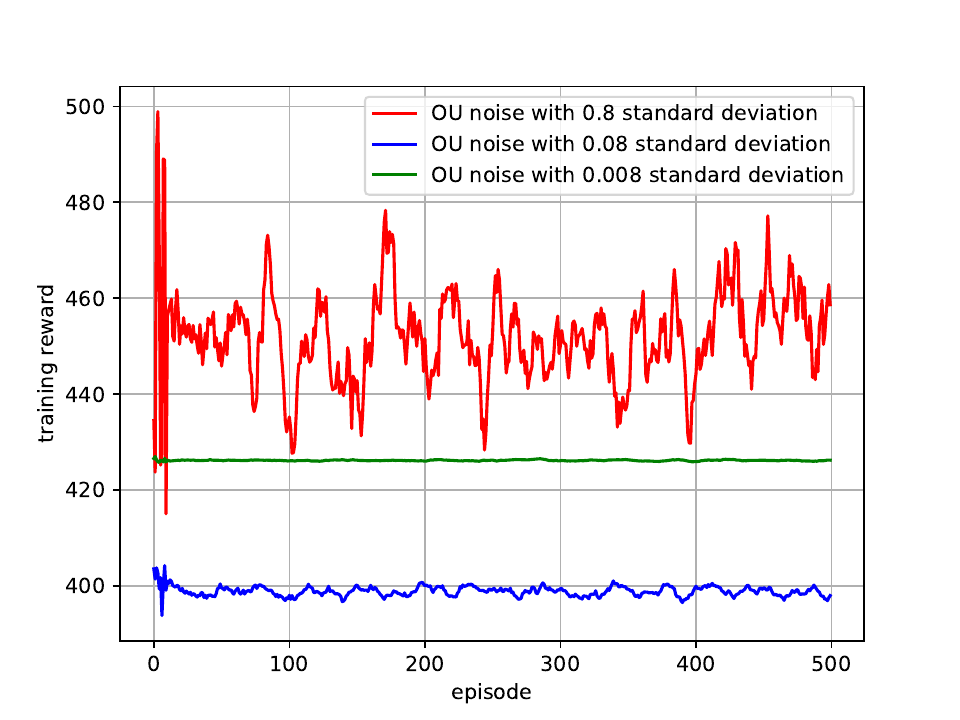} 
    \end{minipage}
    \begin{minipage}{0.23\linewidth}
    \includegraphics[width=1\linewidth]{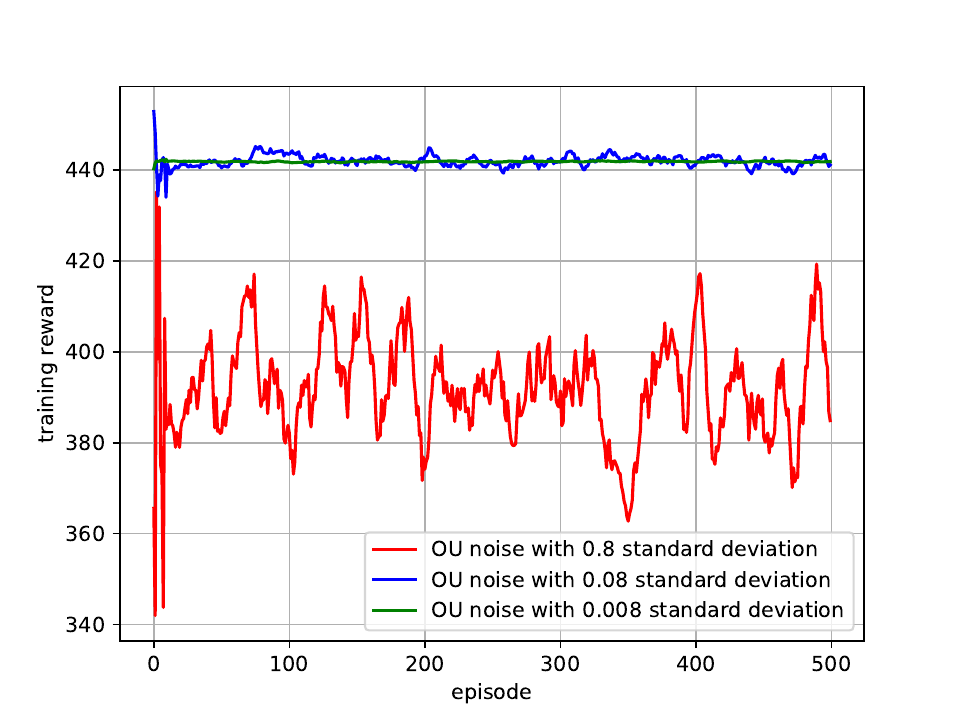} 
    \end{minipage} 
    \begin{minipage}{0.23\linewidth}
    \includegraphics[width=1\linewidth]{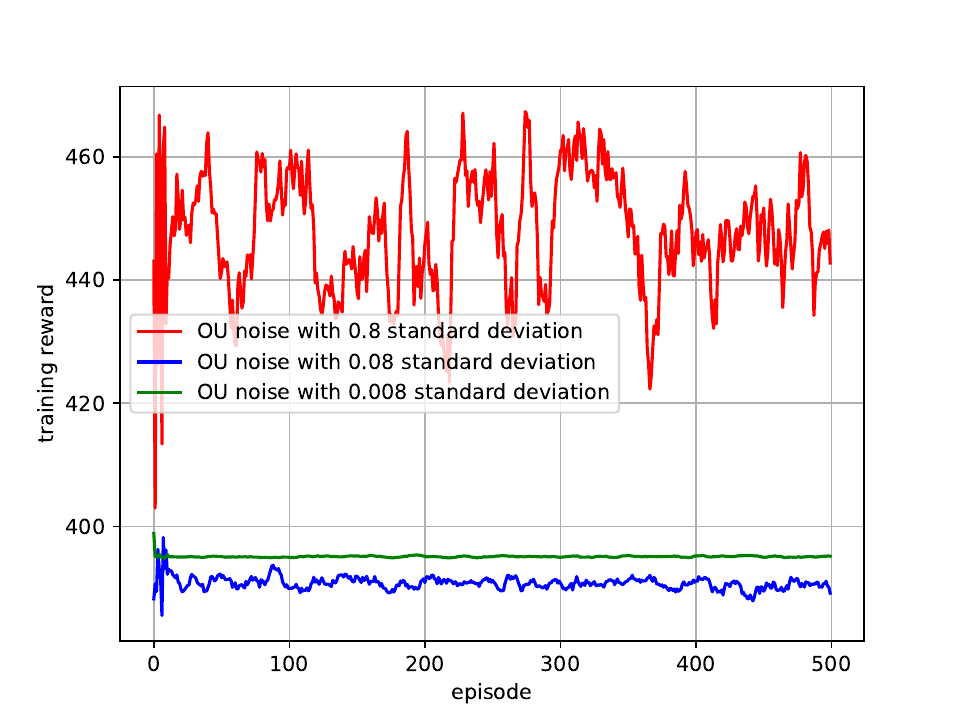} 
    \end{minipage} 
    \caption{
    Exploitability computation training reward with $\alpha_{\text{actor}}=0.05$. Batch size from left to right: 16, 32, 64, 128.
    }
    \label{fig:hyper predator-prey 2D with 4 groups with learning rate 5e-2}
\end{figure}

\clearpage
\subsection{Distribution planning in 2D}
\begin{figure}[!htbp]
    \centering
    \begin{minipage}{0.23\linewidth}
    \includegraphics[width=1\linewidth]{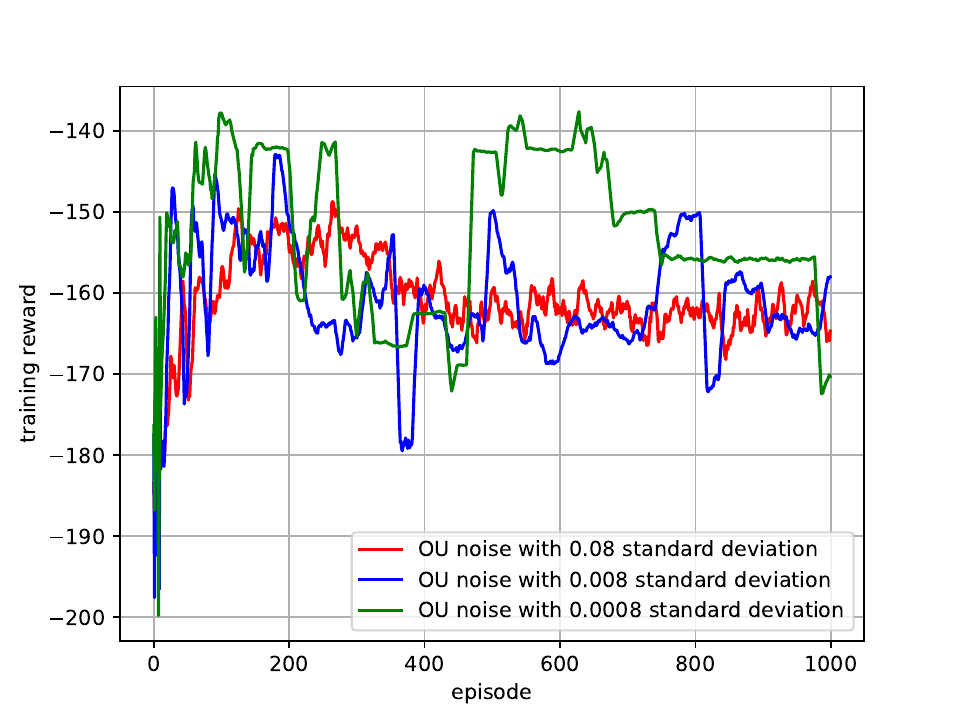} 
    \end{minipage}
    \begin{minipage}{0.23\linewidth}
    \includegraphics[width=1\linewidth]{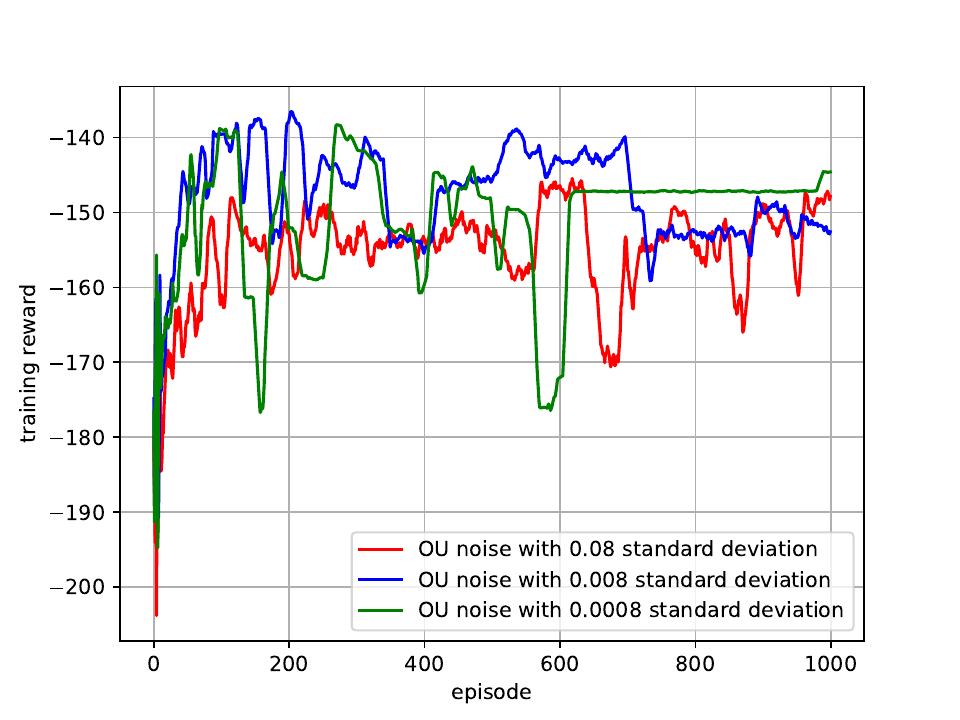} 
    \end{minipage} 
    \begin{minipage}{0.23\linewidth}
    \includegraphics[width=1\linewidth]{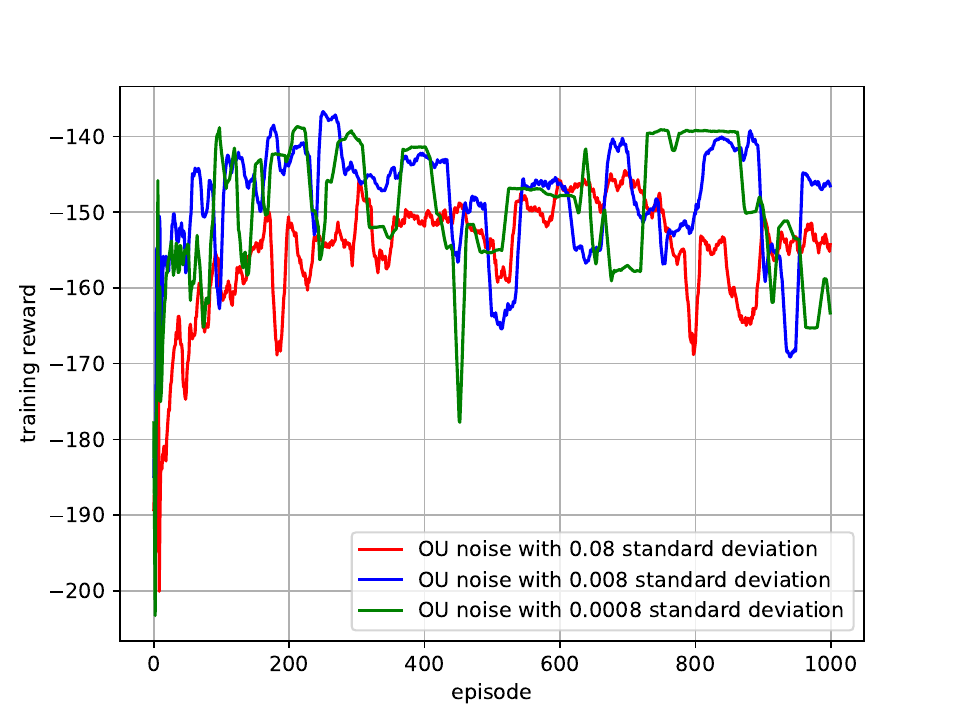} 
    \end{minipage} 
    \begin{minipage}{0.23\linewidth}
    \includegraphics[width=1\linewidth]{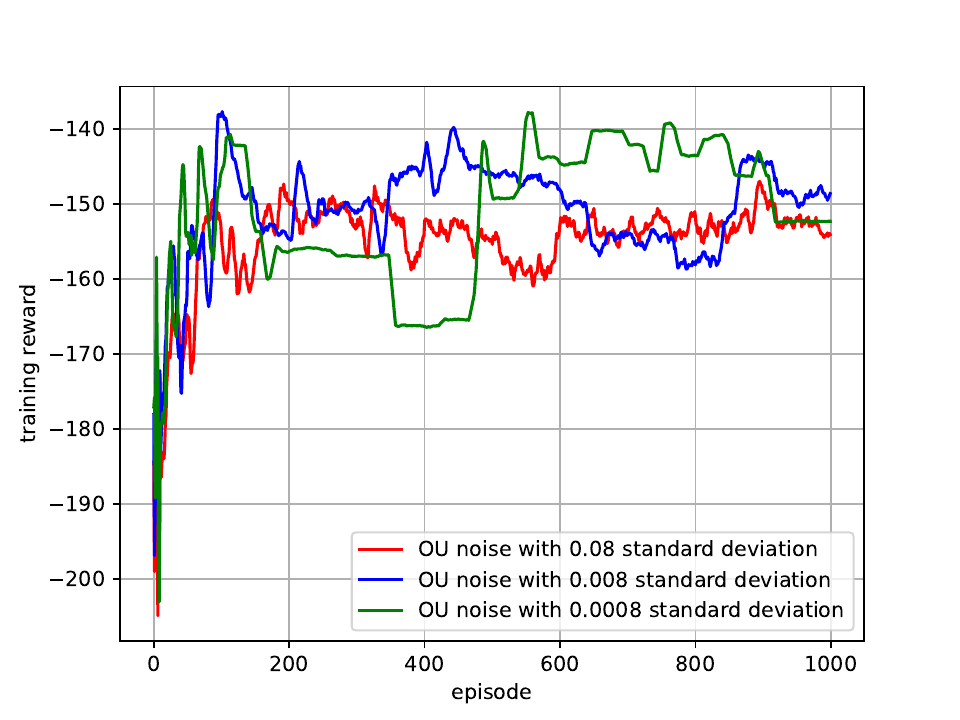} 
    \end{minipage}
    \caption{
    Exploitability computation training reward with $\alpha_{\text{actor}}=0.0005$. Batch size from left to right: 16, 32, 64, 128.
    }
    \label{fig:four-room-hyper 1}
\end{figure}

\begin{figure}[!htbp]
    \centering
    \begin{minipage}{0.23\linewidth}
    \includegraphics[width=1\linewidth]{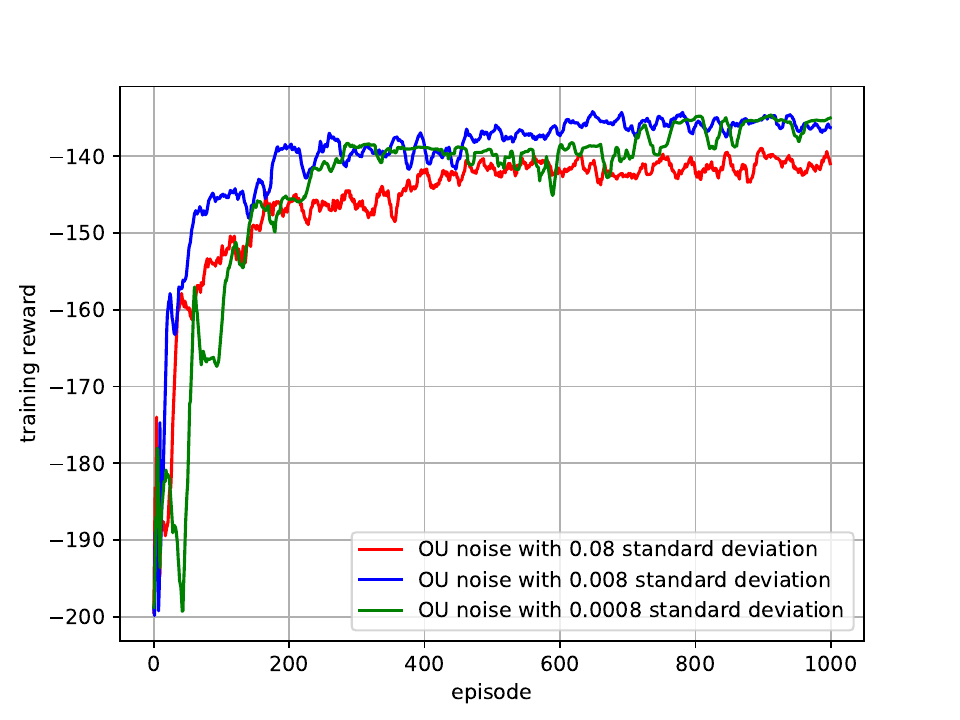} 
    \end{minipage}%
    \begin{minipage}{0.23\linewidth}
    \includegraphics[width=1\linewidth]{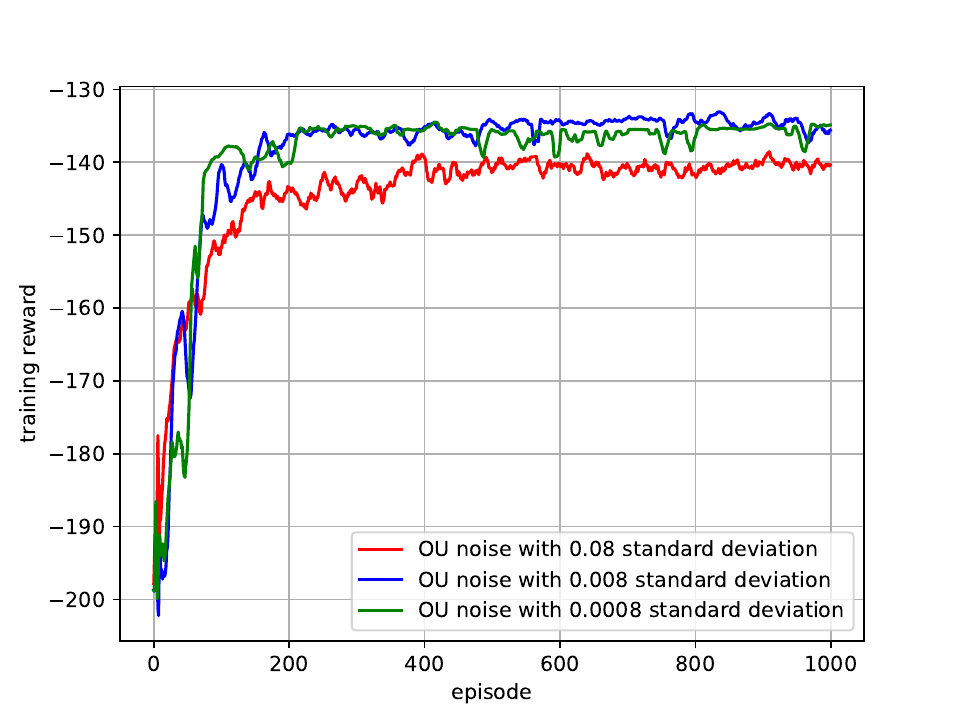} 
    \end{minipage} %
    \begin{minipage}{0.23\linewidth}
    \includegraphics[width=1\linewidth]{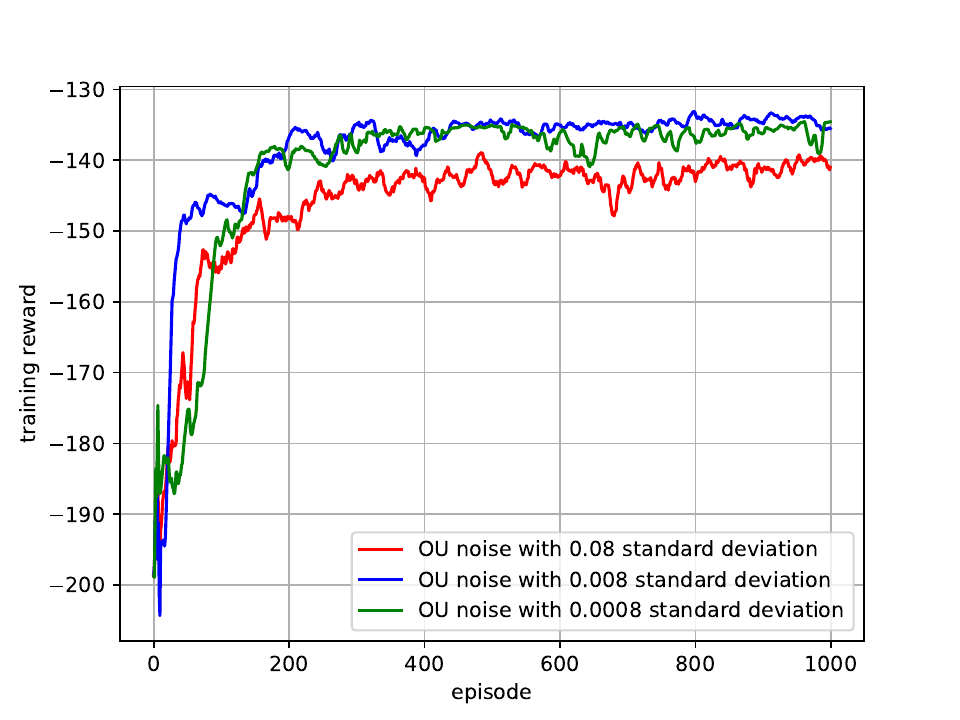} 
    \end{minipage}%
    \begin{minipage}{0.23\linewidth}
    \includegraphics[width=1\linewidth]{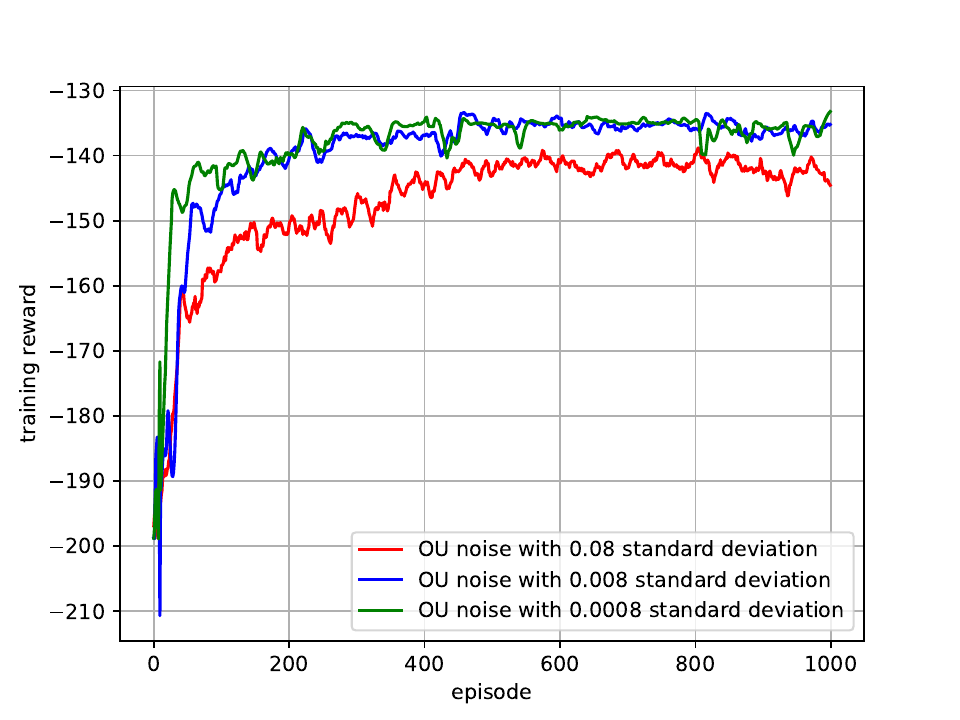} 
    \end{minipage}%
    \caption{
    Exploitability computation training reward with $\alpha_{\text{actor}}=5\times10^{-5}$. Batch size from left to right: 16, 32, 64, 128.
    }
    \label{fig:four-room-hyper 2}
\end{figure}
\begin{figure}[!htbp]
    \centering
    \begin{minipage}{0.23\linewidth}
    \includegraphics[width=1\linewidth]{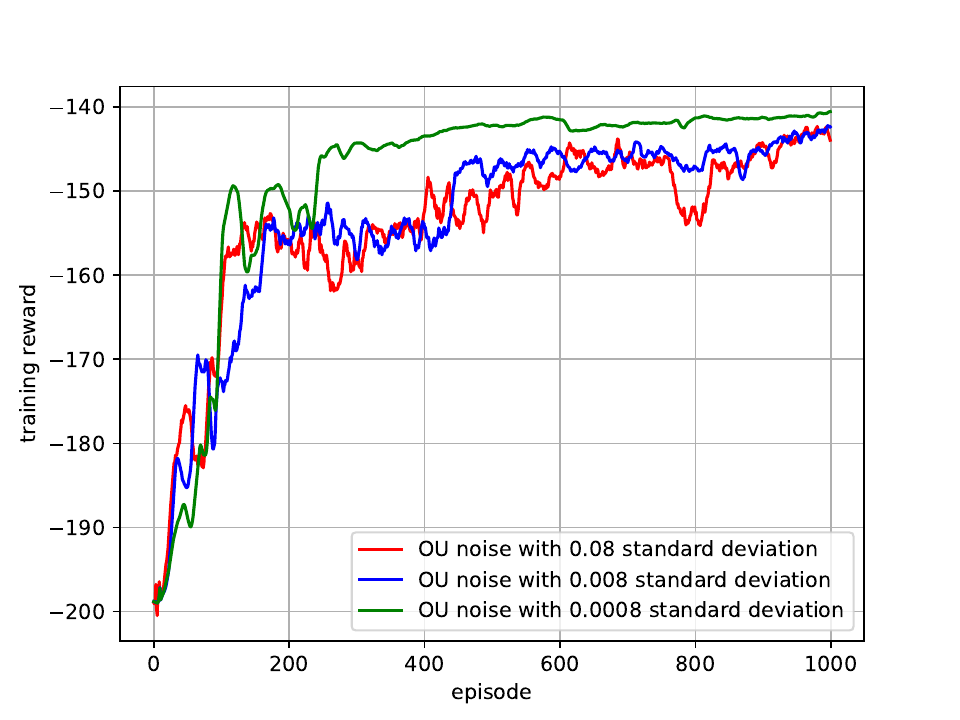} 
    \end{minipage}
    \begin{minipage}{0.23\linewidth}
    \includegraphics[width=1\linewidth]{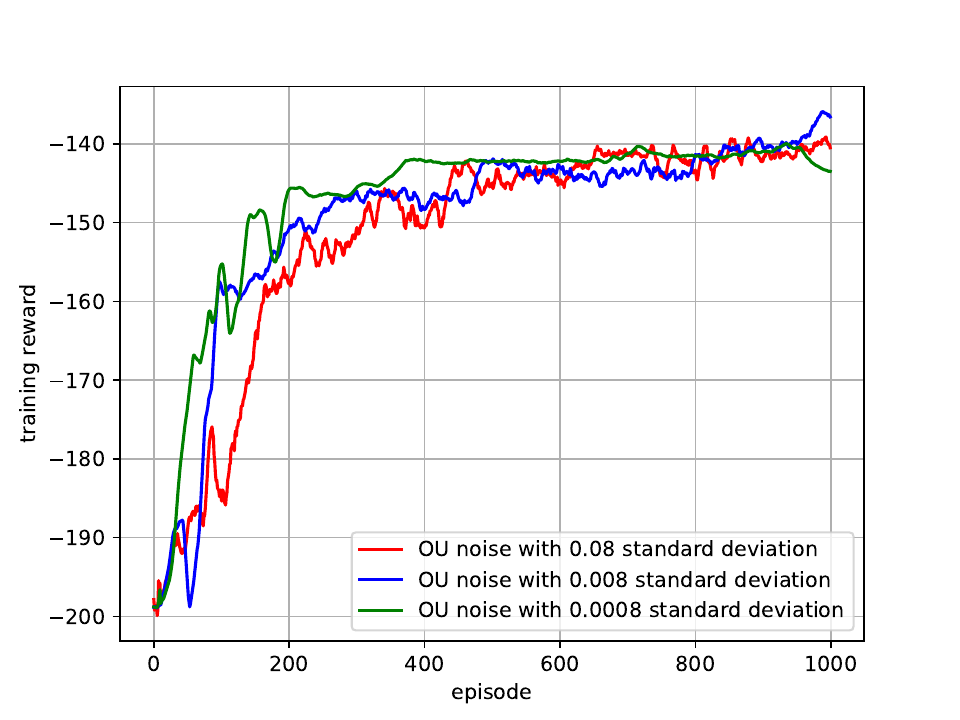} 
    \end{minipage} 
    \begin{minipage}{0.23\linewidth}
    \includegraphics[width=1\linewidth]{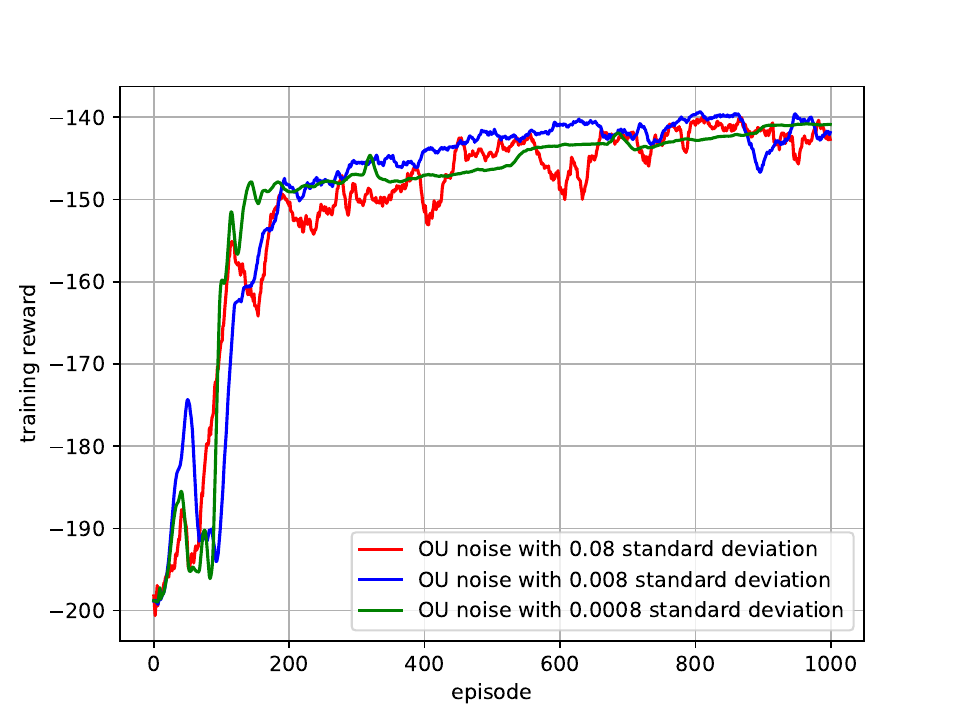} 
    \end{minipage} 
    \begin{minipage}{0.23\linewidth}
    \includegraphics[width=1\linewidth]{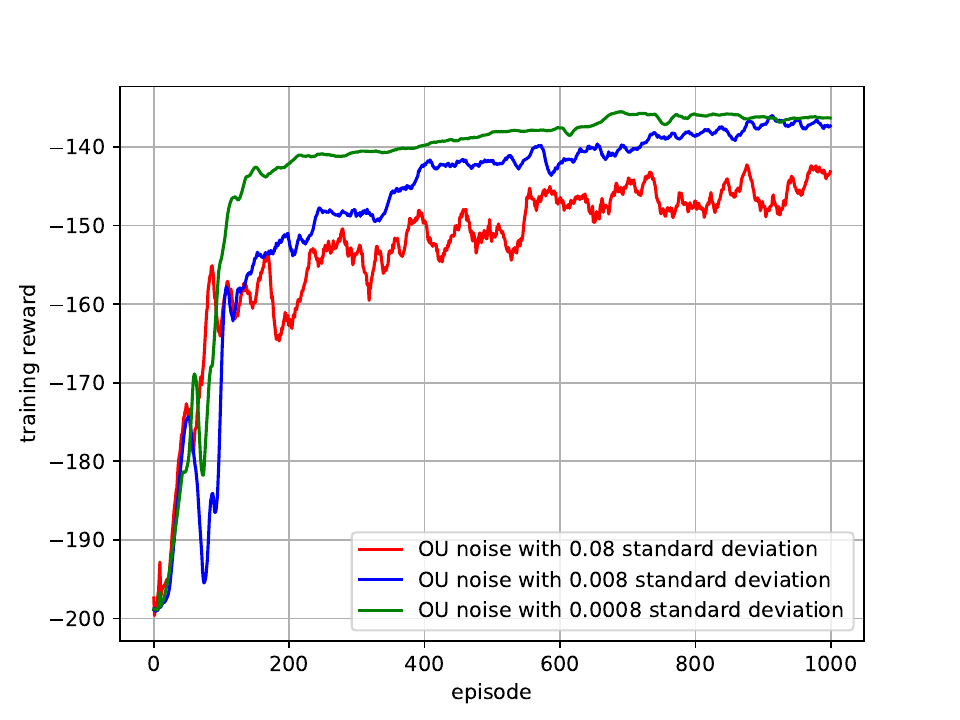} 
    \end{minipage} 
    \caption{
    Exploitability computation training reward with $\alpha_{\text{actor}}=5\times10^{-6}$. Batch size from left to right: 16, 32, 64, 128.
    }
    \label{fig:four-room-hyper 3}
\end{figure}
\clearpage
\subsection{Four-room with crowd aversion}
\begin{figure}[!htbp]
    \centering
    \begin{minipage}{0.23\linewidth}
    \includegraphics[width=1\linewidth]{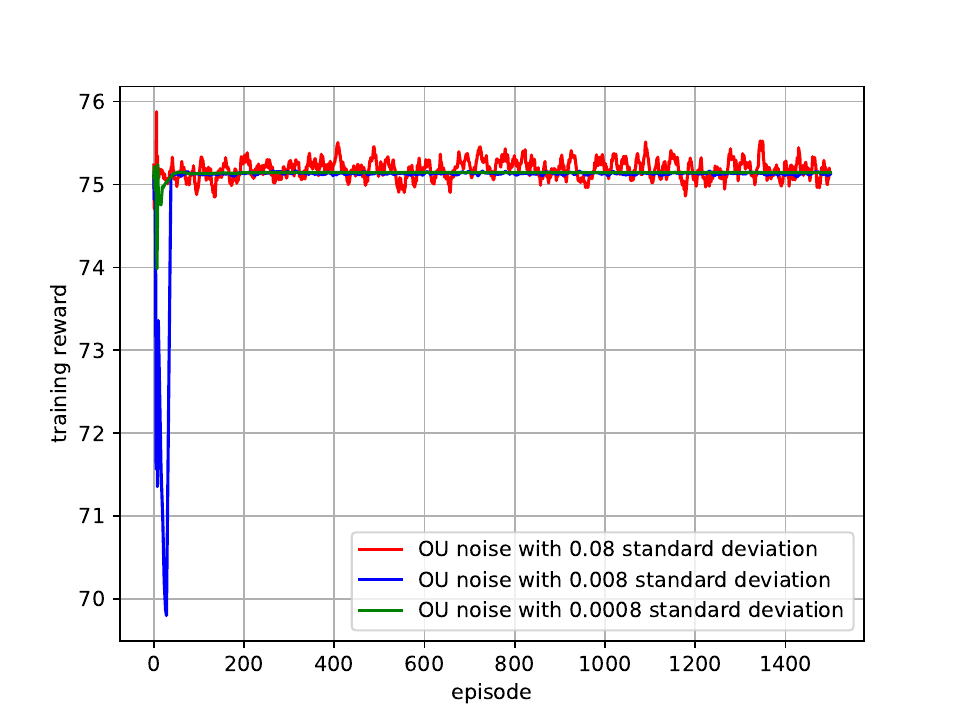} 
    \end{minipage}
    \begin{minipage}{0.23\linewidth}
    \includegraphics[width=1\linewidth]{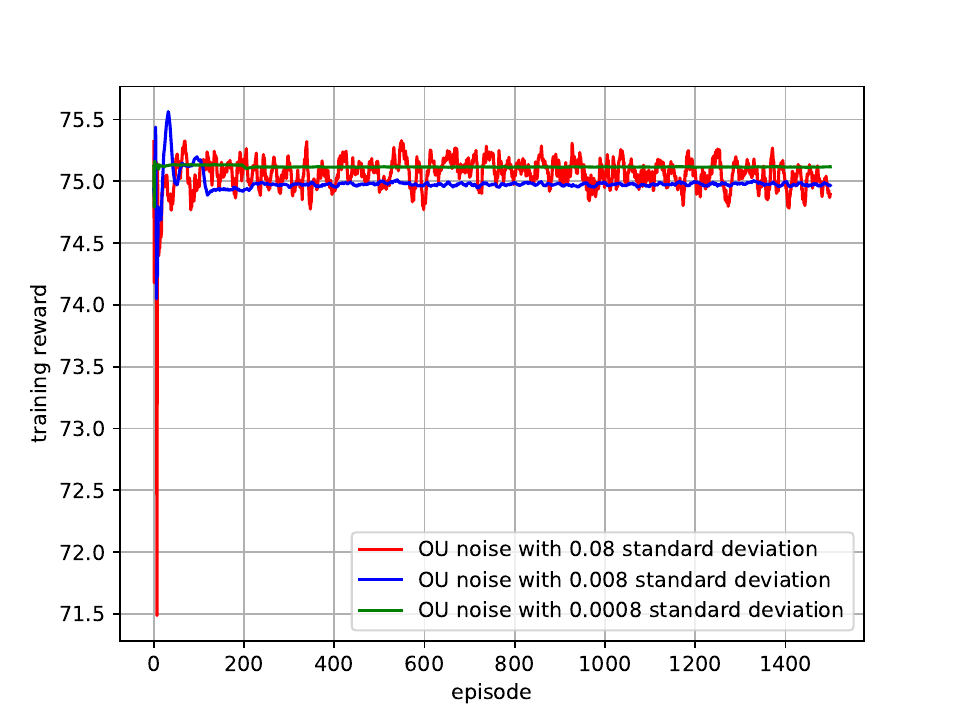} 
    \end{minipage} 
    \begin{minipage}{0.23\linewidth}
    \includegraphics[width=1\linewidth]{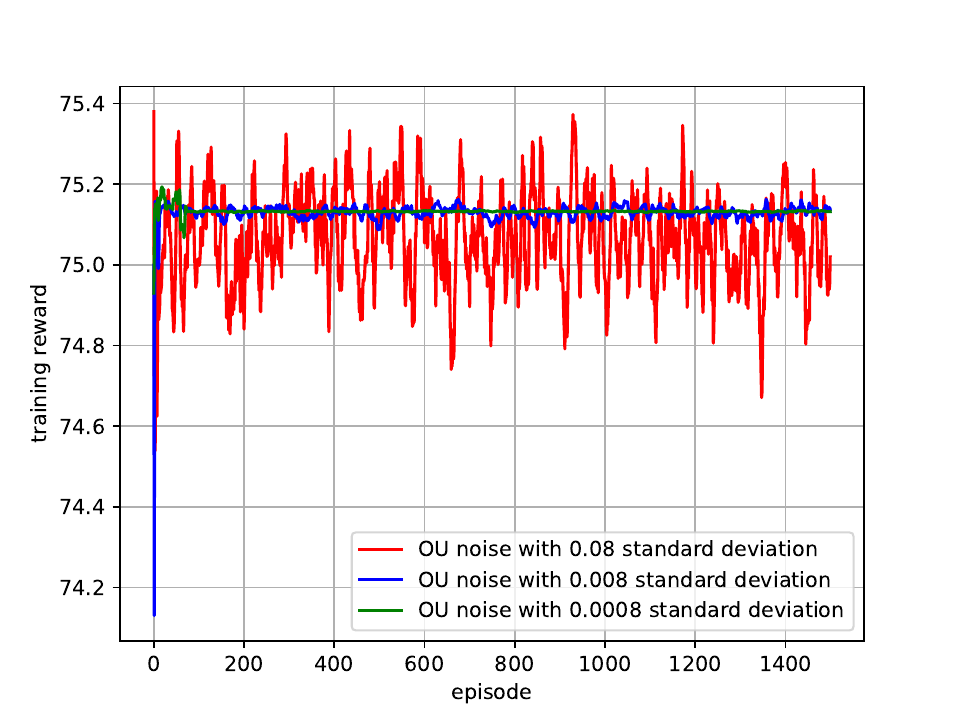} 
    \end{minipage} 
    \begin{minipage}{0.23\linewidth}
    \includegraphics[width=1\linewidth]{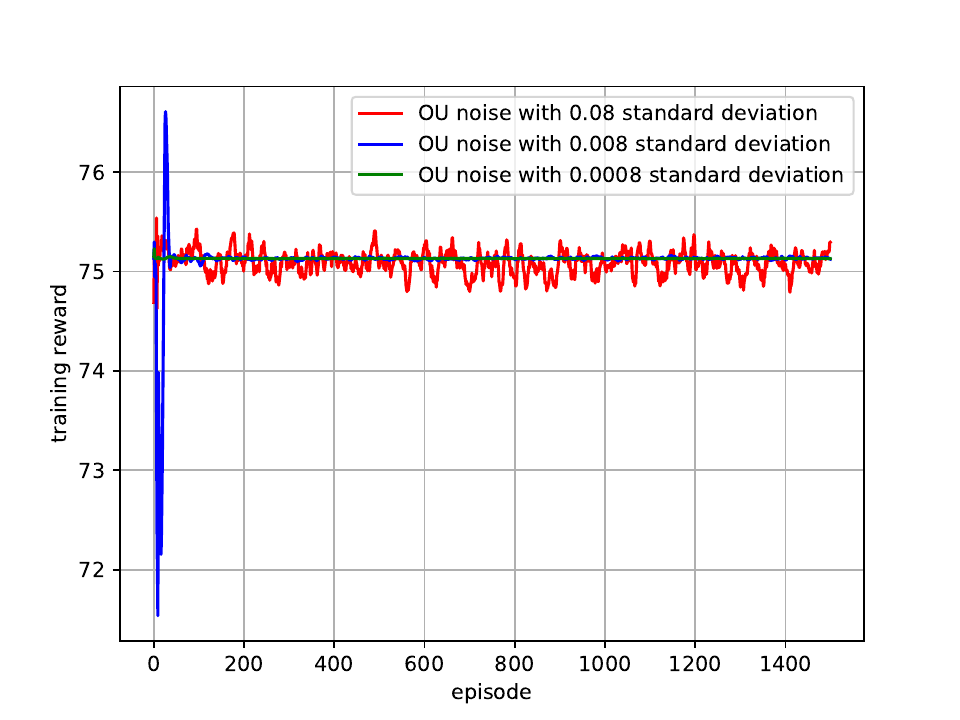} 
    \end{minipage} 
    \caption{
    Exploitability computation training reward with $\alpha_{\text{actor}}=0.005$. Batch size from left to right: 16, 32, 64, 128.
    }
    \label{fig:four-room-hyper 4}
\end{figure}

\begin{figure}[!htbp]
    \centering
    \begin{minipage}{0.23\linewidth}
    \includegraphics[width=1\linewidth]{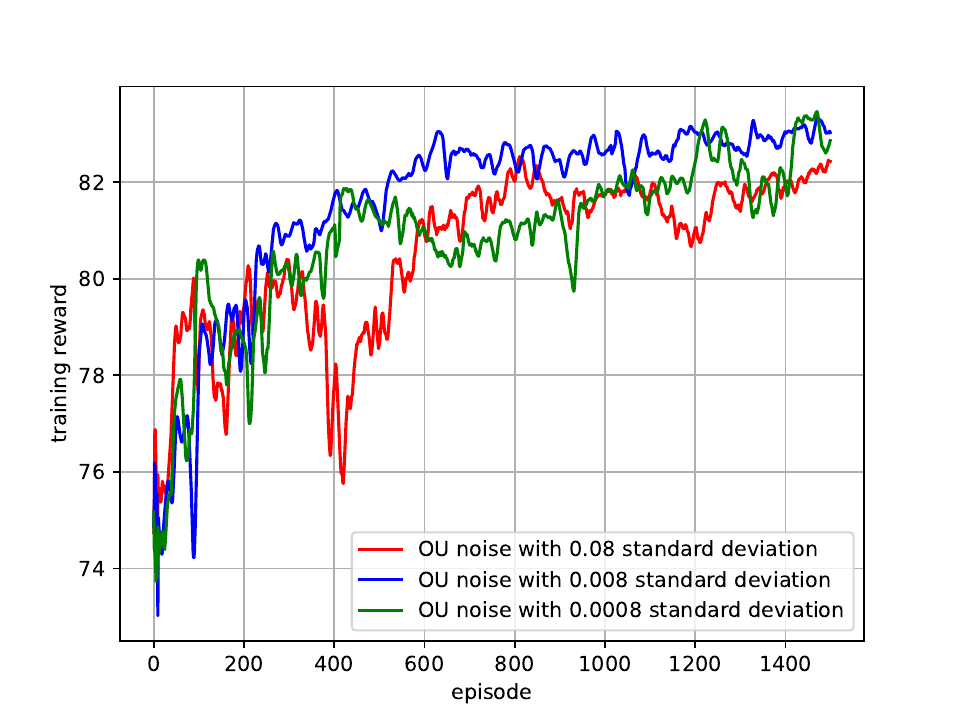} 
    \end{minipage}
    \begin{minipage}{0.23\linewidth}
    \includegraphics[width=1\linewidth]{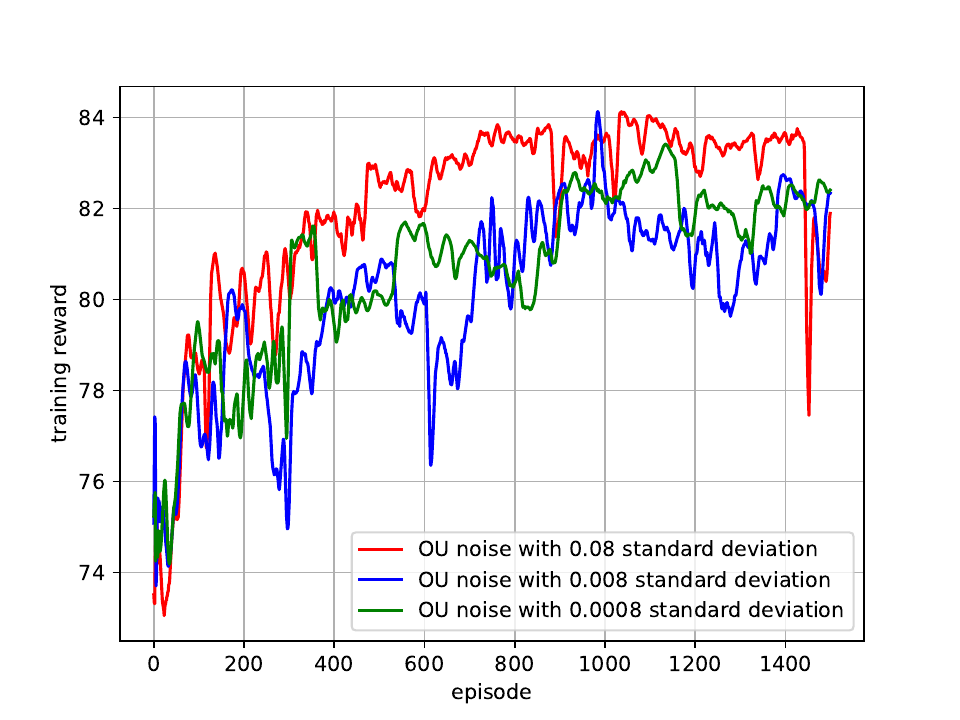} 
    \end{minipage} 
    \begin{minipage}{0.23\linewidth}
    \includegraphics[width=1\linewidth]{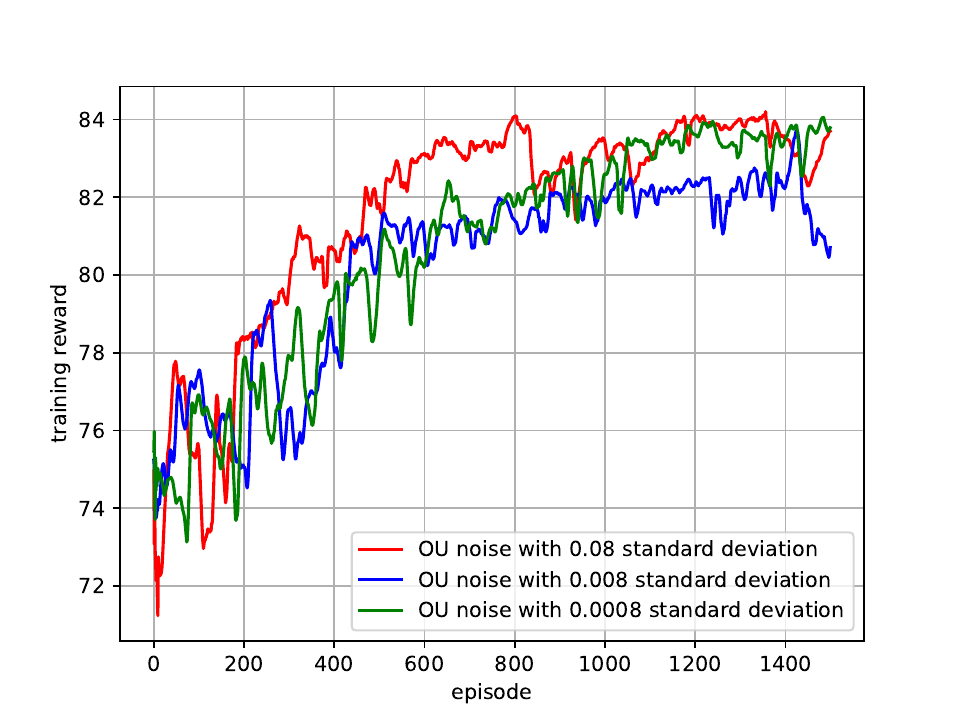} 
    \end{minipage} 
    \begin{minipage}{0.23\linewidth}
    \includegraphics[width=1\linewidth]{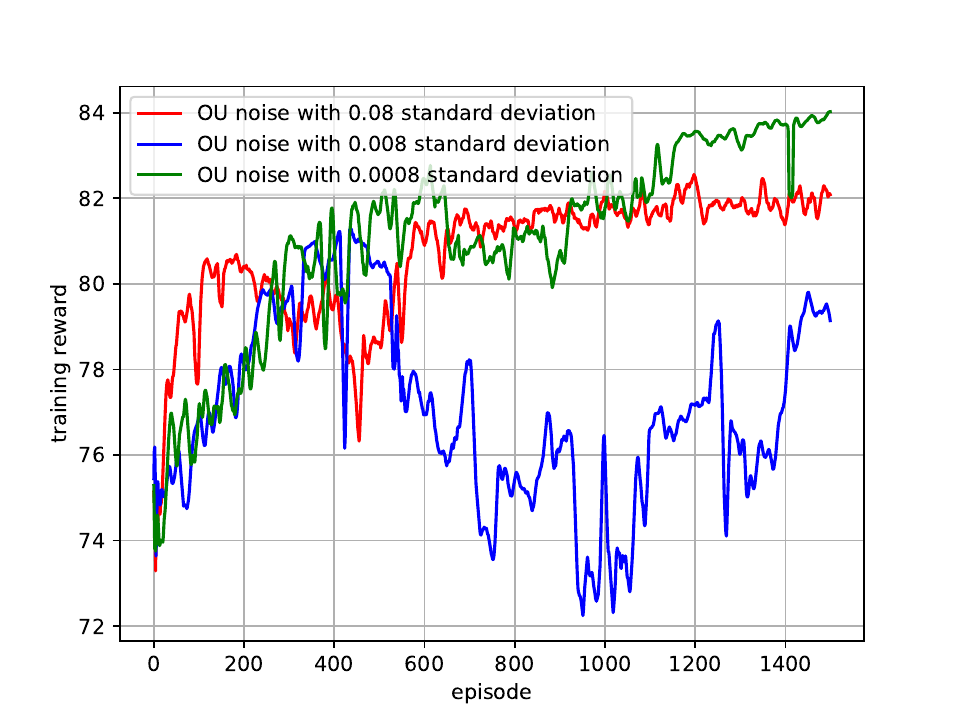} 
    \end{minipage}
    \caption{
    Exploitability computation training reward with $\alpha_{\text{actor}}=0.0005$. Batch size from left to right: 16, 32, 64, 128.
    }
    \label{fig:four-room-hyper 5}
\end{figure}

\begin{figure}[!htbp]
    \centering
    \begin{minipage}{0.23\linewidth}
    \includegraphics[width=1\linewidth]{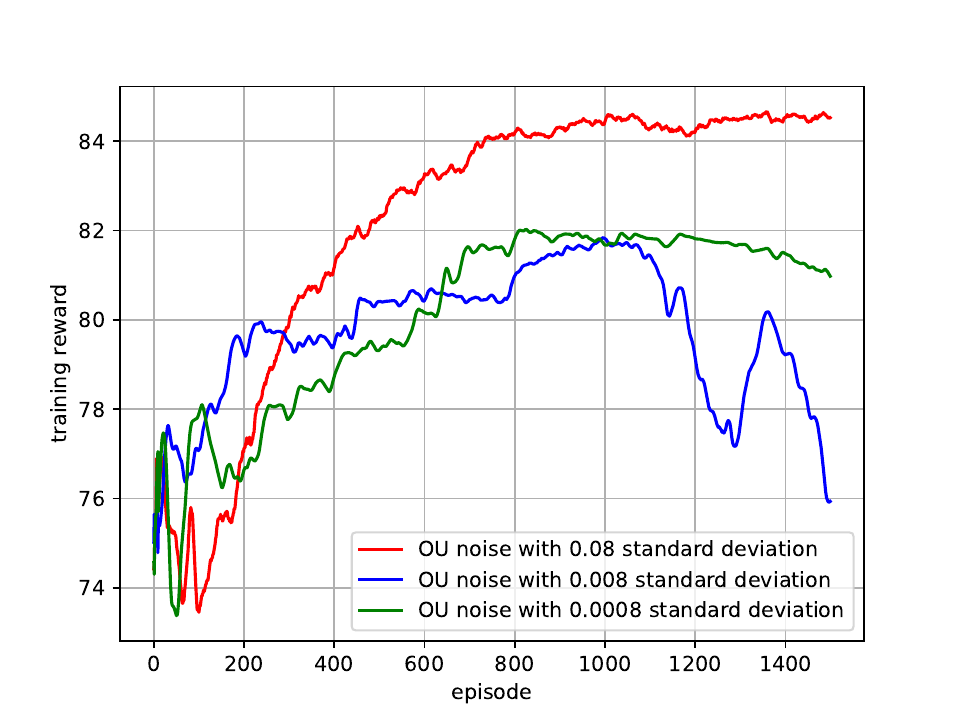} 
    \end{minipage}%
    \begin{minipage}{0.23\linewidth}
    \includegraphics[width=1\linewidth]{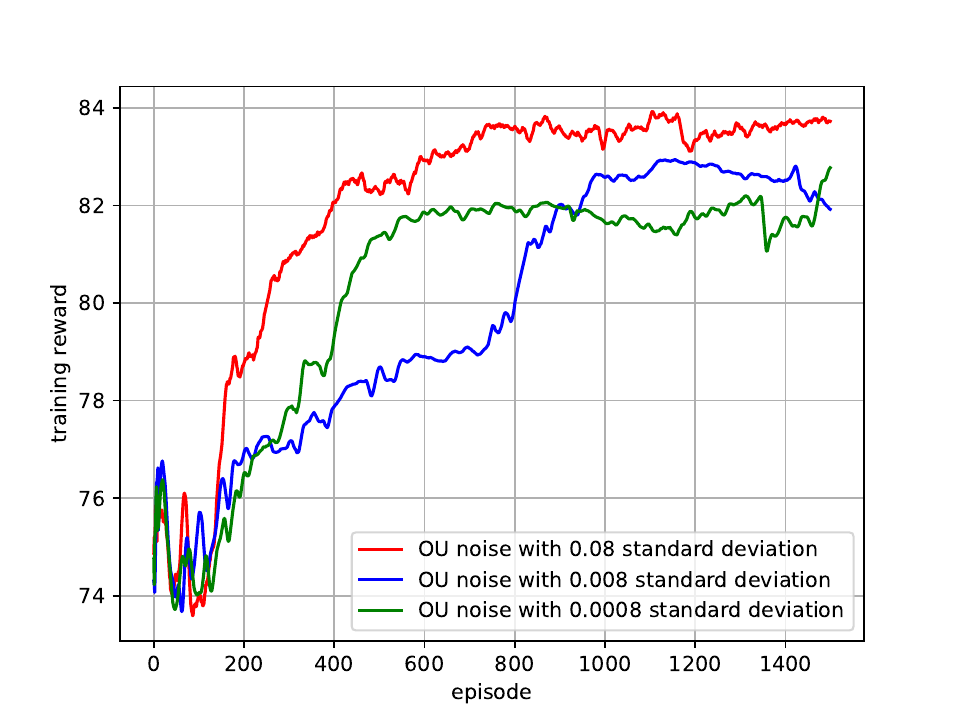} 
    \end{minipage} %
    \begin{minipage}{0.23\linewidth}
    \includegraphics[width=1\linewidth]{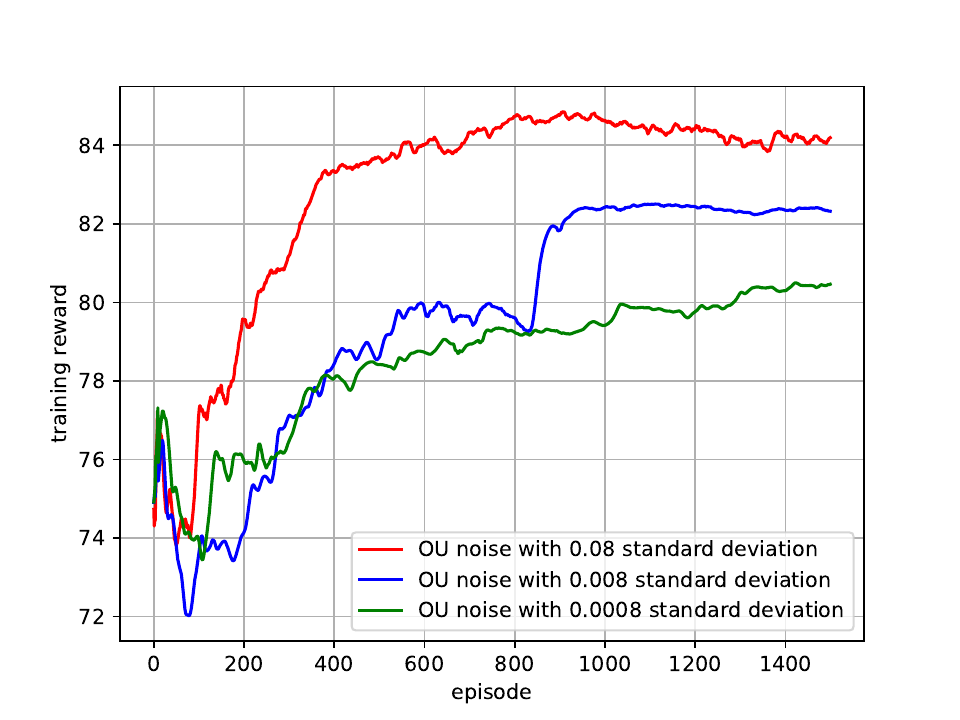} 
    \end{minipage}%
    \begin{minipage}{0.23\linewidth}
    \includegraphics[width=1\linewidth]{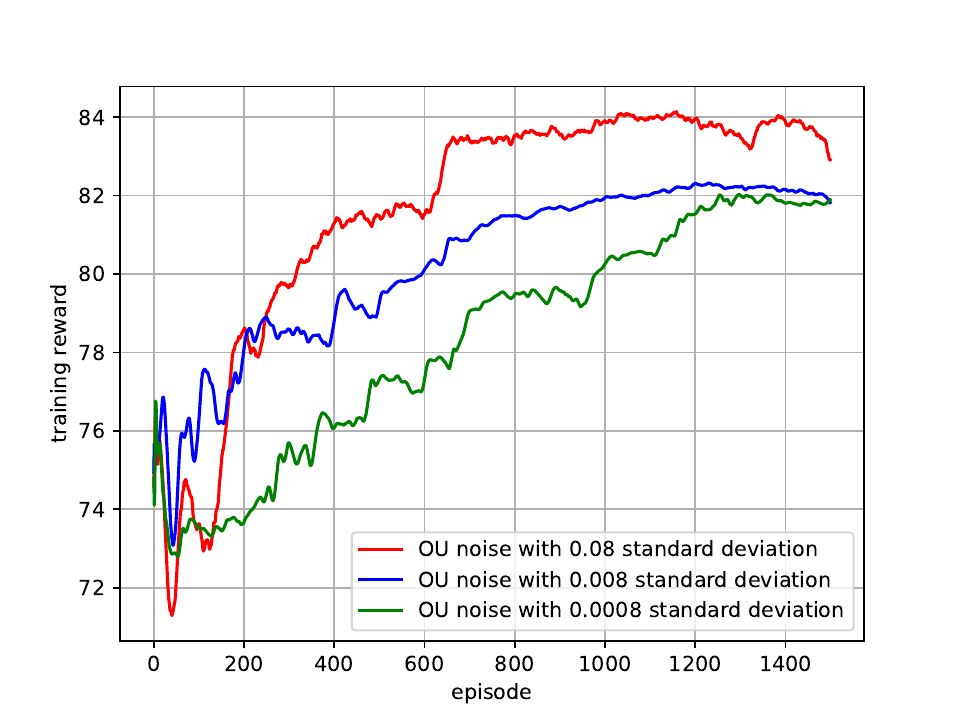} 
    \end{minipage}%
    \caption{
    Exploitability computation training reward with $\alpha_{\text{actor}}=5\times 10^{-5}$. Batch size from left to right: 16, 32, 64, 128.
    }
    \label{fig:four-room-hyper 6}
\end{figure}
\begin{figure}[!htbp]
    \centering
    \begin{minipage}{0.23\linewidth}
    \includegraphics[width=1\linewidth]{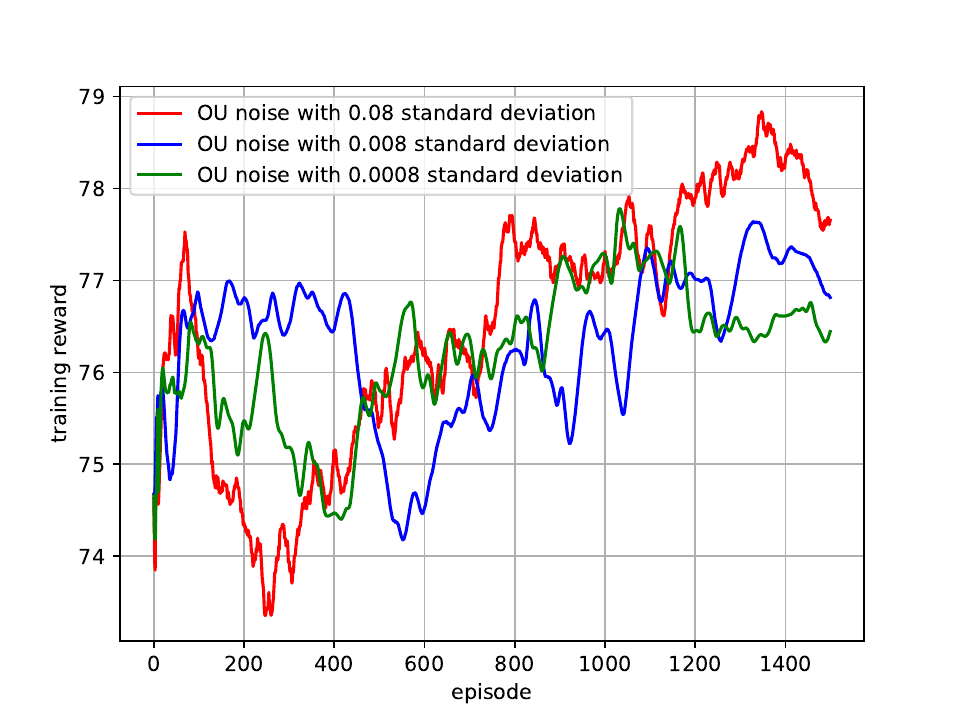} 
    \end{minipage}
    \begin{minipage}{0.23\linewidth}
    \includegraphics[width=1\linewidth]{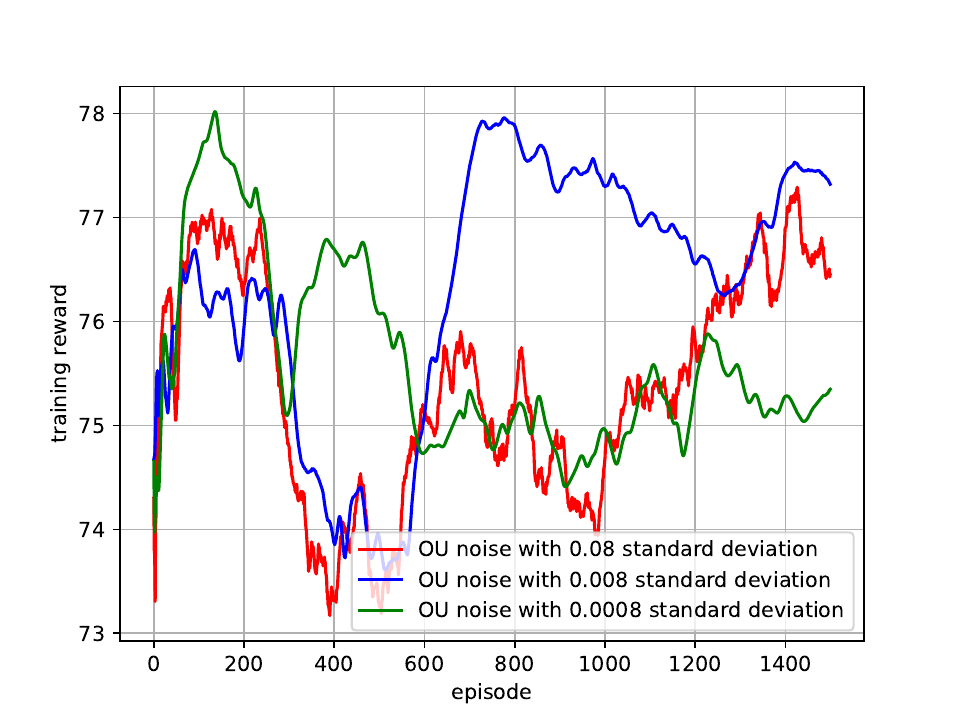} 
    \end{minipage} 
    \begin{minipage}{0.23\linewidth}
    \includegraphics[width=1\linewidth]{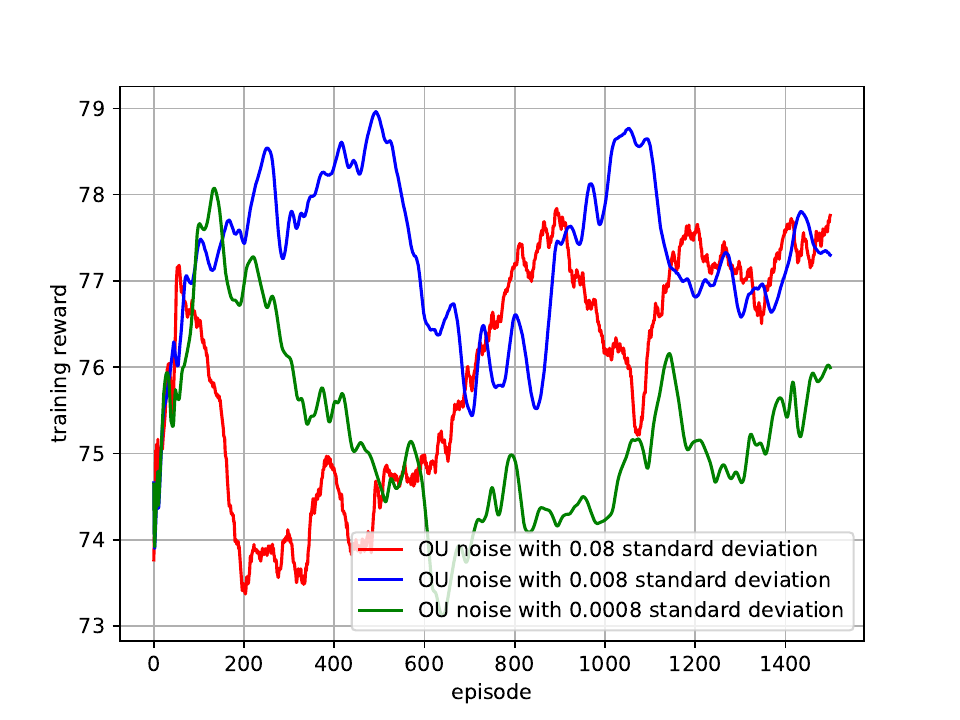} 
    \end{minipage} 
    \begin{minipage}{0.23\linewidth}
    \includegraphics[width=1\linewidth]{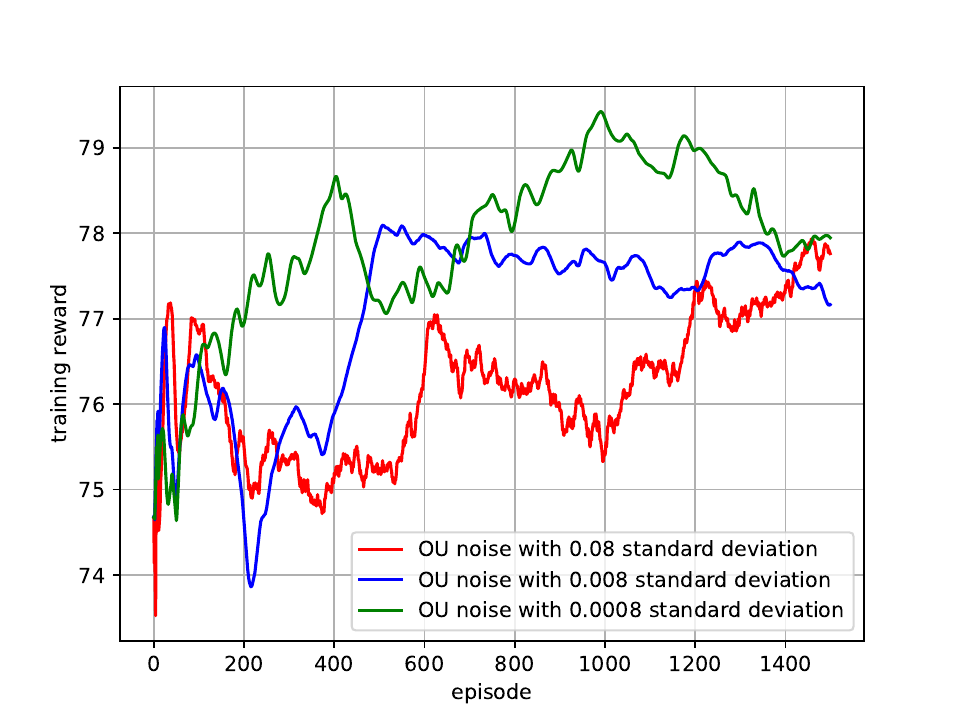} 
    \end{minipage} 
    \caption{
    Exploitability computation training reward with $\alpha_{\text{actor}}=5\times 10^{-6}$. Batch size from left to right: 16, 32, 64, 128.
    }
    \label{fig:four-room-hyper 7}
\end{figure}
\end{document}